\documentclass[journal=jacsat,manuscript=article]{achemso}

\usepackage[T1]{fontenc} % Modern font encoding

\usepackage[table,x11names,dvipsnames]{xcolor}
\usepackage{mdframed}

\usepackage{amsmath}
\usepackage{amsfonts}
\usepackage{chemformula} % Formula subscripts using \ch{}

\usepackage{graphicx}
\usepackage{pdfpages}
\usepackage{float}
\usepackage{overpic} % Overlay text on images

\usepackage{caption}
\usepackage{subcaption}

\usepackage{booktabs}
\usepackage{array}
\usepackage{makecell}
\usepackage{multirow}
\usepackage[table,x11names,dvipsnames]{xcolor}
\usepackage{siunitx}
\usepackage{afterpage}

\makeatother

\mdfsetup{%
	backgroundcolor=yellow!30,
	linecolor=yellow!60!black,
	innertopmargin=10pt,
	innerbottommargin=10pt,
	innerleftmargin=8pt,
	innerrightmargin=8pt,
	skipabove=6pt,
	skipbelow=6pt,
	leftmargin=0pt,
	rightmargin=0pt
}

\usepackage{multirow}
\usepackage{array} % Line breaks in headers
\usepackage{makecell}

\usepackage{cleveref}
\usepackage{comment}

\usepackage{titlesec}
\usepackage[section]{placeins}
\usepackage{setspace}

\author{Manuela L. Kim}
\alsoaffiliation{Department of Materials Chemistry, Faculty of Engineering, Shinshu University, 4-17-1 Wakasato, Nagano 380-8553, Japan}
\altaffiliation{These authors contributed equally to this work.}

\author{Mauricio E. Calvo}
\alsoaffiliation{Instituto de Ciencia de Materiales de Sevilla (ICMS-CSIC-US), c/ Americo Vespucio 49, 41092 Sevilla, Spain}
\altaffiliation{These authors contributed equally to this work.}

\author{Katsuya Teshima}
\affiliation{Institute for Aqua Regeneration, Shinshu University, 4-17-1 Wakasato, Nagano 380-8553, Japan}
\alsoaffiliation{Department of Materials Chemistry, Faculty of Engineering, Shinshu University, 4-17-1 Wakasato, Nagano 380-8553, Japan}
\alsoaffiliation{Research Initiative for Supra-Materials (RISM), Interdisciplinary Cluster for Cutting Edge Research (ICCER), Shinshu University, Nagano 380-8553, Japan}

\author{Fabio La Mattina}
\alsoaffiliation{Empa, Swiss Federal Laboratories for Materials Science and Technology, Überlandstrasse 129, 8600 Dübendorf, Switzerland}

\author{Eugenio H. Otal}
\affiliation{Department of Materials Chemistry, Faculty of Engineering, Shinshu University, 4-17-1 Wakasato, Nagano 380-8553, Japan}
\alsoaffiliation{Research Initiative for Supra-Materials (RISM), Interdisciplinary Cluster for Cutting Edge Research (ICCER), Shinshu University, Nagano 380-8553, Japan}
\alsoaffiliation{Institute for Aqua Regeneration, Shinshu University, 4-17-1 Wakasato, Nagano 380-8553, Japan}
\email{eugenio_otal@shinshu-u.ac.jp}

\phone{+81 (0)268-21-5499}
\fax{+81 (0)268-21-5499}

\title[Beyond Panchromatic Absorption]
{Beyond Panchromatic Absorption: Deciphering the Excited-State Maze from Light Absorption to Photocatalysis in Dye-Sensitized MOFs}

	
\begin{document}
	
\newpage
% \linenumbers  
	
\begin{abstract}
	Metal--organic frameworks (MOFs) are promising photocatalysts whose visible-light absorption can be extended through linker functionalization; however, a red-shifted absorption edge does not guarantee enhanced efficiency. Here, we establish a multi-spectroscopic framework to decipher the photophysical fate of photoexcited states in diazo-sensitized UiO-66 using 2D photoluminescence (PL/PLE) mapping and wavelength-resolved continuous-wave X-band photo-EPR. By correlating visible absorption with photo-EPR and PLE action spectra, we distinguish a \textit{productive red shift} from a \textit{non-productive emissive red shift}.
	
	In highly active UiO-66-Anisole (97\% activity relative to \ch{TiO2}), photo-EPR tracks the new absorption band, confirming that excitation populates a charge-transfer pathway yielding persistent, spin-separated states. Conversely, poorly active UiO-66-$\beta$-naphthol (4\%) exhibits an extended visible absorption tracked by PLE but not photo-EPR. This reflects excitation trapping in a localized state caused by an \textit{ortho}-$\mathrm{OH}$ group forming a rigid intramolecular hydrogen bond, which locks the keto-hydrazone tautomer and disrupts the conjugated azo bridge.
	
	A new optical overlap descriptor ($S_{\mathrm{Exc}}$) quantitatively captures this trade-off across the series. We demonstrate that photosensitizer design must suppress rigid tautomeric traps and target specific charge-transfer manifolds ($\lambda \le 500$~nm) rather than merely maximizing apparent panchromatic absorption breadth.
\end{abstract}
	
\section{Introduction}

The conversion of solar energy into chemical work through visible-light photocatalysis is central to sustainable environmental remediation, yet the efficiency of this process in heterogeneous systems rarely approaches theoretical limits.\cite{linsebigler_photocatalysis_1995} In dye-sensitized architectures, overall performance depends not merely on the spectral breadth of light absorption, but fundamentally on the photophysical pathways followed by the photoexcited population.\cite{ORegan1991,Hagfeldt2010} Charge separation must compete directly against multiple fast deactivation channels---fluorescence, internal conversion, intersystem crossing, defect trapping, and back-electron transfer---whose branching ratios and lifetimes are dictated by the molecular and local framework environment.\cite{Hagfeldt2010,Boschloo2009} Maximizing the yield of long-lived charge-separated states therefore requires precise molecular-level control over these competing relaxation pathways, a longstanding challenge in dye-sensitized systems.\cite{Boschloo2009}

Metal--organic frameworks (MOFs) offer a structurally precise platform to address this challenge: their crystalline, porous scaffolds organize metal-oxide secondary building units (SBUs) and organic linkers with atomic-level regularity, enabling rational control over both light absorption and charge-transfer pathways.\cite{Eddaoudi2002} Among them, UiO-66-NH$_2$ is particularly attractive: its $\text{Zr}_6\text{O}_4(\text{OH})_4$ nodes combine exceptional thermal and chemical stability\cite{cavka_new_2008} with a discrete, molecular-like electronic structure that facilitates the spectroscopic tracking of photoexcited charge carriers.\cite{Wang2018} Post-synthetic modification (PSM) of the amino handles via diazonium coupling enables the covalent grafting of extended $\pi$-conjugated chromophores onto the linkers, red-shifting the optical absorption edge well into the visible range without compromising framework crystallinity.\cite{Cohen2012,KalajCohen2020,Mondal2023} This modular strategy decouples the design of the light-harvesting antenna from framework stability, permitting systematic variation of the sensitizer's electronic structure within a single, well-defined Zr-oxo scaffold.

The central design challenge, however, is that broadening light absorption does not guarantee efficient photocatalysis. Charge injection from the photoexcited linker into the framework catalytic nodes competes directly with localized radiative and non-radiative dissipation. Recent photophysical studies combining time-resolved photoluminescence (TRPL) and TDDFT calculations have demonstrated that productive charge-transfer (CT) configurations are predominantly populated in the higher-energy visible window ($\lambda \le 500$~nm), in agreement with theoretical predictions that productive CT states lie at higher transition energies than the lowest localized excitons confined to the azo bridge.\cite{kultaeva_mechanistic_2026,nasalevich_electronic_2016} Consequently, preserving an electronically coupled donor--$\pi$--acceptor (D--$\pi$--A) pathway is essential to ensure that photoexcitation feeds the charge-separating CT manifold rather than rapidly relaxing into non-productive emissive traps.\cite{roys_efficiency_2025} In this context, the relative position of auxiliary hydroxyl ($-\mathrm{OH}$) substituents plays a decisive role: intramolecular hydrogen bonding can trigger azo/hydrazone (enol/keto) tautomerism that directly disrupts the conjugated $-\mathrm{N=N}-$ double bond, turning an intended charge-transfer bridge into a localized non-conductive sink.

While direct ligand-to-metal charge transfer (LMCT) is well established as the operative visible-light mechanism in Ti-based frameworks such as NH$_2$-MIL-125(Ti),\cite{Zhang2015} the high energy of the empty $4d$ orbitals in $\mathrm{Zr^{IV}}$ renders direct visible-light LMCT inaccessible in Zr-based analogues.\cite{Karimi2017} In diazo-sensitized UiO-66, charge separation cannot rely on metal reduction and must instead proceed through linker-based charge-transfer configurations ($^3\mathrm{CT}$) that generate persistent, spin-protected paramagnetic intermediates. Circumventing the intrinsic limitations of Zr-oxo nodes therefore requires a precise understanding of the photogenerated radical pathways and the structural factors that prevent excitons from becoming trapped in localized non-conductive states \cite{kultaeva_mechanistic_2026}.

Here, we report a wavelength-resolved, multi-spectroscopic investigation of a series of diazo-coupled UiO-66 materials---incorporating anisole, $\alpha$-naphthol (AN), $\beta$-naphthol (BN), diphenylamine (DPA), and $N,N$-dimethylaniline (NNDMA)---combining diffuse reflectance spectroscopy (DRS), steady-state 2D fluorescence excitation--emission mapping, wavelength-resolved continuous-wave X-band photo-EPR (400--700~nm), and methylene blue (MB) photodegradation measurements. By correlating the functionalization-induced visible absorption with both the photoluminescence excitation (PLE) profiles and the photo-EPR action spectra, this study establishes a competitive photophysical branching framework to evaluate how photoexcited carriers partition among productive charge separation, radiative deactivation, and non-radiative decay. 

The findings yield three complementary design guidelines for dye-sensitized Zr-MOFs (and related wide-bandgap architectures lacking direct visible LMCT): (i)~\textbf{Minimization of radiative sinks:} Optical overlap between absorption and excitation profiles ($S_{\mathrm{Exc}} \to 0$) is a necessary criterion to prevent trapping of photoexcited carriers in localized non-conductive channels;(ii)~\textbf{Control of conformational and tautomeric equilibria:} Substituent geometry must avoid rigid tautomeric locks that disrupt the conjugated azo double bond, break donor--$\pi$--acceptor electronic connectivity, and open radiative channels; and (iii)~\textbf{Exploitation of the high-energy CT window:} Because charge separation in Zr-MOFs relies on populating linker-based charge-transfer states rather than direct node reduction, photosensitization must specifically target the short-wavelength visible region ($\lambda \le 500$~nm) where persistent, catalytically active carrier populations accumulate.

Furthermore, we examine the background radical species associated with post-synthetic modification conditions and demonstrate that their EPR signatures can be reliably distinguished from light-induced paramagnetic intermediates. To ensure clarity and mechanistic flow, our primary discussion focuses on three distinct archetypes \cite{leone_il_1967} that embody the diverse photophysical pathways: (i)~\textit{The Good}: UiO-66-Anisole, which exhibits a moderate visual color shift but achieves the highest photocatalytic activity (97\%) through productive charge separation; (ii)~\textit{The Bad}: UiO-66-BN ($\beta$-naphthol), which displays extended visible absorption but negligible photocatalytic activity (4\%) due to rapid excitation trapping in its locked keto-hydrazone form; and (iii)~\textit{The Unexpected}: the parent UiO-66-NH$_2$, which reveals a photooxidation-driven spectral broadening that modulates its radical generation under prolonged irradiation. Intermediate cases ($\alpha$-naphthol, DPA, NNDMA) are systematically mapped within this branching landscape.

\section{Results and discussion}

\subsection{Structural stability and surface-governed reactivity}

Post-synthetic modification (PSM) of UiO-66-NH$_2$ was carried out via diazotization of the 2-aminoterephthalate linkers followed by electrophilic azo coupling with electron-rich aromatic substrates, yielding a systematic series of donor--$\pi$--acceptor (D--$\pi$--A) sensitized MOFs (\Cref{fig:ligands_ordered}): UiO-66-Anisole, UiO-66-$\alpha$-naphthol (AN), UiO-66-$\beta$-naphthol (BN), UiO-66-diphenylamine (DPA), and UiO-66-$N,N$-dimethylaniline (NNDMA).\cite{otal_panchromatic_2016} Powder X-ray diffraction (pXRD) patterns (\Cref{fig:XRD}) confirm that the face-centered cubic framework of UiO-66 is fully preserved across the entire series, without detectable phase degradation, loss of crystallinity, or collapse of long-range structural order.

Thermogravimetric analysis (TGA, \Cref{tab:tga_summary_narrow,tab:tga_mr_yield_narrow}) provides a model-dependent bulk estimate of the apparent diazo-coupling yield, modeled as a two-component mixture of dye-functionalized and hydrolyzed ($\mathrm{-OH}$) linkers.  Within this model, UiO-66-Anisole exhibits a lower apparent bulk functionalization degree ($\sim 18$\%) than the bulkier naphthol and aniline derivatives (up to $\sim 45$\%), consistent with the sterically demanding nature of the larger coupling partners. Nitrogen physisorption isotherms at 77~K (\Cref{tab:SI-BET}) show a moderate decrease in Brunauer--Emmett--Teller (BET) specific surface area from 980~m$^2$/g for pristine UiO-66-NH$_2$ to 620--780~m$^2$/g for the functionalized materials, confirming that the grafted chromophores occupy pore space while maintaining accessible porosity.

Because methylene blue (MB, molecular dimensions $\approx 14.3 \times 6.1 \times 4.0$~\AA) is sterically excluded from entering the narrow triangular pore windows ($\approx 6$~\AA) of the UiO-66 framework, photocatalytic dye degradation occurs predominantly at the outer crystal surface and pore apertures. Consequently, the high catalytic performance of UiO-66-Anisole does not require complete bulk functionalization; rather, the compact methoxy group facilitates clean surface coupling and unobstructed substrate access, whereas bulkier donors may introduce local steric congestion and defect-associated recombination sites near the external crystal boundary.

\subsection{The Good: UiO-66-Anisole}

\begin{figure}[H]
	\centering
	\begin{subfigure}[b]{0.48\linewidth}
		\centering
		\begin{overpic}[width=\linewidth]{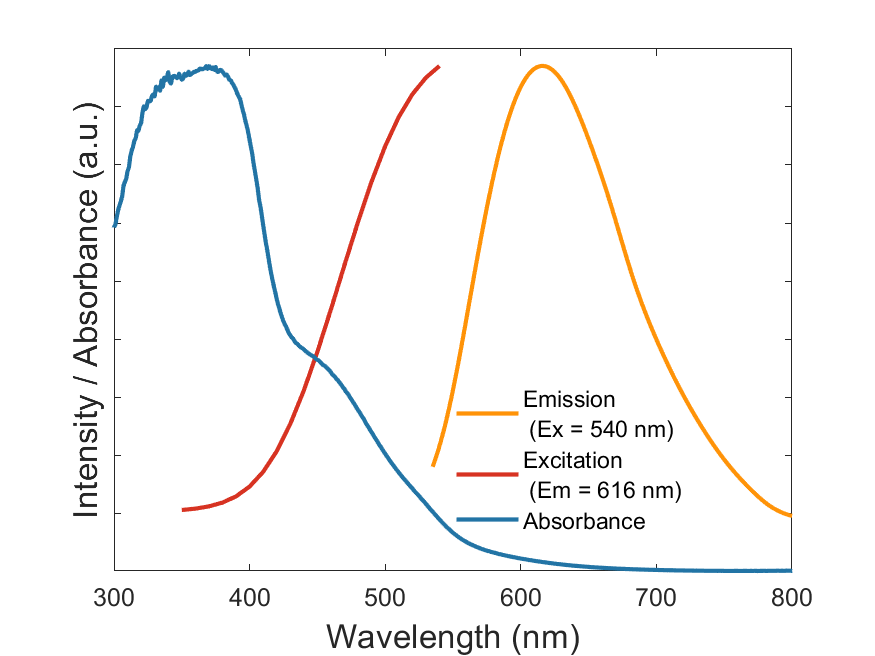}
			\put(0,65){\color{black}\textbf{(A)}}
		\end{overpic}
	\end{subfigure}
	\begin{subfigure}[b]{0.48\linewidth}
		\centering
		\begin{overpic}[width=\linewidth]{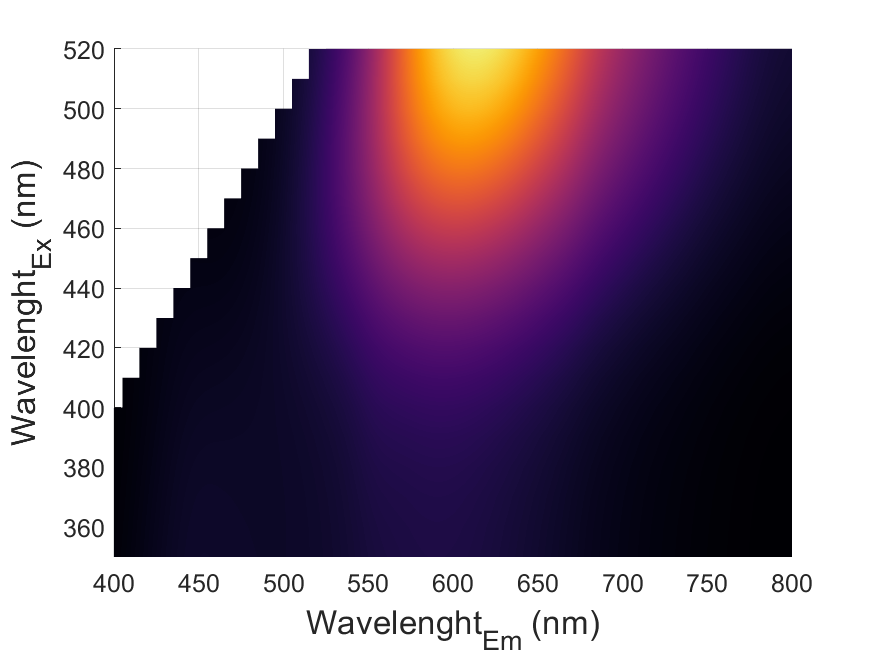}
			\put(0,65){\color{black}\textbf{(B)}}
		\end{overpic}
	\end{subfigure}
	
	\begin{subfigure}[b]{0.48\linewidth}
		\centering
		\begin{overpic}[width=\linewidth]{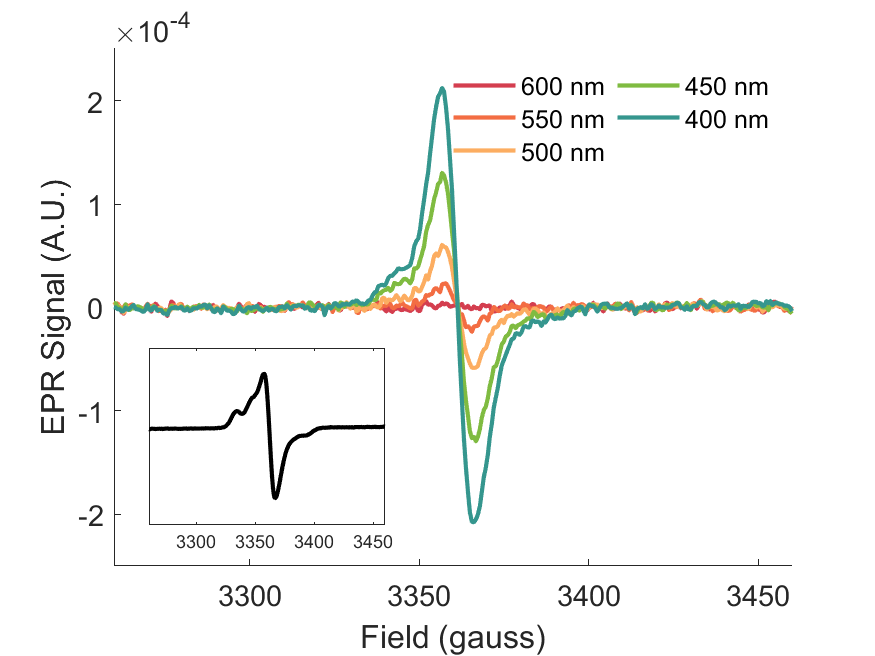}
			\put(0,65){\color{black}\textbf{(C)}}
		\end{overpic}
	\end{subfigure}
	\begin{subfigure}[b]{0.48\linewidth}
		\centering
		\begin{overpic}[width=\linewidth]{Spectra_UiO-66-Anisole_linear}
			\put(0,65){\color{black}\textbf{(D)}}
		\end{overpic}
	\end{subfigure}
		\caption{Spectroscopic characterization of UiO-66-Anisole: \textbf{(A)} Blue: Kubelka--Munk transformed diffuse reflectance spectrum; red: fluorescence excitation spectrum; orange: fluorescence emission spectrum recorded at the indicated wavelengths. \textbf{(B)} Steady-state fluorescence excitation--emission correlation map. \textbf{(C)} Continuous-wave X-band EPR differential spectra recorded under monochromatic illumination (the inset shows the reference spectrum recorded in the dark). \textbf{(D)} Integrated continuous-wave X-band photo-EPR response under monochromatic illumination overlaid with the optical absorption spectrum.}
	\label{fig:Patchwork-Anisole}
\end{figure}

UiO-66-Anisole displays the highest relative photocatalytic performance in the sensitized series (97\% methylene blue degradation relative to the \ch{TiO2} benchmark), despite its pale-yellow color and an absorption edge confined near 500~nm (\Cref{fig:Patchwork-Anisole}A). The sub-gap photoluminescence excitation (PLE) spectrum monitored at 616~nm reveals a weak, localized excitation band confined strictly to the low-energy tail ($>500$~nm, \Cref{fig:Patchwork-Anisole}B) that exhibits an approximately mirror-related profile with the 616~nm emission band. This optical profile is characteristic of an isolated, localized molecular-like or defect-related emissive center with a small Stokes shift. Crucially, because this localized sub-gap pathway is spectrally decoupled from the higher-energy absorption region ($<500$~nm) that drives charge separation, radiative recombination from this channel does not compete significantly with the charge-transfer pathway or limit photocatalysis in UiO-66-Anisole.

The integrated photo-EPR response is largest under 400~nm excitation and decreases progressively toward longer wavelengths (\Cref{fig:Patchwork-Anisole}C,D), closely tracking the additional visible absorption produced by functionalization across the sampled wavelengths. This spectral correspondence directly links excitation of the newly installed optical band to the accumulation of an EPR-active population, supporting the assignment of the functionalization-induced band to a charge-transfer transition. Together with the high photocatalytic activity and time-resolved PL evidence showing ultrafast sub-nanosecond singlet quenching ($\tau_1 \approx 0.4$~ns) into a resonant triplet charge-transfer ($^3\mathrm{CT}$) state where spin-forbidden recombination extends carrier lifetimes up to $\sim 1500$~ns,\cite{kultaeva_mechanistic_2026} this result identifies a \textit{productive red shift}: visible light absorption generates persistent, spin-protected unpaired-electron configurations that survive sufficiently to accumulate under continuous illumination and participate in interfacial redox catalysis.

In the dark, a background EPR signal with a similar overall line shape is already present near $g \approx 2.004$ (\Cref{fig:Patchwork-Anisole}C inset and \Cref{fig:EPR_SI}-B). Illumination therefore appears mainly to increase an existing or spectrally overlapping paramagnetic population, rather than generating a clearly resolved new EPR component. Contributions from a delocalized $\mathrm{N\text{--}O}$-centered or iminoxyl-like radical may be considered on the basis of chemical precedent from acidic diazotization conditions,\cite{challis_chemistry_1971,cudic_transformations_2000,krylov_oxime_2020,sturgeon_tyrosine_2001} but this assignment remains tentative because no resolved $^{14}\mathrm{N}$ hyperfine pattern or independent chemical identification is accessible in the solid state.\cite{alif_photochemistry_1991,lakkaraju_epr_1994} The preserved line shape under illumination is also compatible with closely related radical intermediates or residual $\text{UiO-66-NH}_2$ contributions.\cite{biktagirov_unveiling_2025}

From a molecular standpoint, the methoxy ($-\mathrm{OCH}_3$) substituent in anisole possesses no exchangeable enolic protons and cannot undergo azo/hydrazone (enol/keto) tautomeric equilibria (\Cref{fig:keto_enol-SI}). This structural feature preserves the intact, conjugated $-\mathrm{N=N}-$ double bond and maintains the electronically connected donor--$\pi$--acceptor pathway, providing a clear chemical rationale for efficient charge injection from the appended donor into the framework.

\subsection{The Bad: UiO-66-$\beta$-Naphthol and the Intermediate Case of UiO-66-$\alpha$-Naphthol}

Two naphthol derivatives were investigated to examine the effect of extended aromatic systems and the role of hydroxyl ($-\mathrm{OH}$) substitution geometry: UiO-66-$\alpha$-naphthol (UiO-66-AN) and UiO-66-$\beta$-naphthol (UiO-66-BN). We focus primarily on UiO-66-BN as the archetype of non-productive emissive trapping, and contrast it with UiO-66-AN (the full spectroscopic characterization of UiO-66-AN is provided in \Cref{fig:Patchwork-AN}).

\begin{figure}[H]
	\centering
	\begin{subfigure}[b]{0.48\linewidth}
		\centering
		\begin{overpic}[width=\linewidth]{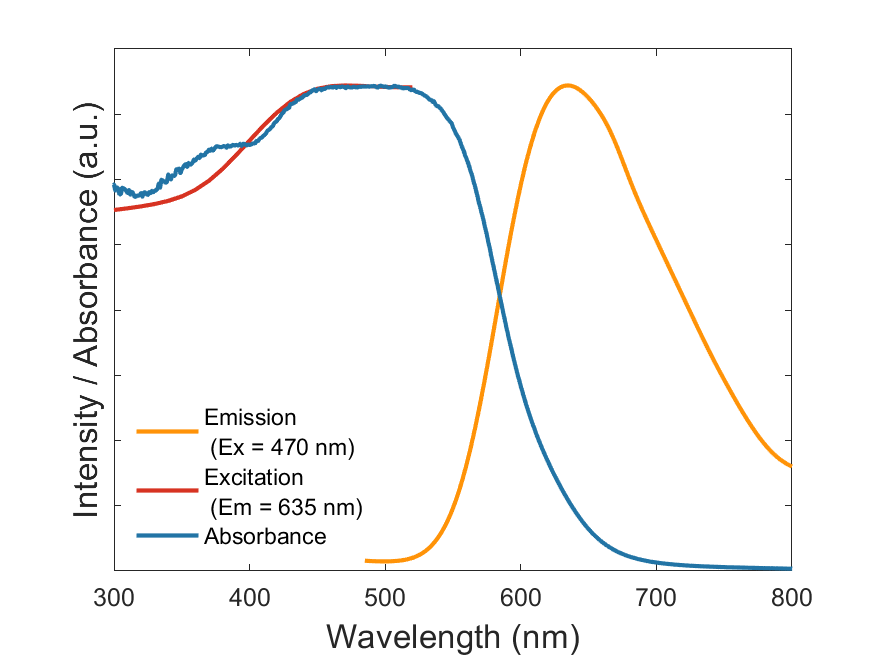}
			\put(0,65){\color{black}\textbf{(A)}}
		\end{overpic}
	\end{subfigure}
	\begin{subfigure}[b]{0.48\linewidth}
		\centering
		\begin{overpic}[width=\linewidth]{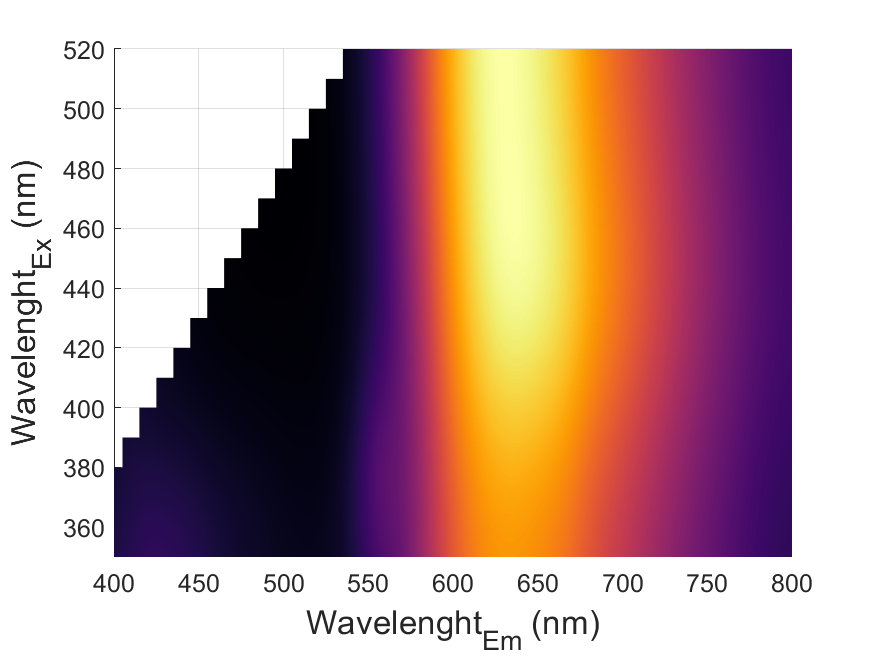}
			\put(0,65){\color{black}\textbf{(B)}}
		\end{overpic}
	\end{subfigure}
	
	\vspace{0.5cm}
	\begin{subfigure}[b]{0.48\linewidth}
		\centering
		\begin{overpic}[width=\linewidth]{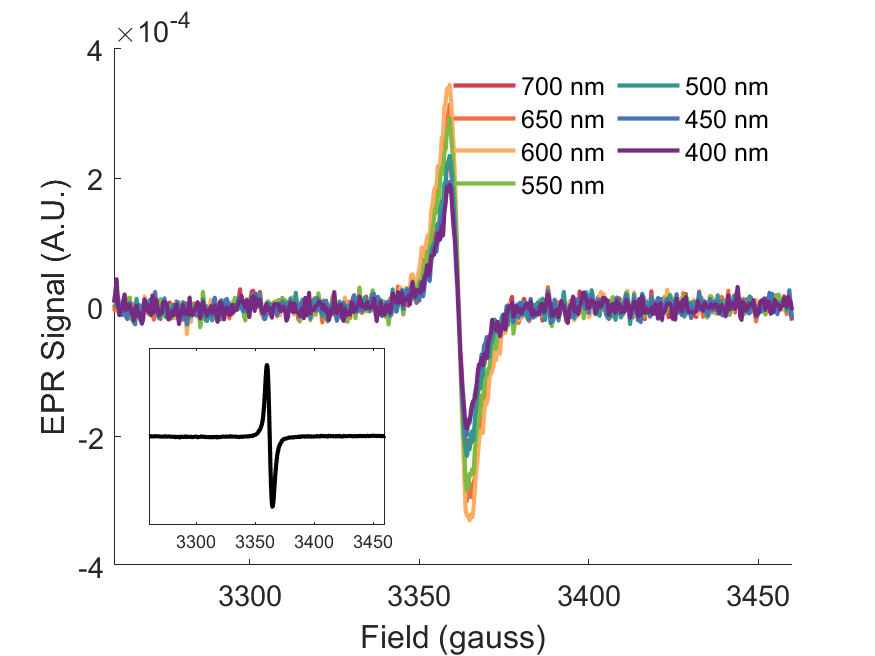}
			\put(0,65){\color{black}\textbf{(C)}}
		\end{overpic}
	\end{subfigure}
	\begin{subfigure}[b]{0.48\linewidth}
		\centering
		\begin{overpic}[width=\linewidth]{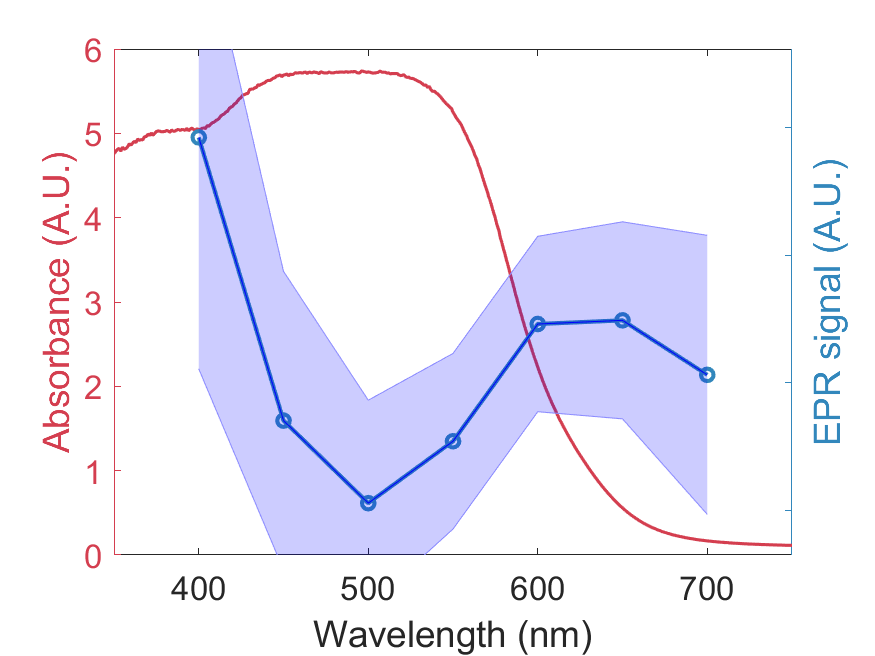}
			\put(0,65){\color{black}\textbf{(D)}}
		\end{overpic}
	\end{subfigure}
			\caption{Spectroscopic characterization of UiO-66-$\beta$-Naphthol: \textbf{(A)} Blue: Kubelka--Munk transformed diffuse reflectance spectrum; red: fluorescence excitation spectrum; orange: fluorescence emission spectrum recorded at the indicated wavelengths. \textbf{(B)} Steady-state fluorescence excitation--emission correlation map. \textbf{(C)} Continuous-wave X-band EPR differential spectra recorded under monochromatic illumination (the inset shows the reference spectrum recorded in the dark). \textbf{(D)} Integrated continuous-wave X-band photo-EPR response under monochromatic illumination overlaid with the optical absorption spectrum.}
	\label{fig:Patchwork-BN}
\end{figure}

For UiO-66-BN (\Cref{fig:Patchwork-BN}A,B), the photoluminescence excitation (PLE) spectrum monitored at the 630~nm emission band closely reproduces the broad functionalization-induced visible absorption across the 400--600~nm window. While this spectral matching does not imply a high absolute photoluminescence quantum yield (PLQY), it demonstrates that optical excitation across this broad visible manifold directly feeds a localized, non-conductive emissive trap rather than populating charge-transfer states. Because the photoexcited population is trapped in this localized manifold, excitation relaxes through local radiative and non-radiative channels before persistent charge-separated states can accumulate, providing a clear photophysical rationale for the absence of photo-EPR enhancement and the near-zero photocatalytic activity of UiO-66-BN compared to UiO-66-Anisole.

The continuous-wave X-band EPR spectrum of UiO-66-BN is dominated by a narrow signal near $g \approx 2.004$ that is already present in the dark (\Cref{fig:Patchwork-BN}C). This dark signal is tentatively assigned to a minor population of ligand-centered phenoxyl/naphthoxyl-like radicals formed through oxidative single-electron transfer during diazotization with surface-accumulated nitrite (\Cref{sec:NOplus_SET,sec:NOplus_formation} \cite{taboada-puig_activation_2013,tang_evaluation_2024}). Under monochromatic illumination, however, the differential and integrated photo-EPR signals (\Cref{fig:Patchwork-BN}C,D) remain within experimental noise and fail to reproduce the red-shifted absorption edge. The absence of light-induced EPR accumulation confirms that the extended visible absorption in UiO-66-BN constitutes a \textit{non-productive (emissive) red shift}: absorbed photons are trapped in localized chromophore states without generating long-lived, catalytically active charge separation.

In azo dyes bearing auxiliary hydroxyl groups, the photophysical response is governed by the relative position of the $-\mathrm{OH}$ group with respect to the azo bridge (\Cref{fig:keto_enol-SI}).\cite{joshi_temperature_2001,gegiou_emission_1971} In 1-phenylazo-2-naphthol (UiO-66-BN), the \textit{ortho}-$\mathrm{OH}$ configuration enables a rigid, six-membered intramolecular hydrogen bond ($\mathrm{O\cdots H\text{--}N}$).\cite{haessner_1h-nmr-spektroskopische_1985} This intramolecular hydrogen bond strongly biases the equilibrium toward the phenylhydrazone/keto form, directly disrupting the conjugated $-\mathrm{N=N}-$ double bond and breaking the donor--$\pi$--acceptor pathway required for charge injection into the framework.\cite{roys_efficiency_2025} As a consequence, the sensitizer is locked into a localized emissive trap at 630~nm, collapsing photocatalytic activity to only 4\% MB degradation.

The validity of this $-\mathrm{OH}$ positional model is directly corroborated by comparing the two naphthol structural isomers. In UiO-66-AN (where the azo linkage is grafted at the 4-position of 1-naphthol), the altered substitution geometry weakens and flexibilizes the intramolecular hydrogen bonding interaction. As a result, UiO-66-AN maintains a dynamic equilibrium where a substantial fraction of linkers retains the conjugated azo-enol form alongside the keto-hydrazone tautomer. This dynamic partitioning produces an intermediate regime characterized by partial photo-EPR carrier accumulation and moderate photocatalytic activity (58\% MB degradation, \Cref{fig:Patchwork-AN}), confirming that avoiding a rigid tautomeric lock allows the framework to balance localized relaxation and productive charge transfer.

\subsubsection{The Unexpected: UiO-66-NH$_2$}

The seemingly anomalous behavior of pristine UiO-66-NH$_2$ reveals an intricate photophysical and photochemical landscape, yielding unexpected insights that decisively clarify and confirm our general framework connecting optical absorption, fluorescence quenching, and paramagnetic intermediate accumulation.

\begin{figure}[H]
	\centering
	\begin{subfigure}[b]{0.48\linewidth}
		\centering
		\begin{overpic}[width=\linewidth]{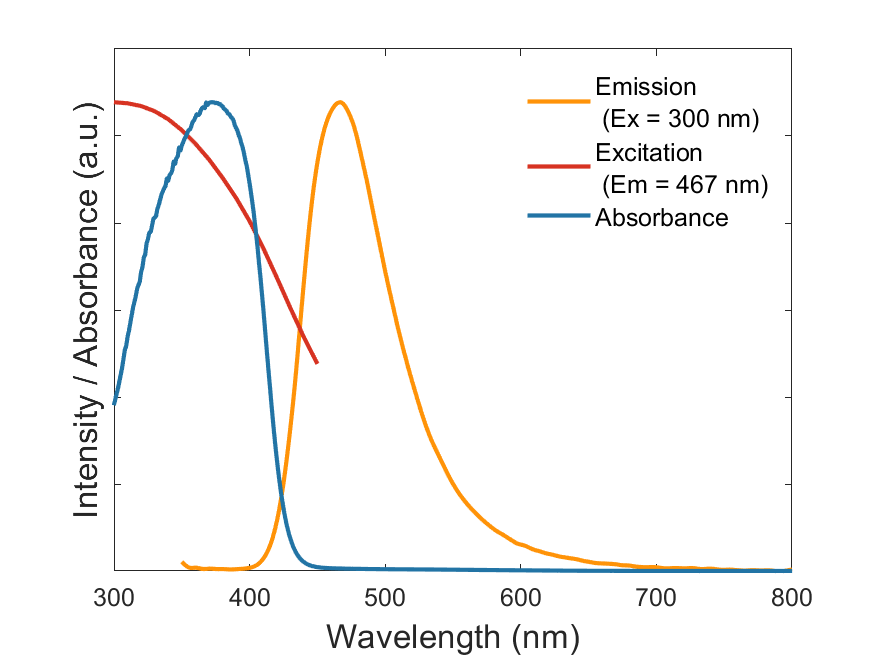}
			\put(0,65){\color{black}\textbf{(A)}}
		\end{overpic}
	\end{subfigure}
	\begin{subfigure}[b]{0.48\linewidth}
		\centering
		\begin{overpic}[width=\linewidth]{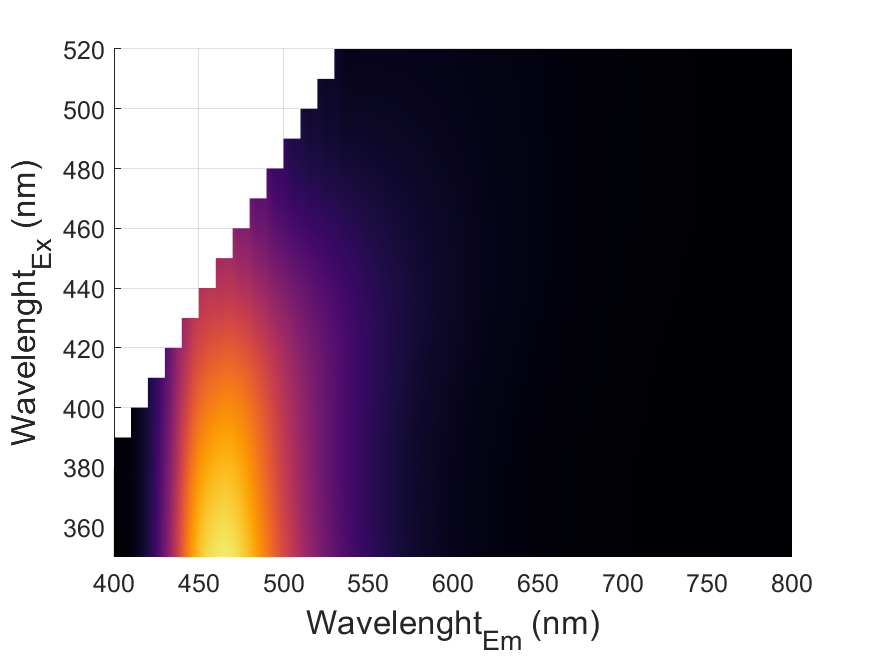}
			\put(0,65){\color{black}\textbf{(B)}}
		\end{overpic}
	\end{subfigure}
	
	\begin{subfigure}[b]{0.48\linewidth}
		\centering
		\begin{overpic}[width=\linewidth]{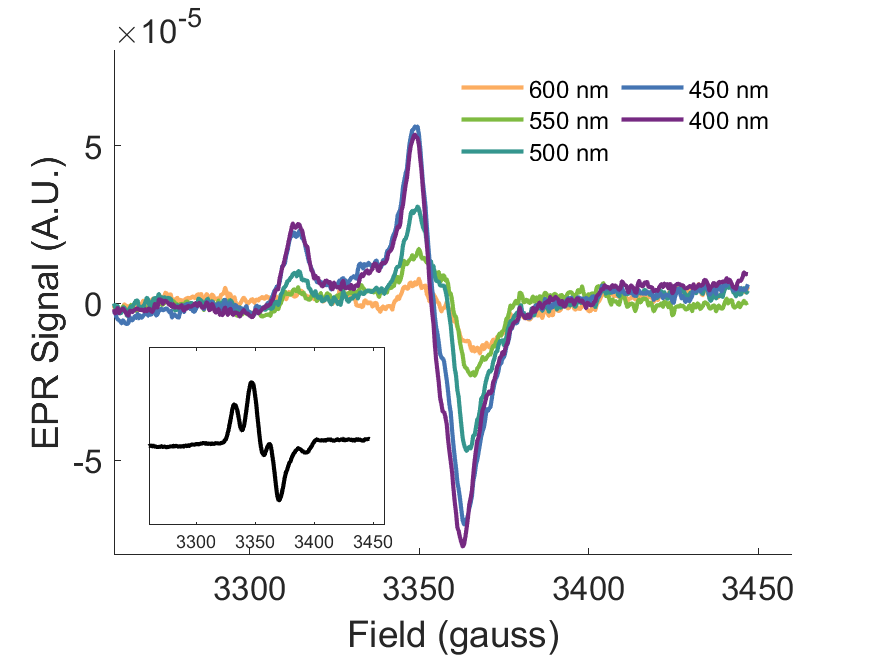}
			\put(0,65){\color{black}\textbf{(C)}}
		\end{overpic}
	\end{subfigure}
	\begin{subfigure}[b]{0.48\linewidth}
		\centering
		\begin{overpic}[width=\linewidth]{Spectra_UiO-66-NH2_linear}
			\put(0,65){\color{black}\textbf{(D)}}
		\end{overpic}
	\end{subfigure}
	\caption{Spectroscopic characterization of UiO-66-NH$_2$: \textbf{(A)} Blue: Kubelka--Munk transformed diffuse reflectance spectrum; red: fluorescence excitation spectrum; orange: fluorescence emission spectrum recorded at the indicated wavelengths. \textbf{(B)} Steady-state fluorescence excitation--emission correlation map. \textbf{(C)} Continuous-wave X-band EPR differential spectra recorded under monochromatic illumination (the inset shows the reference spectrum recorded in the dark). \textbf{(D)} Integrated continuous-wave X-band photo-EPR response under monochromatic illumination overlaid with the optical absorption spectrum.}
	\label{fig:Patchwork-NH2}
\end{figure}

In the pristine material, the photoluminescence excitation (PLE) spectrum overlaps substantially with the intrinsic absorption band (\Cref{fig:Patchwork-NH2}A,B), providing a radiative pathway for the characteristic blue fluorescence emission at 465~nm. 

Unlike the modified materials that exhibit unresolved singlet-like EPR lines, the continuous-wave X-band EPR spectrum of UiO-66-NH$_2$ (\Cref{fig:Patchwork-NH2}C) displays a distinct multi-line pattern previously assigned to a linker-centered $\mathrm{NH^\bullet}$ radical rather than to the Zr metal nodes.\cite{biktagirov_unveiling_2025} This signal arises from anisotropic hyperfine coupling with one $^{14}\mathrm{N}$ nucleus ($I = 1$) and one $^{1}\mathrm{H}$ nucleus ($I = 1/2$), yielding an apparent splitting of $\approx 2$~mT. The extracted $g$-tensor components ($g_x, g_y, g_z = 2.0027, 2.0062, 2.0092$) are fully consistent with a $\pi$-type organic radical localized on the 2-aminoterephthalate linker.\cite{biktagirov_unveiling_2025}

Comparison between the optical absorption spectrum and the wavelength-resolved integrated photo-EPR response of UiO-66-NH$_2$ reveals two distinct regimes (\Cref{fig:Patchwork-NH2}D): (i) below 450~nm, within the intrinsic absorption manifold, photoexcitation simultaneously generates native fluorescence and an increased population of persistent paramagnetic intermediates; (ii) above 450~nm, an appreciable photo-EPR response is detected despite minimal pristine optical absorption. 

The sub-gap EPR activation is rationalized by \textit{in situ} photoinduced chemical transformation during prolonged irradiation: under light exposure, UiO-66-NH$_2$ generates reactive oxygen species ($\mathrm{O_2^{\bullet-}}$ and $^\bullet\mathrm{OH}$),\cite{long_amine-functionalized_2012} which partially oxidize amino handles into nitroso- ($-\mathrm{NO}$) and nitro- ($-\mathrm{NO}_2$) functionalities.\cite{ahn_fluorescence_nodate} These photooxidation products act as internal quenchers that progressively suppress native fluorescence and induce red/NIR optical absorption tails (\Cref{fig:photoxidation_SI}), thereby accounting for the light-induced EPR response at longer wavelengths. Ultimately, deciphering this unexpected behavior consolidates our central premise: localized fluorescence acts as a terminal sink for photogenerated charges. Unmasking these concealed dynamics proves the diagnostic power of our multi-spectroscopic framework as an essential tool to navigate the excited-state maze from light absorption to photocatalysis in dye-sensitized MOFs.

 Because pristine UiO-66-NH$_2$ undergoes continuous photochemical evolution during measurement, it is treated separately from the stable diazo-sensitized series in the following cross-correlation analyses.

\subsubsection{Aniline derivatives: DPA and NNDMA}

Serving as compelling intermediate cases, the aniline derivatives UiO-66-DPA and UiO-66-NNDMA provide crucial insights that further bridge the gap between productive and non-productive excitation. While their comprehensive spectroscopic mapping is detailed in the Supporting Information (\Cref{fig:Patchwork-DFA,fig:Patchwork-NNDMA}), their wavelength-resolved behavior is a key piece of the mechanistic puzzle.

For UiO-66-DPA, the excitation spectrum lies within the absorption envelope without fully covering it (\Cref{fig:Patchwork-DFA}A), resolving three distinct spectral regimes: (i) below approximately 450~nm, substantial absorption occurs without a strong fluorescence response, coinciding with a prominent integrated photo-EPR signal (\Cref{fig:Patchwork-DFA}D); (ii) between approximately 450 and 550~nm, absorption remains significant but is accompanied by rising PLE emission, while the integrated EPR response becomes weak; and (iii) above approximately 570~nm, neither significant fluorescence nor an appreciable integrated EPR response is observed despite residual absorption. 

Crucially, this wavelength-resolved photo-EPR action spectrum provides direct experimental validation for earlier TDDFT predictions for Zr-MOFs.\cite{kultaeva_mechanistic_2026,nasalevich_electronic_2016} It definitively confirms that productive charge-transfer configurations are selectively accessed only in the short-wavelength visible region ($\lambda \le 500$~nm). Given the bandwidth of the optical band-pass filters ($\lambda_0 \pm 20$~nm), the experimental response reflects a continuous transition toward localized, non-productive exciton states as excitation extends to longer wavelengths.

\subsection{Photocatalytic response and the spectral-overlap index} \label{section:PC}

\begin{figure}[H]
	\centering
	\begin{subfigure}[b]{0.48\linewidth}
		\centering
		\begin{overpic}[width=\linewidth]{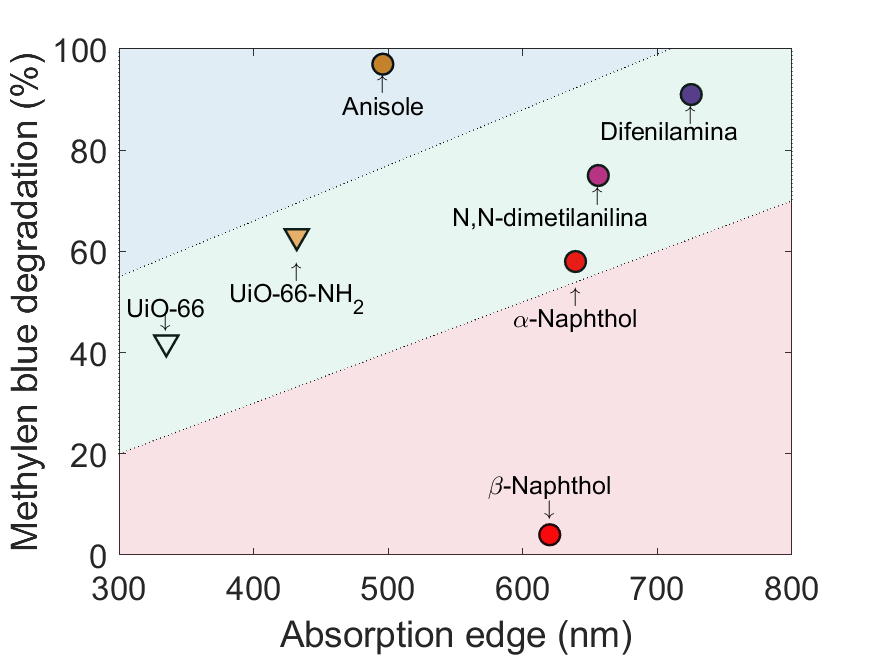}
			\put(-10,65){\color{black}\textbf{(A)}}
		\end{overpic}
	\end{subfigure}
	\begin{subfigure}[b]{0.48\linewidth}
		\centering
		\begin{overpic}[width=\linewidth]{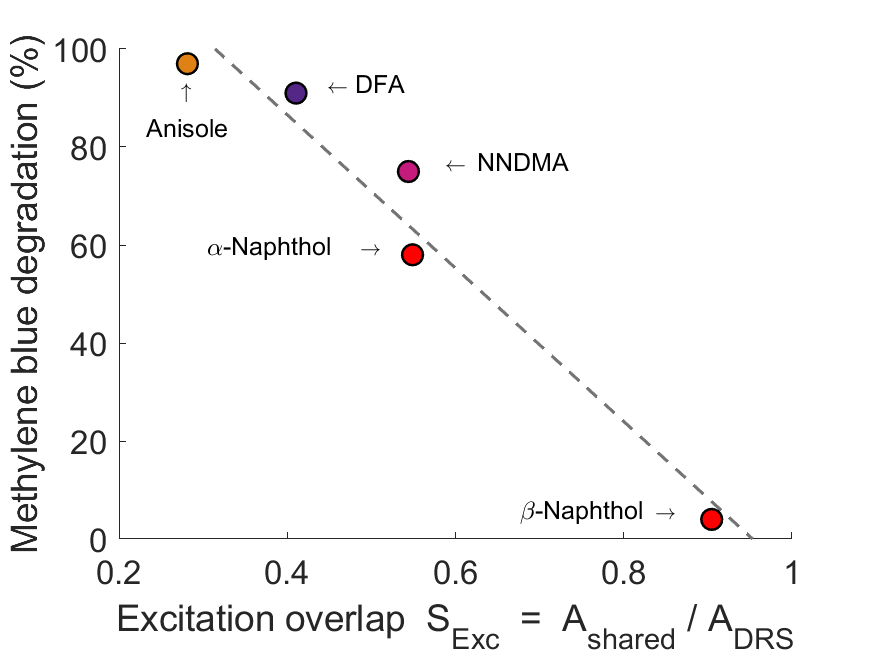}
			\put(-10,65){\color{black}\textbf{(B)}}
		\end{overpic}
	\end{subfigure}
	\caption{\textbf{(A)} Absorption edge and \ch{TiO2}-normalized MB response. Sensitized samples are indicated by circles; separately evaluated samples (UiO-66 and UiO-66-NH$_2$) are indicated by triangles. \textbf{(B)} Normalized absorption--PLE spectral-overlap index, $S_{\mathrm{Exc}}$, compared with the \ch{TiO2}-normalized MB response.}
	\label{fig:Correlations}
\end{figure}

The cross-series comparison shows that extending the optical absorption edge to longer wavelengths does not ensure high photocatalytic performance (\Cref{fig:Correlations}A). UiO-66-BN exhibits extended visible absorption up to $\approx 620$~nm but achieves only 4\% of the \ch{TiO2} reference response, whereas UiO-66-Anisole absorbs up to $\approx 496$~nm and achieves 97\% of the \ch{TiO2} reference response. The aniline derivatives (DPA and NNDMA) likewise exhibit high relative performance despite pronounced spectral differences.

Pristine UiO-66 is omitted from the sensitized-material correlation because it lacks substantial visible absorption. UiO-66-NH$_2$, as previously mentioned, is treated separately because irradiation may change its optical and paramagnetic response during the experiment.\cite{ahn_fluorescence_nodate}

To quantify the optical coupling between the absorption envelope and the monitored emission across the series, the normalized absorption--PLE overlap index is defined as:

\begin{equation}
	S_{\mathrm{Exc}} =
	\frac{
		\displaystyle
		\int_{\lambda_{\min}}^{\lambda_{\max}}
		\min \left[ PLE_{\mathrm{norm}}(\lambda), Abs_{\mathrm{norm}}(\lambda) \right] \, d\lambda
	}{
		\displaystyle
		\int_{\lambda_{\min}}^{\lambda_{\max}}
		Abs_{\mathrm{norm}}(\lambda) \, d\lambda
	}
	\label{eq:SExc_combined}
\end{equation}

$S_{\mathrm{Exc}}$ is a dimensionless spectral-shape descriptor ranging from zero to one. It quantifies the fraction of the normalized Kubelka--Munk absorption envelope covered by the normalized PLE profile used for the selected emission (see \Cref{sec:SExc_definition} for details and \Cref{fig:Abs-Exc_SI} for visualizations). Because diffuse-reflectance Kubelka--Munk spectra and PLE profiles for powder samples are recorded in relative intensity units, both are area-normalized over the integration window. Crucially, this normalization means $S_{\mathrm{Exc}}$ must be treated strictly as a descriptor of spectral profile shape and optical coupling, rather than an absolute fluorescence quantum yield or radiative loss fraction.

Evaluation of these two metrics clarifies the structural determinants of photocatalytic performance. While the raw optical absorption edge yields a scattered distribution that fails to correlate with the observed \ch{TiO2}-normalized MB response (\Cref{fig:Correlations}A), the $S_{\mathrm{Exc}}$ descriptor provides a much clearer trend across the sensitized series. The observed inverse correlation (\Cref{fig:Correlations}B) illustrates that extensive optical coupling to localized emissive traps ($S_{\mathrm{Exc}} \to 1$) effectively suppresses photocatalytic activity by promoting \textit{non-productive emissive red shift}. Ultimately, this relationship provides a practical framework to visualize the fundamental competition between charge-transfer and radiative recombination, capturing photophysical information not contained in the absorption edge alone.

Complementing the quantitative $S_{\mathrm{Exc}}$ descriptor, the wavelength-resolved photo-EPR comparison provides a key qualitative distinction, since a rigorous quantitative interpretation of photo-EPR requires accounting for complex variables such as CT energy levels. In UiO-66-Anisole, the photo-EPR action spectrum closely follows the functionalization-induced absorption, whereas in UiO-66-$\beta$-naphthol, the new absorption follows PLE. These qualitative observations assist in establishing our current interpretative framework: functionalization can create either a \textit{productive red shift} that sustains a paramagnetic population or a \textit{non-productive red shift} dominated by a spectroscopically broad but rapidly recombining excitation pathway.

\section{Conclusions}

This work establishes a multi-spectroscopic framework that deciphers the photophysical fate of photoexcited states in dye-sensitized MOFs, resolving the longstanding misconception that broadening optical absorption into the visible necessarily enhances photocatalytic efficiency. By integrating diffuse reflectance spectroscopy, steady-state 2D fluorescence excitation--emission mapping, wavelength-resolved continuous-wave photo-EPR, and photocatalytic testing, we demonstrate that the destination of photoexcited carriers is dictated by molecular structure and tautomeric equilibria rather than absorption breadth alone.

Across the diazo-functionalized UiO-66 series, the normalized absorbance--excitation overlap descriptor ($S_{\mathrm{Exc}}$) quantitatively captures the competition between localized deactivation and photocatalytic performance: extensive coupling to an emissive manifold systematically suppresses chemical reactivity. In parallel, wavelength-resolved photo-EPR action spectra provide direct evidence that productive charge separation requires optical transitions that accumulate persistent, spin-protected paramagnetic configurations ($^3\mathrm{CT}$), whereas lower-energy excitation often feeds localized excitons or non-radiative decay. Furthermore, structural isomerism in the naphthol-functionalized frameworks reveals that an \textit{ortho}-$\mathrm{OH}$ configuration enables a rigid six-membered intramolecular hydrogen bond ($\mathrm{O\cdots H\text{--}N}$) that shifts the equilibrium to the keto-hydrazone tautomer, disrupting the conjugated $-\mathrm{N=N}-$ double bond and creating an intense localized trap at 630~nm (4\% activity in $\beta$-naphthol). In contrast, altering the substitution geometry flexibilizes this interaction, sustaining a dynamic tautomeric equilibrium with partial charge-transfer capability (58\% activity in $\alpha$-naphthol).

Collectively, these findings establish three hierarchical design guidelines for dye-sensitized Zr-MOFs (and related wide-bandgap frameworks where direct visible LMCT is energetically unfavorable): (i)~\textbf{Suppression of radiative sinks:} Minimizing optical overlap with emissive pathways ($S_{\mathrm{Exc}} \to 0$) is essential to prevent rapid loss of photoexcited carriers into localized non-conductive states; (ii)~\textbf{Control of conformational and tautomeric equilibria:} Substituent geometry must be engineered to avoid rigid tautomeric locks that disrupt the conjugated azo double bond, destroy donor--$\pi$--acceptor electronic connectivity, and open radiative channels; and (iii)~\textbf{Selective targeting of the CT manifold:} Because charge separation in Zr-MOFs relies on populating linker-based charge-transfer states rather than direct node reduction, photosensitization must specifically target the short-wavelength visible region ($\lambda \le 500$~nm) where persistent, catalytically active carrier populations accumulate.

This framework provides a solid foundation for rational photosensitizer design in porous crystalline scaffolds, setting the stage for future time-resolved and ultrafast kinetic studies to further map the intricate landscape of excited-state charge separation.
	
\begin{acknowledgement}
% This work was supported by ...
We thank our colleagues and collaborators for fruitful discussions.
\end{acknowledgement}

% \begin{suppinfo}
% The Supporting Information is available free of charge on the ACS Publications website at DOI: 10.1021/jacs.xxxxxxx.
% Detailed experimental procedures, TGA, XRD, BET surface area determinations, 2D fluorescence correlation diagrams, full EPR spectra, tautomeric equilibrium schemes, and overlap descriptor mathematical derivations.
% \end{suppinfo}

\clearpage
% ---- REINICIAR CONTADORES PARA LA SI ----
\setcounter{page}{1}
\setcounter{section}{0}
\setcounter{figure}{0}
\setcounter{table}{0}
\renewcommand{\thefigure}{S\arabic{figure}}
\renewcommand{\thetable}{S\arabic{table}}
\renewcommand{\thesection}{S\arabic{section}}
\renewcommand{\theequation}{S\arabic{equation}}

% ---- PORTADA MANUAL DE LA SI ----
\begin{center}
	\LARGE\textbf{Supporting Information}\\[1em]
	\Large\textbf{Beyond Panchromatic Absorption: Deciphering the Excited-State Maze from Light Absorption to Photocatalysis in Dye-Sensitized MOFs}
\end{center}
\vspace{1cm}

% ---- ÍNDICE DE LA SI ----
\tableofcontents
\clearpage

\section{Experimental Details}

\subsection{Sample preparation} 

UiO-66-NH$_2$ was prepared by a solvothermal method using zirconyl chloride octahydrate (ZrOCl$_4\cdot$8H$_2$O) and 2-aminoterephthalic acid (NH$_2$-BDC) as the inorganic and organic precursors, respectively, following a modified method reported by Katz et al. \cite{katz_facile_2013}. In a typical preparation, ZrOCl$_4\cdot$8H$_2$O (174~mg, 0.54~mmol) was dissolved in one third of the total N,N-dimethylformamide (DMF, 15~mL total) together with concentrated hydrochloric acid (1~mL). The mixture was sonicated for 20~minutes until complete dissolution of the metal salt. Subsequently, NH$_2$-BDC (134~mg, 0.75~mmol) and the remaining DMF were added, and the suspension was further sonicated for 20~minutes. The reaction was heated at 80~\textdegree C overnight under static conditions. 

After cooling to room temperature, the resulting solid was separated by filtration and thoroughly washed twice with DMF (2~$\times$~30~mL) and twice with ethanol (2~$\times$~30~mL). The material was centrifuged to remove residual solvent. Solids were dried at 150~\textdegree C overnight. 

the MOF postfunctionalization by a diazotization reaction was performed using acetic acid solutions of N,N-dimethylaniline, $\alpha$-naphtol, $\beta$-naphtol, diphenylamine, N,N-dimethylaniline and anisole, as coupling agents following the procedure reported by Otal et al. The formula of UiO-66-NH2 was assumed to be $Zr_{24}O_{120}C_{192}H_{144}N_{24}$ g/mol) with 24 NH$_2$ groups per mole and 7\% solvent content (TGA based). A solution of NaNO$_2$ with six times the equivalent amount to -NH$_2$ groups was cooled down to 0°C with ice and the UiO-66-NH$_2$ was suspended in it. A solution 0.03mol.L$^{-1}$ was added to the yellowish suspension until a 3 fold excess of HCl respect to nitrite was achieved. The suspension was stirred 10 minutes until a change of colour from pale yellow to brownish orange is observed. The diazo coupling agents were added dropwise with a twofold excess respect to nitrite. The suspension exhibited a fast colour change once the first drops were added. The suspension was kept in an ice bath and left to warm up overnight. The solids were isolated by centrifugation, washed with acetic acid 3 times and twice with acetone. The solid were dried at room temperature. To determine the specific reaction with –NH${_2}$ groups at the MOF structure, the reaction was repeated in the above mentioned conditions with 1-naphthol and UiO-66. No colour was developed.

\begin{figure}[H]
	\centering
	
	% --- Subfigura A ---
	\begin{subfigure}[b]{0.48\linewidth}
		\centering
		\begin{overpic}[height=4cm]{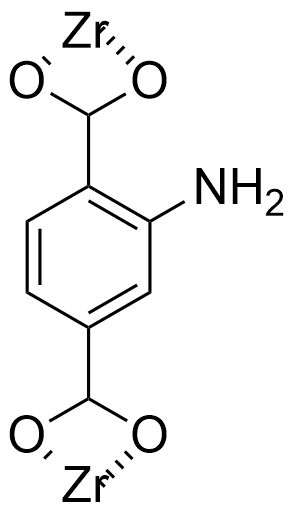}
			\put(-30,90){\color{black}\textbf{(A)}}
		\end{overpic}
	\end{subfigure}
	% --- Subfigura B ---
	\begin{subfigure}[b]{0.48\linewidth}
		\centering
		\begin{overpic}[height=4cm]{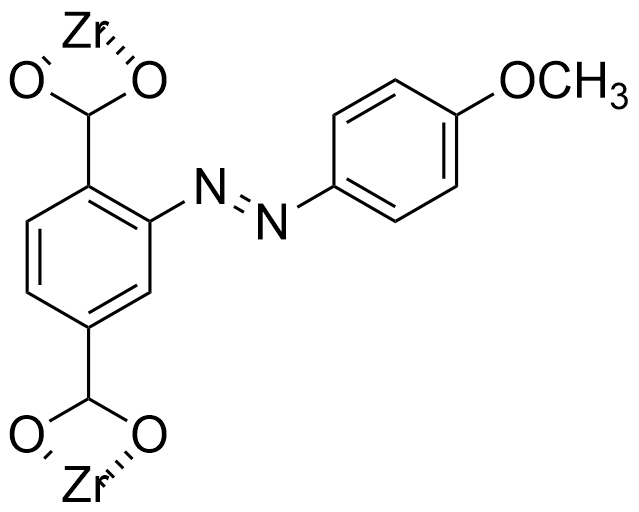}
			\put(-20,80){\color{black}\textbf{(B)}}
		\end{overpic}
	\end{subfigure}
	
	\vspace{1.5cm}
	
	% --- Subfigura C ---
	\begin{subfigure}[b]{0.48\linewidth}
		\centering
		\begin{overpic}[height=4cm]{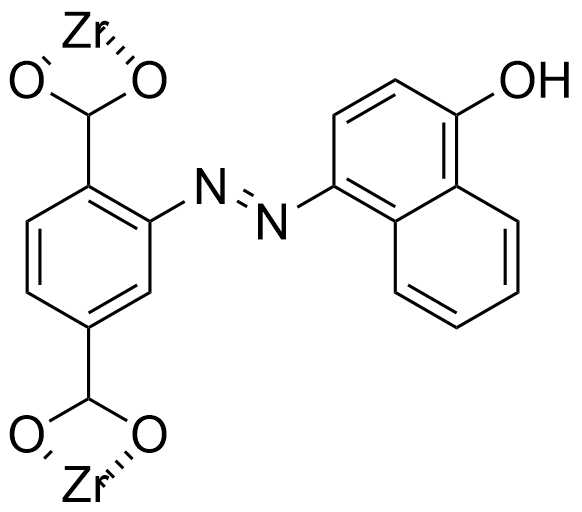}
			\put(-20,90){\color{black}\textbf{(C)}}
		\end{overpic}
	\end{subfigure}
	% --- Subfigura D ---
	\begin{subfigure}[b]{0.48\linewidth}
		\centering
		\begin{overpic}[height=4cm]{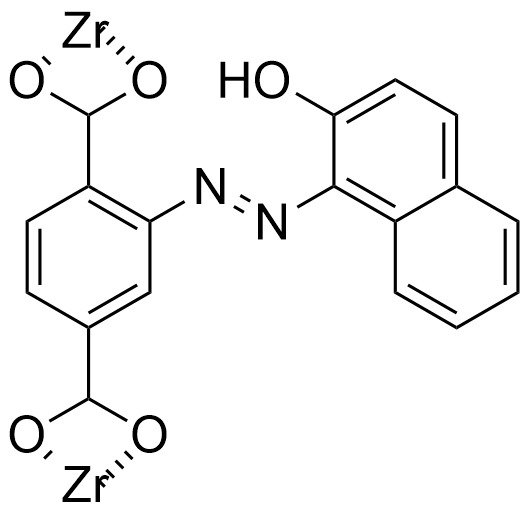}
			\put(-20,90){\color{black}\textbf{(D)}}
		\end{overpic}
	\end{subfigure}
	
	\vspace{1.5cm}
	
	% --- Subfigura E ---
	\begin{subfigure}[b]{0.48\linewidth}
		\centering
		\begin{overpic}[height=4cm]{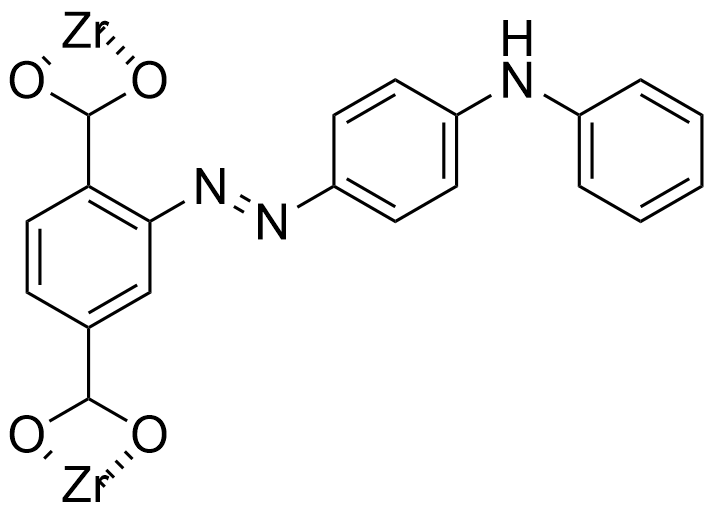}
			\put(-20,70){\color{black}\textbf{(E)}}
		\end{overpic}
	\end{subfigure}
	\begin{subfigure}[b]{0.48\linewidth}
		\centering
		\begin{overpic}[height=4cm]{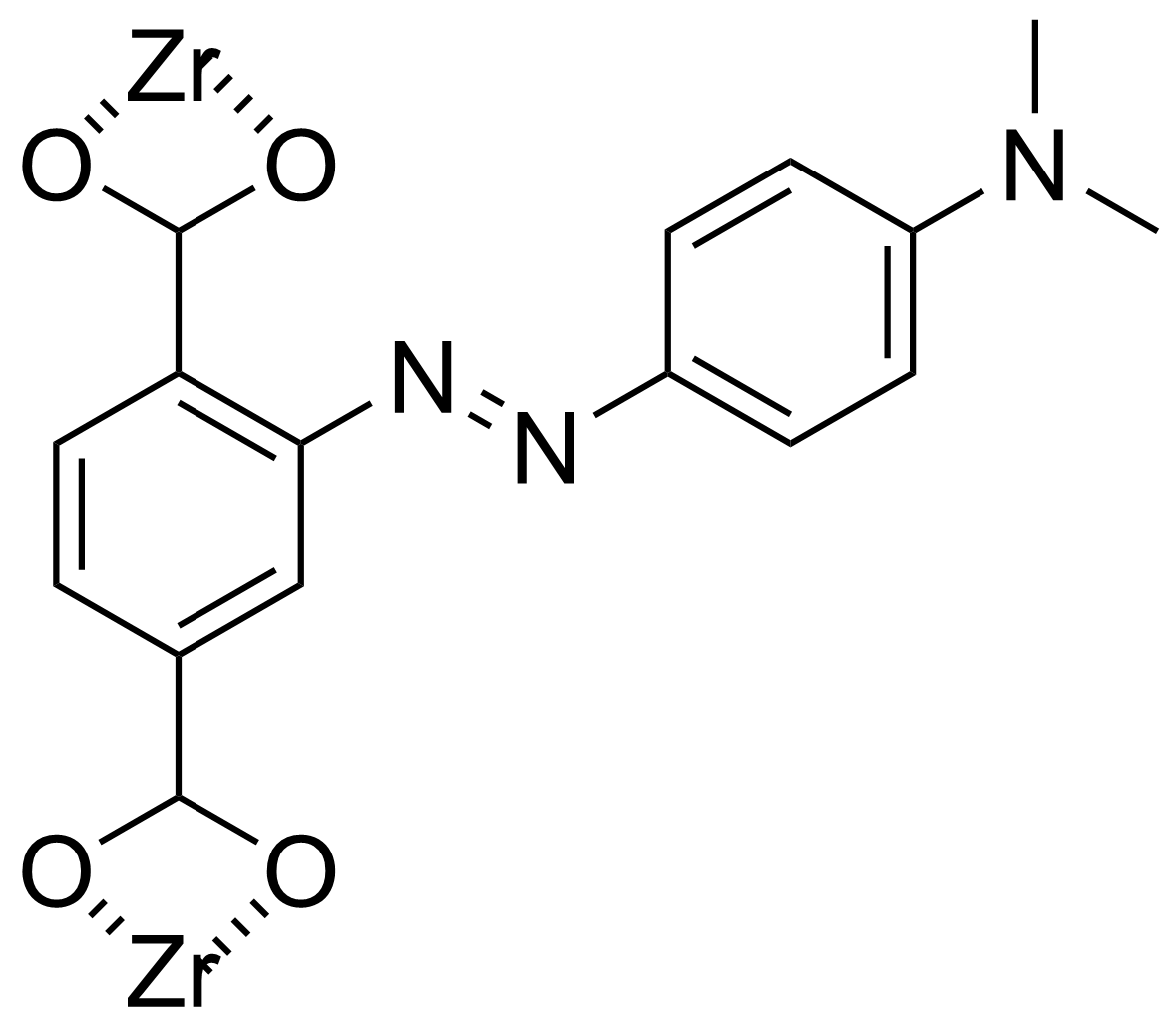}
			\put(-20,70){\color{black}\textbf{(F)}}
		\end{overpic}
	\end{subfigure}

	\caption{Ligand and reference structures: 
		\textbf{(A)} UiO-66-NH$_2$, 
		\textbf{(B)} anisole, 
		\textbf{(C)} AN (alpha-naphthol), 
		\textbf{(D)} BN (beta-naphthol),
		\textbf{(E)} DPA (diphenylamine) 
		\textbf{(\textbf{F})} NNDMA (N,N-dimethylaniline).}
	\label{fig:ligands_ordered}
\end{figure}

Modified ligand structures is shown in \Cref{fig:ligands_ordered}

\subsection{XRD}

X-Ray diffraction (XRD) measurements were performed in a Panalytical Diffractometer, Model Empyrean with PIXCEL3D detector using Cu K$_\alpha$ radiation (K$_{\alpha 1}$ = 1.54056\AA).

\subsection{TGA}

The experiments were performed with a NETZSCH STA 409 CD thermobalance. Approximately 25~mg of each MOF was heated in an alumina crucible with 5~\textdegree C min$^{-1}$ up to 950~\textdegree C in synthetic air (20.5 \% O$_2$ in N$_2$) flow of 50 mL min$^{-1}$.

Thermogravimetric analysis (TGA) was used to estimate the amount of solvent adsorbed in the pores, the average molar mass of the functionalized MOFs, and the diazo-coupling yield. The following quantities were defined from the TGA traces:

\begin{itemize}
	\item $m_1$: first mass-loss step (\%), assigned to solvent removal,
	\item $m_2$: second mass-loss step (\%), assigned to decomposition of the organic fraction,
	\item $m_f$: final mass (\%), assigned to the inorganic residue.
\end{itemize}

The decomposition steps were identified using coupled MS signals at $m/z = 18$ and $m/z = 44$, corresponding to water and carbon dioxide, respectively. The first mass-loss step was attributed mainly to desorption of pore solvent (predominantly water). The second mass-loss step corresponds to combustion of the organic framework. In this temperature range, the MOF is converted into $\mathrm{ZrO_2}$ and gaseous combustion products.

The second mass-loss step can be obtained from the measured quantities as:
\begin{equation}
	m_2 = 100 - m_1 - m_f
	\label{eq:m2_definition}
\end{equation}

A solvent-corrected organic mass loss can then be defined as:
\begin{equation}
	m_{2,\mathrm{corr}} = \frac{m_2}{100 - m_1}\times 100
	= \frac{m_2}{m_2 + m_f}\times 100
	\label{eq:m2corr}
\end{equation}

However, the values reported in Table~\ref{tab:tga_summary_narrow} correspond to the normalized final mass (normalized residue), defined as:
\begin{equation}
	m_3 = \frac{m_f}{100 - m_1}\times 100
	= \frac{m_f}{m_2 + m_f}\times 100
	\label{eq:m3_norm_residue}
\end{equation}

For pristine UiO-66, the decomposition pathway can be represented as:
\begin{equation}
	\mathrm{Zr}_{24}\mathrm{O}_{120}\mathrm{C}_{192}\mathrm{H}_{96} + x\,\mathrm{O}_2
	\rightarrow 24\,\mathrm{ZrO}_2 + \text{combustion products}
	\label{eq:decomp_uio66}
\end{equation}

Based on Eq.~\ref{eq:decomp_uio66}, the ideal normalized final mass (ideal $\mathrm{ZrO_2}$ residue) is:
\begin{equation}
	m_{3,\mathrm{ideal}} =
	\frac{24\,M_r(\mathrm{ZrO}_2)}{M_r(\mathrm{UiO\mbox{-}66)}}\times 100
	\label{eq:m3_ideal}
\end{equation}

The same approach was applied to the substituted UiO-66 derivatives. The corresponding ideal residue values are listed in Table~\ref{tab:tga_mr_yield_narrow}.

The average molar mass estimated from TGA was calculated from the measured residue according to:
\begin{equation}
	M_{r,\mathrm{av}} =
	\frac{24\,M_r(\mathrm{ZrO}_2)\,(100 - m_1)}{m_f}
	=
	\frac{24\,M_r(\mathrm{ZrO}_2)\times 100}{m_3}
	\label{eq:mr_av_tga}
\end{equation}

The TGA-derived molar masses differ from the values calculated from the ideal molecular formulas. To estimate the diazo-coupling yield, the average composition was modeled as a weighted average of two limiting structures: (i) the successfully diazo-coupled MOF, and (ii) the hydrolysis product of the diazonium salt, represented by an OH-substituted linker environment. In this model, the hydrolysis product is approximated by the formula $\mathrm{Zr}_{24}\mathrm{O}_{144}\mathrm{C}_{192}\mathrm{H}_{96}$.

\begin{equation}
	M_{r,\mathrm{av}} = x_1 M_{r,1} + x_2 M_{r,2}
	\label{eq:weighted_mr}
\end{equation}

with
\begin{equation}
	x_1 + x_2 = 1
	\label{eq:xsum}
\end{equation}

Thus,
\begin{equation}
	x_1 = \frac{M_{r,\mathrm{av}} - M_{r,2}}{M_{r,1} - M_{r,2}}
	\label{eq:x1}
\end{equation}

where $M_{r,1}$ is the molar mass of the diazo-coupled MOF, $M_{r,2}$ is the molar mass of the unreacted MOF, and $x_1$ and $x_2$ are the corresponding weighting factors. Within this model, $x_1$ is interpreted as the diazo-coupling yield (or degree of diazo coupling).

\subsection{Diffuse reflection}

Reflectance spectra were measured using a UV-visible scanning spectrophotometer (SHIMADZU UV-2101PC) attached to an integrating sphere. Sample powders were mounted into the integrating sphere using a quartz slide. Reflectance was transformed to absorbance using the Kubelka—Munk model.

\subsubsection{Kubelka Munk transformation}
In diffuse reflectance spectroscopy (DRS), the Kubelka--Munk function is used to convert reflectance data into a form that is proportional to the absorption and scattering coefficients of a material.

Let \( R \) denote the reflectance, defined as the ratio between the diffusely reflected intensity and the incident light intensity. The Kubelka--Munk function is expressed as:

\begin{equation}
	F(R) = \frac{(1 - R)^2}{2R}
\end{equation}

The function \( F(R) \) is proportional to the ratio of the absorption coefficient \( K \) to the scattering coefficient \( S \):

\begin{equation}
	F(R) = \frac{K}{S}
\end{equation}

Hence, \( F(R) \) can be interpreted as a pseudo--absorbance suitable for analyzing diffuse reflectance data.

\subsubsection{Reflectance in Percent Form}

If the reflectance is provided as a percentage value \( R_{\%} \), it must first be converted into a fractional value:

\begin{equation}
	R = \frac{R_{\%}}{100}
\end{equation}

The Kubelka--Munk transformation then becomes:

\begin{equation}
	F(R) = \frac{\left( 1 - \frac{R_{\%}}{100} \right)^2}{2 \left( \frac{R_{\%}}{100} \right)}
\end{equation}

\subsection{Luminescence}

Steady-state photoluminescence (PL) measurements were performed using an Edinburgh FLS1000 spectrometer equipped with a 450 W ozone-free Xe arc lamp, excitation and emission monochromators, and a PMT-900 red-sensitive photomultiplier tube detector. Emission spectra were recorded over the 350-800~nm spectral range using different excitation wavelengths, selected according to the specific measurement requirements. To minimize detector saturation caused by Rayleigh scattering, a spectral offset of 15~nm was applied between the excitation wavelength and the starting wavelength of the emission scan whenever necessary

\subsection{Continuous-Wave Photo-EPR and Optical Normalization}
\label{sec:SI_EPR_methods}

\subsubsection{EPR Instrumentation and Measurement Parameters}
Continuous-wave electron paramagnetic resonance (CW-EPR) measurements were performed at room temperature ($T = 298$~K) on a Bruker EMX X-band spectrometer operating at $\sim 9.5$~GHz, equipped with a standard rectangular $\mathrm{TE}_{102}$ microwave cavity featuring an optical irradiation grid. The experimental parameters were: microwave power $P_{\mathrm{mw}} = 1.0$~mW (within the linear, non-saturating regime), modulation frequency $\nu_{\mathrm{mod}} = 100$~kHz, modulation amplitude $B_{\mathrm{mod}} = 0.1$~mT, time constant $\tau = 40.96$~ms, and conversion time $t_{\mathrm{conv}} = 81.92$~ms.

\subsubsection{In Situ Monochromatic Irradiation}
In situ photoexcitation was carried out using a stabilized light source coupled to a series of calibrated narrow bandpass optical interference filters centered at $\lambda_0 = 400, 450, 500, 550, 600, 650,$ and $700$~nm (full-width at half-maximum, $\mathrm{FWHM} = \pm 20$~nm). The optical power $P(\lambda)$ impinging on the sample inside the cavity was determined at each wavelength with a calibrated optical power meter.

The corresponding incident photon flux $\Phi_{\mathrm{ph}}(\lambda)$ (in photons~s$^{-1}$) was calculated as:
\begin{equation}
	\Phi_{\mathrm{ph}}(\lambda) = \frac{P(\lambda)\,\lambda}{h\,c}
	\label{eq:SI_photon_flux}
\end{equation}
where $h = 6.62607 \times 10^{-34}$~J$\cdot$s is Planck's constant and $c = 2.99792 \times 10^8$~m$\cdot$s$^{-1}$ is the speed of light. The measured optical powers and calculated photon fluxes are summarized in Table~\ref{tab:lamp_power_SI}.

\begin{table}[htbp]
	\centering
	\caption{Incident optical power $P$ and calculated photon flux $\Phi_{\mathrm{ph}}$ at each irradiation wavelength $\lambda$.}
	\label{tab:lamp_power_SI}
	\small
	\begin{tabular}{@{}cccccccc@{}}
		\toprule
		$\lambda$ (nm) & 400 & 450 & 500 & 550 & 600 & 650 & 700 \\
		\midrule
		$P$ (mW) & 46.2 & 63.7 & 83.1 & 88.6 & 81.3 & 66.8 & 43.8 \\
		$\Phi_{\mathrm{ph}}$ ($10^{15}$~photons~s$^{-1}$) & 0.93 & 1.44 & 2.09 & 2.45 & 2.46 & 2.19 & 1.54 \\
		\bottomrule
	\end{tabular}
\end{table}

\subsubsection{Differential Photo-EPR Spectra and Action Spectra Normalization}
Prior to illumination, a dark reference first-derivative spectrum $S_{\mathrm{dark}}(B)$ was acquired. Under continuous monochromatic irradiation at each wavelength $\lambda$, the illuminated spectrum $S_{\mathrm{light}}(B,\lambda)$ was recorded after reaching steady-state. 

The net photoinduced differential spectrum $\Delta S(B,\lambda)$ was obtained by baseline subtraction:
\begin{equation}
	\Delta S(B,\lambda) = S_{\mathrm{light}}(B,\lambda) - S_{\mathrm{dark}}(B)
	\label{eq:diff_EPR_SI}
\end{equation}

The relative concentration of photogenerated paramagnetic species is proportional to the double integral ($\mathrm{DI}$) of the differential first-derivative signal over the magnetic field range $[B_{\min}, B_{\max}]$:
\begin{equation}
	\mathrm{DI}[\Delta S(B,\lambda)] = \int_{B_{\min}}^{B_{\max}} \left( \int_{B_{\min}}^{B} \Delta S(B',\lambda)\, dB' \right) dB
	\label{eq:DI_EPR_SI}
\end{equation}

To construct the wavelength-resolved photo-EPR action spectra, the integrated response was normalized by the incident optical power $P(\lambda)$:
\begin{equation}
	I_{\mathrm{EPR}}(\lambda) = \frac{\mathrm{DI}[\Delta S(B,\lambda)]}{P(\lambda)}
	\label{eq:norm_power_EPR_SI}
\end{equation}
or alternatively expressed on a per-photon basis by normalization against the incident photon flux $\Phi_{\mathrm{ph}}(\lambda)$:
\begin{equation}
	I_{\mathrm{EPR,\Phi}}(\lambda) = \frac{\mathrm{DI}[\Delta S(B,\lambda)]}{\Phi_{\mathrm{ph}}(\lambda)} = \frac{h\,c}{P(\lambda)\,\lambda}\,\mathrm{DI}[\Delta S(B,\lambda)]
	\label{eq:norm_flux_EPR_SI}
\end{equation}

\subsection{Photocatalytic tests}
For the photocatalytic test, a stock solution of 2mM of Methylene Blue (MB) was prepared dissolving 0.0639 g of MB in 100mL of distilled water. Dilution 1:10 were performed in order to get 0.2mM of MB. 15 mg of each MOF was placed in a Pyrex glass test tube with a magnetic stirrer and 10 mL of 0.2mM MB was added. TiO2 (Titanium (IV) oxide, anatase, 99.9\% (metals basis), Alfa Aesar. 45m2/g) was also tested as a positive test for photocatalysis.
Suspensions were stirred in dark conditions in order to quantify the adsorption of MB onto the solids. After 1 hour, aliquots of 3 mL were filtered with 20$\mu$m nylon filters and absorbance at 664nm was measured.
A Xenon Lamp was used as light source for photocatalysis, and the remaining suspensions were irradiated for 1.5 hs under continuous stirring. At which time, aliquots of 3 mL were filtered with 20 $\mu$m nylon filters and absorbance at 664 nm was measured.

\section{Supplementary results}

\subsection{XRD}

\begin{figure}[H]
	\centering
	\includegraphics[width=\linewidth]{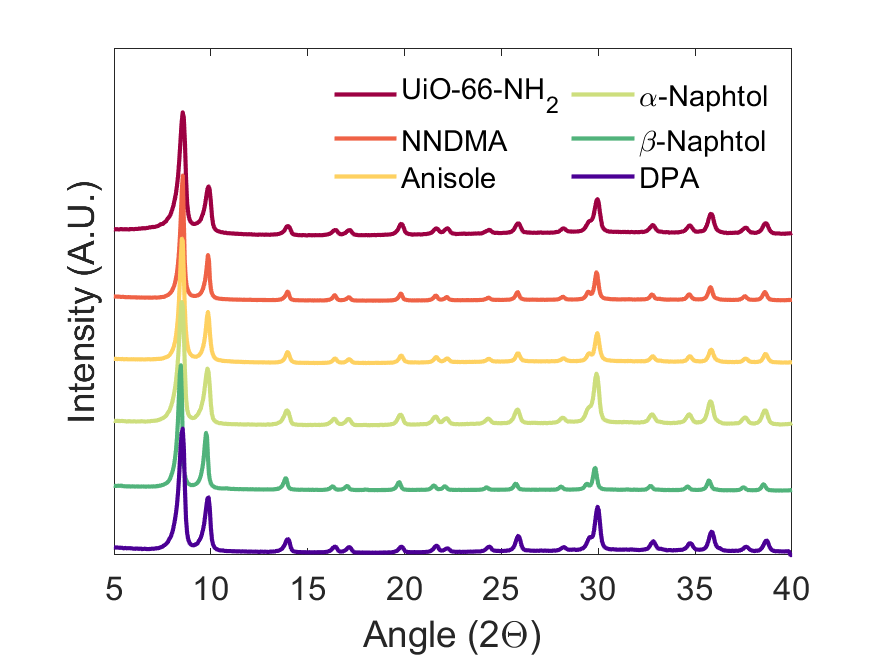} % or XRD.jpg
	\caption{Powder X-ray diffraction (XRD) patterns of UiO-66-NH$_2$ and the aromatic coupling agents used in the PSM (NNDMA: N,N-dimethylaniline, anisole, $\alpha$-naphthol, $\beta$-naphthol, and DPA: Diphenylamine). Patterns for NNDMA, $\alpha$-naphthol, $\beta$-naphthol and DPA were adapted from our previous report.\cite{otal_panchromatic_2016}. Patterns for UiO-66-NH$_2$ and Anisole were measured in the present work. Traces are normalized and vertically offset for clarity.}

	\label{fig:XRD}
\end{figure}

\subsection{BET}

\begin{table}[H]
	\centering
	\caption{BET specific surface areas of UiO-66-NH$_2$ and the aromatic coupling agents investigated in this work. Surface areas were calculated from N$_2$ adsorption isotherms using the Brunauer--Emmett--Teller (BET) model. Isotherms for UiO-66-NH$_2$, NNDMA, $\alpha$-naphthol, $\beta$-naphthol and DPA were adapted from our previous report.\cite{otal_panchromatic_2016}. Isotherm for Anisole was measured in the present work.}
	\label{tab:SI-BET}
	\begin{tabular}{l c}
		\hline
		Sample & BET surface area (m$^2$ g$^{-1}$) \\
		\hline
		UiO-66-NH$_2$            & 650 \\
		$\alpha$-naphthol       & 630 \\
		$\beta$-naphthol        & 360 \\
		N,N-dimethylaniline     & 540 \\
		Diphenylamine           & 620 \\
		Anisole                 & 554 \\
		\hline
	\end{tabular}
\end{table}

\subsection{TGA} \label{sec:TGA}

\begin{table}[htbp]
	\centering
	\caption{TGA summary, solvent loss, final mass, and normalized final mass (residue).}
	\label{tab:tga_summary_narrow}
	\scriptsize
	\setlength{\tabcolsep}{4pt}
	\renewcommand{\arraystretch}{1.15}
	\begin{tabular}{>{\raggedright\arraybackslash}p{3.0cm}
			>{\raggedright\arraybackslash}p{2.6cm}
			>{\raggedright\arraybackslash}p{2.5cm}
			>{\raggedright\arraybackslash}p{2.8cm}}
		\toprule
		Sample &
		TGA solvent loss (\%) $m_1$ &
		TGA final mass (\%) $m_f$ &
		Normalized final mass (\%) $m_3$ \\
		\midrule
		UiO-66              & 13.6 & 38.4 & 44.5 \\
		UiO-66-NH$_2$       &  6.5 & 38.5 & 41.1 \\
		% UiO-66-OH           & N/A  & N/A  & N/A  \\
		N,N-dimethylaniline &  9.8 & 37.4 & 41.4 \\
		1-naphthol          & 10.5 & 38.2 & 38.3 \\
		2-naphthol          &  7.9 & 36.5 & 39.6 \\
		Diphenylamine       &  7.1 & 38.8 & 41.7 \\
		Anisole		        &11.77 & 37.8 & 42.8 \\
		\bottomrule
	\end{tabular}
\end{table}

\begin{table}[H]
	\centering
	\caption{Molecular formulas, ideal residues, TGA-derived average molar masses, and estimated diazo-coupling yields.}
	\label{tab:tga_mr_yield_narrow}
	\scriptsize
	\setlength{\tabcolsep}{3pt}
	\renewcommand{\arraystretch}{1.12}
	\begin{tabular}{>{\raggedright\arraybackslash}p{3.5cm}
			>{\raggedright\arraybackslash}p{3.9cm}
			>{\raggedright\arraybackslash}p{1.7cm}
			>{\raggedright\arraybackslash}p{1.5cm}
			>{\raggedright\arraybackslash}p{2.2cm}
			>{\raggedright\arraybackslash}p{1.8cm}}
		\toprule
		Sample &
		MOF molecular formula &
		MOF molar mass (g/mol) &
		Ideal residue (\%) &
		Avg. molar mass from TGA (g/mol) &
		Diazo-coupling yield (\%) \\
		\midrule
		UiO-66 &
		$\mathrm{Zr}_{24}\mathrm{O}_{120}\mathrm{C}_{192}\mathrm{H}_{96}$ &
		6509 & 45.4 & N/A  & N/A \\
		
		UiO-66-NH$_2$ &
		$\mathrm{Zr}_{24}\mathrm{O}_{120}\mathrm{C}_{192}\mathrm{H}_{144}\mathrm{N}_{24}$ &
		6869 & 43.0 & N/A  & N/A \\
		
		% UiO-66-OH &
		% $\mathrm{Zr}_{24}\mathrm{O}_{144}\mathrm{C}_{192}\mathrm{H}_{96}$ &
		% 6893 & 43.0  & N/A  & N/A \\
		
		$\alpha$-Naphthol &
		$\mathrm{Zr}_{24}\mathrm{O}_{144}\mathrm{C}_{432}\mathrm{H}_{240}\mathrm{N}_{48}$ &
		10589 & 27.9 & 7725 & 22.5 \\
		
		$\beta$-Naphthol &
		$\mathrm{Zr}_{24}\mathrm{O}_{144}\mathrm{C}_{432}\mathrm{H}_{240}\mathrm{N}_{48}$ &
		10589 & 27.9 & 7465 & 15.4 \\
		
		N,N-dimethylaniline &
		$\mathrm{Zr}_{24}\mathrm{O}_{120}\mathrm{C}_{384}\mathrm{H}_{288}\mathrm{N}_{72}$ &
		10013 & 29.5 & 7138 & 7.8 \\
		
		Diphenylamine &
		$\mathrm{Zr}_{24}\mathrm{O}_{144}\mathrm{C}_{480}\mathrm{H}_{312}\mathrm{N}_{72}$ &
		11573 & 25.6 & 7084 & 4.2 \\
		
		Anisole &
		$\mathrm{Zr}_{24}\mathrm{O}_{144}\mathrm{C}_{360}\mathrm{H}_{264}\mathrm{N}_{48}$ &
		9756 & 30.3 & 6910 & 0.6 \\

		\bottomrule
	\end{tabular}
\end{table}

\section{Optical characterization}

\subsubsection{Photoxidation of UiO-66NH$_2$}

\begin{figure}[H]
	\centering
	% --- Subfigura A ---
	\begin{subfigure}[b]{0.70\linewidth}
		\centering
		\begin{overpic}[width=\linewidth]{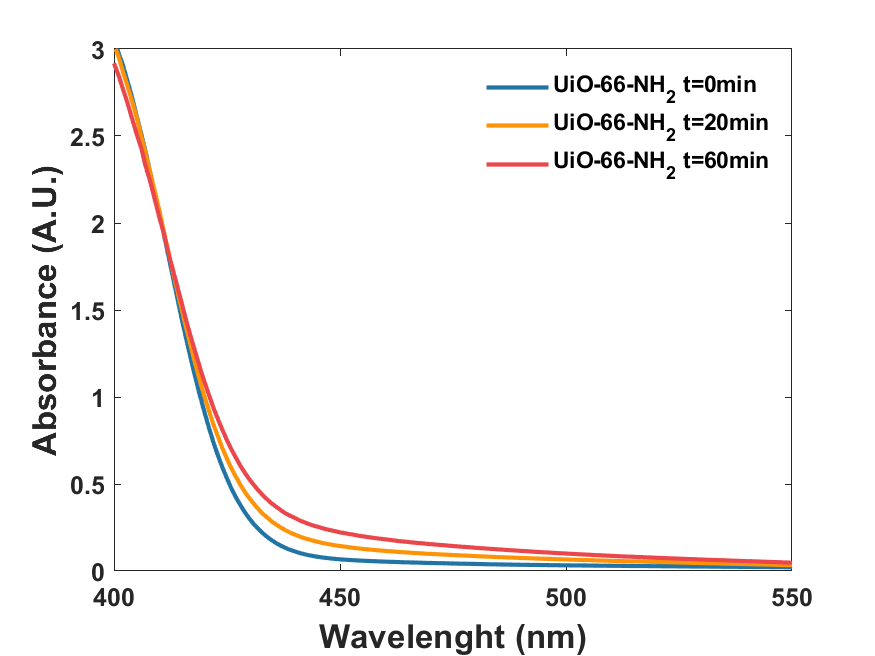}
			\put(-5,65){\color{black}\textbf{(A)}}
		\end{overpic}
	\end{subfigure}
	\caption{Diffuse reflectance UV-Vis spectra of UiO-66-NH$_2$ recorded before and after visible-light irradiation, demonstrating photooxidation of amino groups and subsequent absorption broadening into the NIR.}
	\label{fig:photoxidation_SI}
\end{figure}

\newpage
\subsubsection{Absorbance, Emission and Excitation Spectra}

\begin{figure}[H]
	\centering
	% --- Subfigura A ---
	\begin{subfigure}[b]{0.48\linewidth}
		\centering
		\begin{overpic}[width=\linewidth]{Correlation_NH2_flat.png}
			\put(-5,65){\color{black}\textbf{(A)}}
		\end{overpic}
	\end{subfigure}
	% --- Subfigura A ---
	\begin{subfigure}[b]{0.48\linewidth}
		\centering
		\begin{overpic}[width=\linewidth]{Correlation_Anisole_flat.png}
			\put(-5,65){\color{black}\textbf{(B)}}
		\end{overpic}
	\end{subfigure}
	
	\vspace{0.5cm}
	
	% --- Subfigura B ---
	\begin{subfigure}[b]{0.48\linewidth}
		\centering
		\begin{overpic}[width=\linewidth]{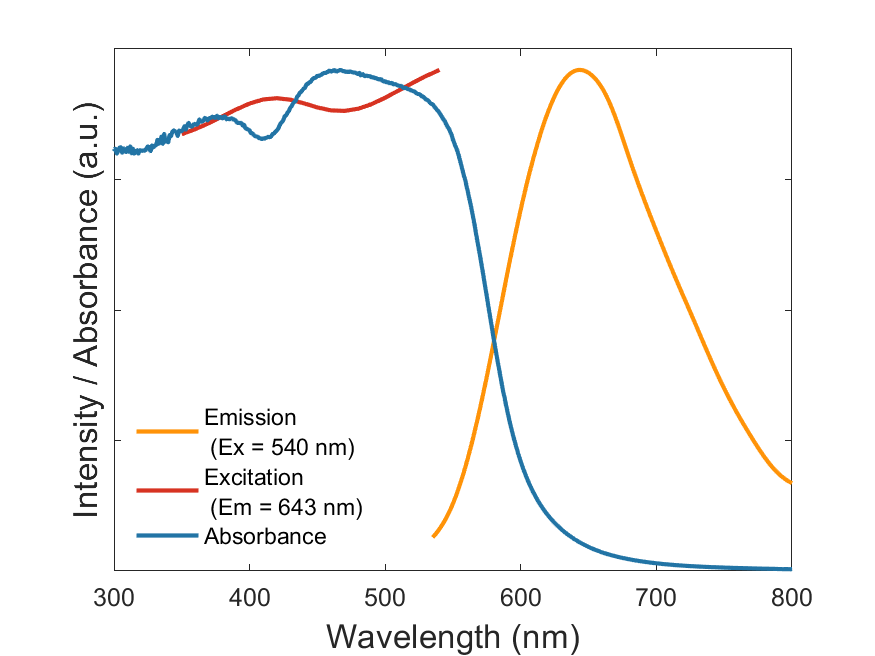}
			\put(-5,65){\color{black}\textbf{(C)}}
		\end{overpic}
	\end{subfigure}
	% --- Subfigura C ---
	\begin{subfigure}[b]{0.48\linewidth}
		\centering
		\begin{overpic}[width=\linewidth]{Correlation_BN_flat.png}
			\put(-5,65){\color{black}\textbf{(D)}}
		\end{overpic}
	\end{subfigure}
	
	% --- Subfigura C ---
	
	\vspace{0.5cm}
	
	\begin{subfigure}[b]{0.48\linewidth}
		\centering
		\begin{overpic}[width=\linewidth]{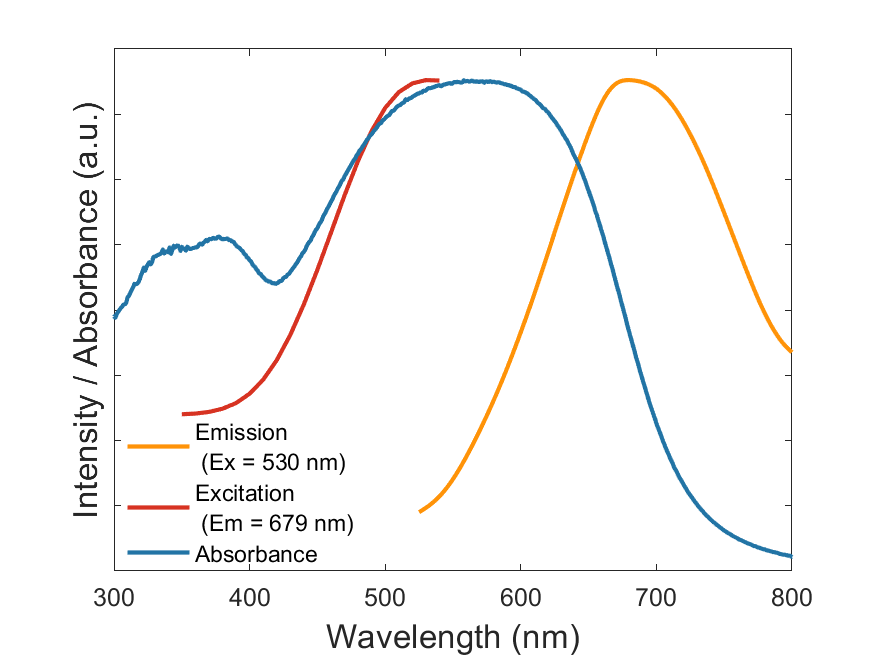}
			\put(-5,65){\color{black}\textbf{(E)}}
		\end{overpic}
	\end{subfigure}
	% --- Subfigura C ---
	\begin{subfigure}[b]{0.48\linewidth}
		\centering
		\begin{overpic}[width=\linewidth]{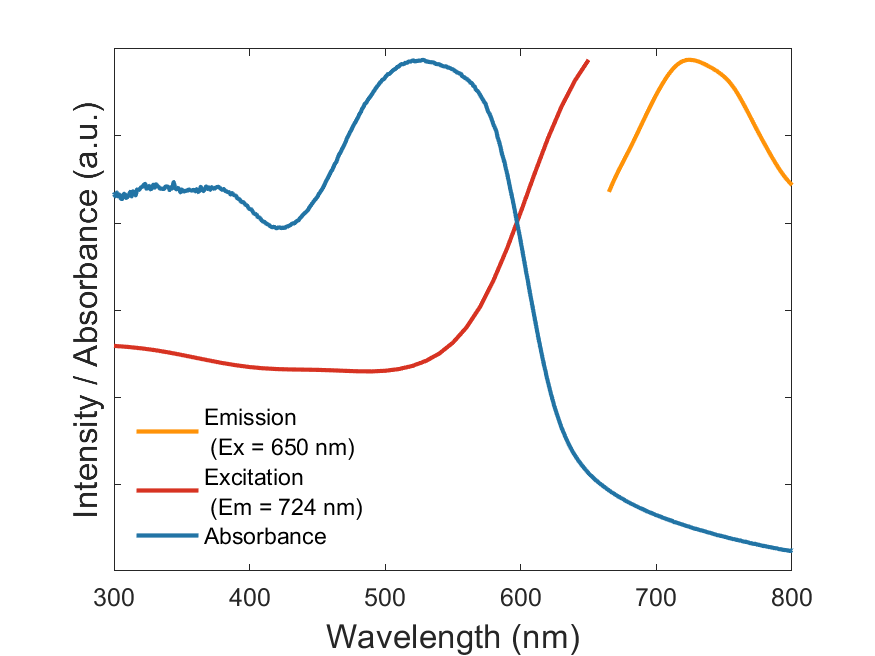}
			\put(-5,65){\color{black}\textbf{(F)}}
		\end{overpic}
	\end{subfigure}
	
	\caption{Blue: absorbance spectrum obtained from Kubelka--Munk transformations of the diffuse reflectance spectra; Red: fluorescence excitation spectrum and  Orange: fluorescence emission spectrum derived from steady--state fluorescence measurements recorded at the emission wavelengths indicated on the legend. \textbf{(A)} UiO--66--NH$_2$, \textbf{(B)} UiO--66--Anisole, \textbf{(C)} UiO--66--AN ($\alpha$--naphthol), \textbf{(D)} UiO--66--BN ($\beta$--naphthol), \textbf{(E)} UiO--66--DFA (diphenylamine), and \textbf{(F)} UiO--66--NNDMA (N,N-dimethyaniline).}
	\label{fig:Abs-Emm-Exc_SI}
\end{figure}

\newpage
\subsubsection{Absorbance and Excitation Spectra Overlap}

\begin{figure}[H]
	\centering
	% --- Subfigura A ---
	\begin{subfigure}[b]{0.48\linewidth}
		\centering
		\begin{overpic}[width=\linewidth]{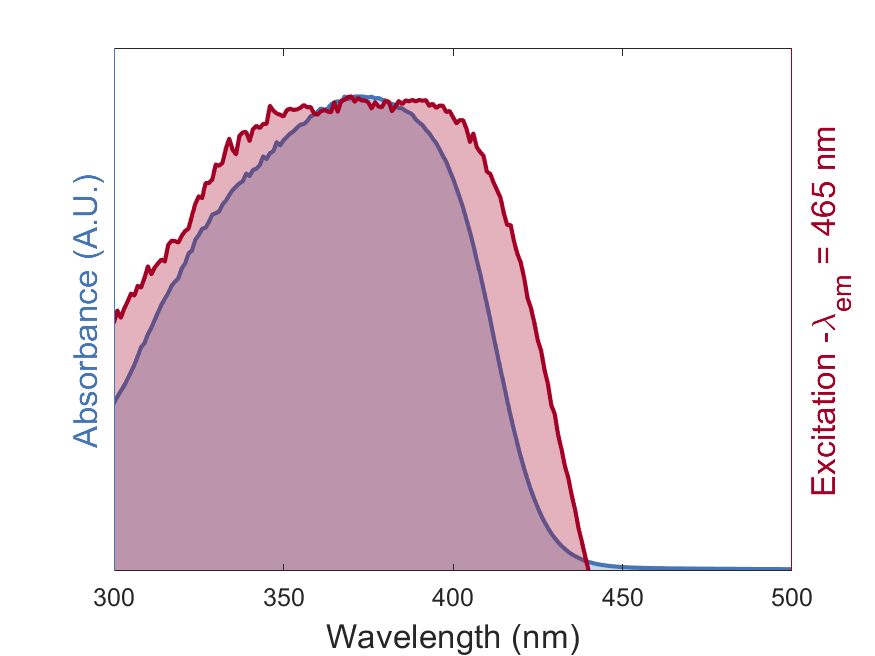}
			\put(-5,65){\color{black}\textbf{(A)}}
		\end{overpic}
	\end{subfigure}
	% --- Subfigura A ---
	\begin{subfigure}[b]{0.48\linewidth}
		\centering
		\begin{overpic}[width=\linewidth]{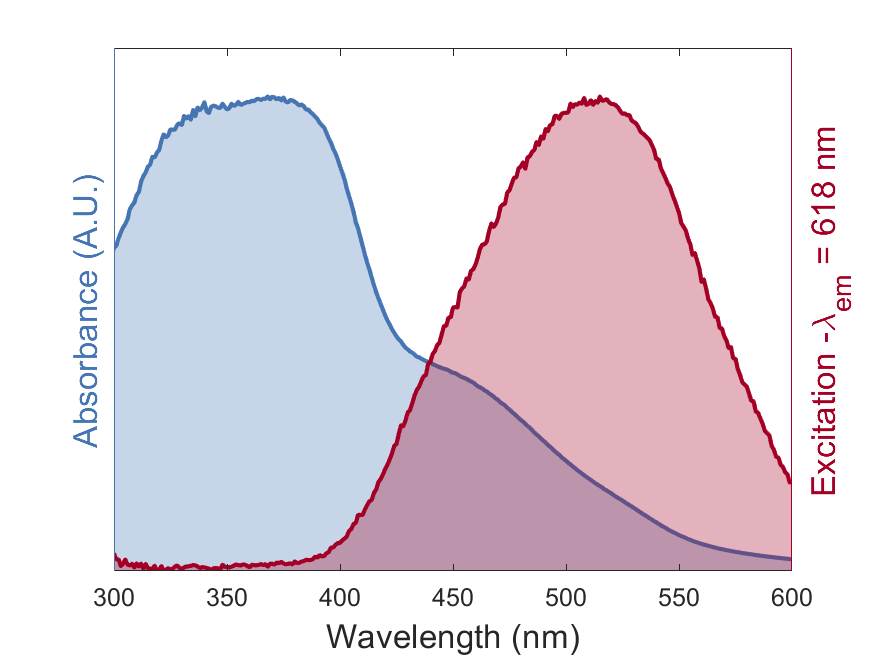}
			\put(-5,65){\color{black}\textbf{(B)}}
		\end{overpic}
	\end{subfigure}
	
	\vspace{0.5cm}
	
	% --- Subfigura B ---
	\begin{subfigure}[b]{0.48\linewidth}
		\centering
		\begin{overpic}[width=\linewidth]{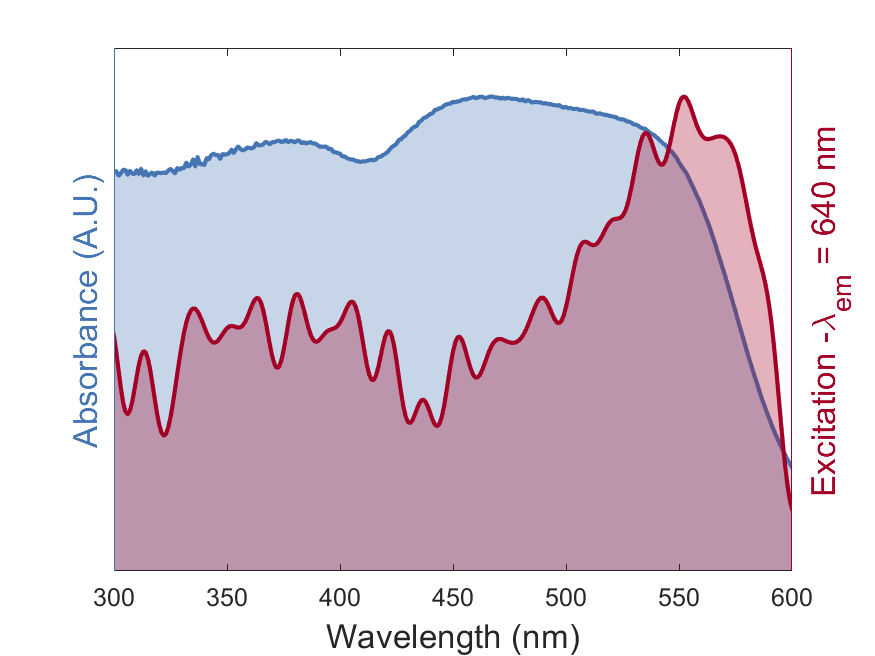}
			\put(-5,65){\color{black}\textbf{(C)}}
		\end{overpic}
	\end{subfigure}
	% --- Subfigura C ---
	\begin{subfigure}[b]{0.48\linewidth}
		\centering
		\begin{overpic}[width=\linewidth]{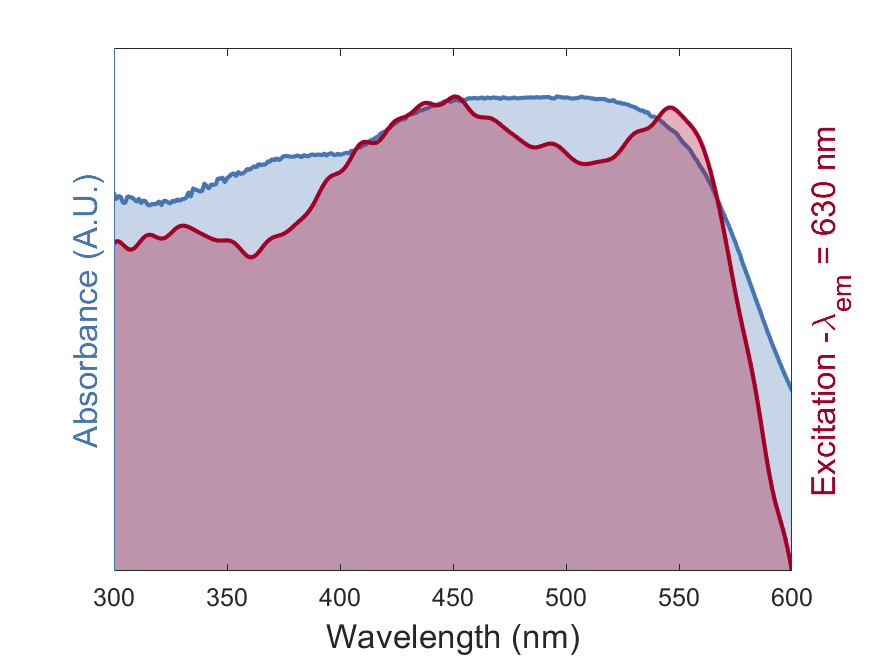}
			\put(-5,65){\color{black}\textbf{(D)}}
		\end{overpic}
	\end{subfigure}
	
	% --- Subfigura C ---
	
	\vspace{0.5cm}
	
	\begin{subfigure}[b]{0.48\linewidth}
		\centering
		\begin{overpic}[width=\linewidth]{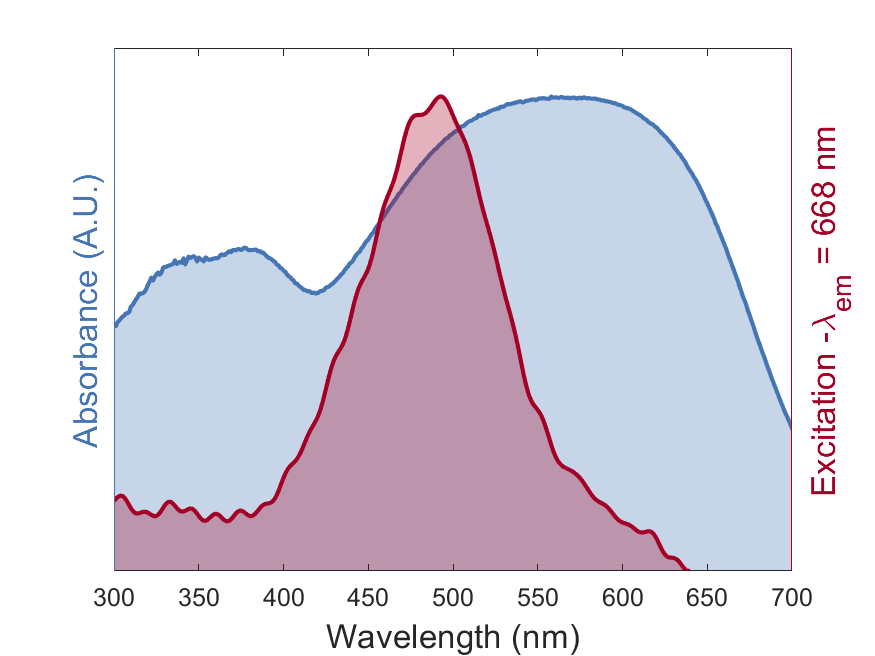}
			\put(-5,65){\color{black}\textbf{(E)}}
		\end{overpic}
	\end{subfigure}
	% --- Subfigura C ---
	\begin{subfigure}[b]{0.48\linewidth}
		\centering
		\begin{overpic}[width=\linewidth]{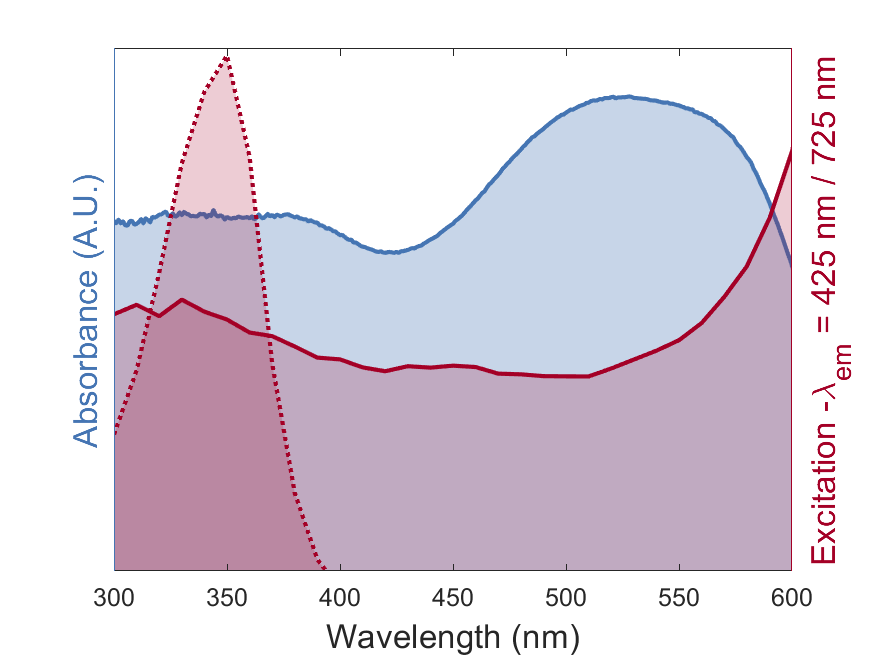}
			\put(-5,65){\color{black}\textbf{(F)}}
		\end{overpic}
	\end{subfigure}
	
	\caption{Blue: absorbance spectrum obtained from Kubelka--Munk transformations of the diffuse reflectance spectra; Red: fluorescence excitation spectrum derived from steady--state fluorescence measurements recorded at the emission wavelengths indicated on the y--axis. \textbf{(A)} UiO--66--NH$_2$, \textbf{(B)} UiO--66--Anisole, \textbf{(C)} UiO--66--AN ($\alpha$--naphthol), \textbf{(D)} UiO--66--BN ($\beta$--naphthol), \textbf{(E)} UiO--66--DFA (diphenylamine), and \textbf{(F)} UiO--66--NNDMA (N,N-dimethyaniline).}
	\label{fig:Abs-Exc_SI}
\end{figure}

\newpage

\subsubsection{Correlation Diagrams}

\begin{figure}[H]
	\centering
	% --- Subfigura A ---
	\begin{subfigure}[b]{0.48\linewidth}
		\centering
		\begin{overpic}[width=\linewidth]{Correlation_NH2.png}
			\put(-5,65){\color{black}\textbf{(A)}}
		\end{overpic}
	\end{subfigure}
	% --- Subfigura A ---
	\begin{subfigure}[b]{0.48\linewidth}
		\centering
		\begin{overpic}[width=\linewidth]{Correlation_Anisole.png}
			\put(-5,65){\color{black}\textbf{(B)}}
		\end{overpic}
	\end{subfigure}
	
	\vspace{0.5cm}
	
	% --- Subfigura B ---
	\begin{subfigure}[b]{0.48\linewidth}
		\centering
		\begin{overpic}[width=\linewidth]{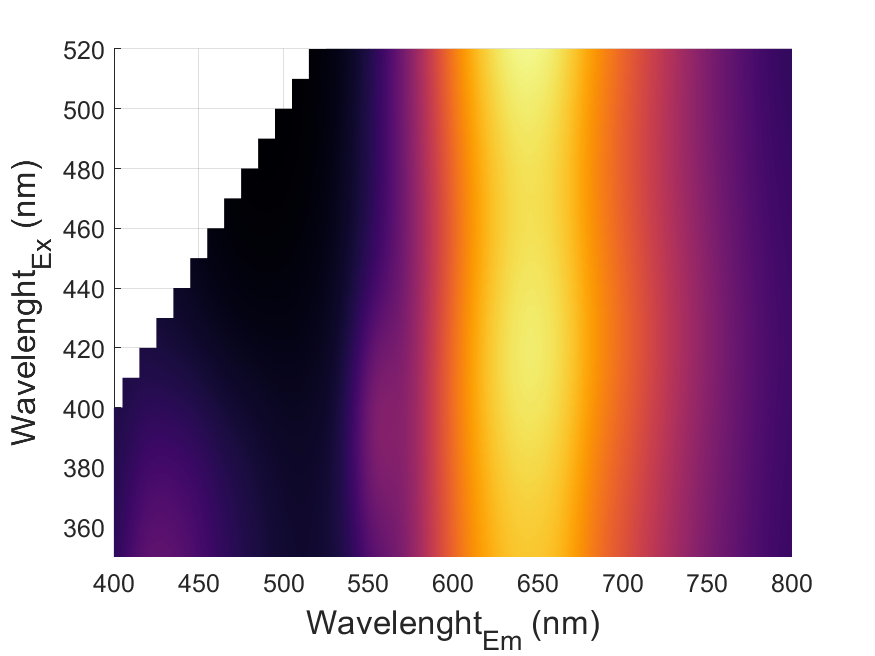}
			\put(-5,65){\color{black}\textbf{(C)}}
		\end{overpic}
	\end{subfigure}
	% --- Subfigura C ---
	\begin{subfigure}[b]{0.48\linewidth}
		\centering
		\begin{overpic}[width=\linewidth]{Correlation_BN.png}
			\put(-5,65){\color{black}\textbf{(D)}}
		\end{overpic}
	\end{subfigure}
	% --- Subfigura C ---
	
	\vspace{0.5cm}
	
	\begin{subfigure}[b]{0.48\linewidth}
		\centering
		\begin{overpic}[width=\linewidth]{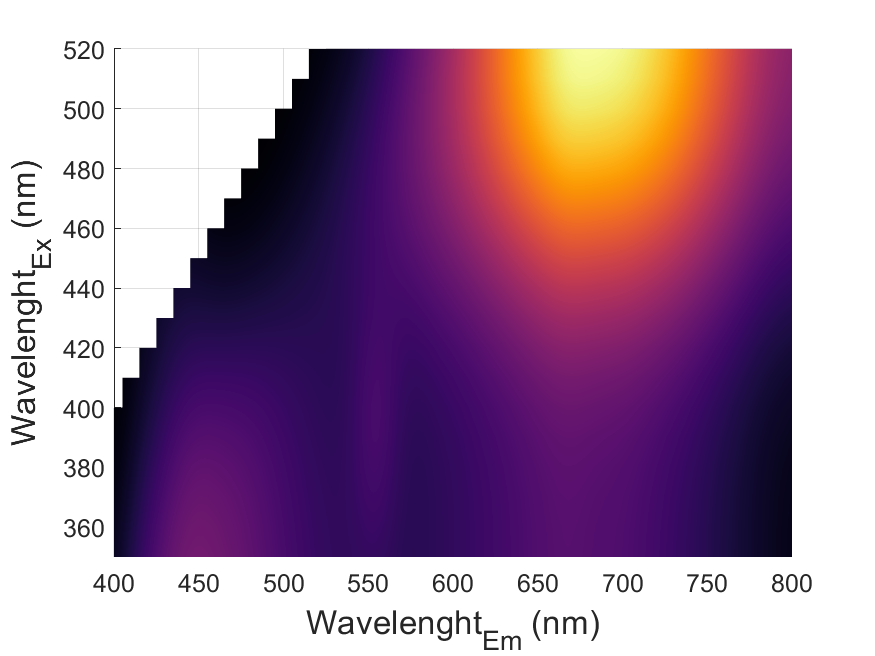}
			\put(-5,65){\color{black}\textbf{(C)}}
		\end{overpic}
	\end{subfigure}
	% --- Subfigura C ---
	\begin{subfigure}[b]{0.48\linewidth}
		\centering
		\begin{overpic}[width=\linewidth]{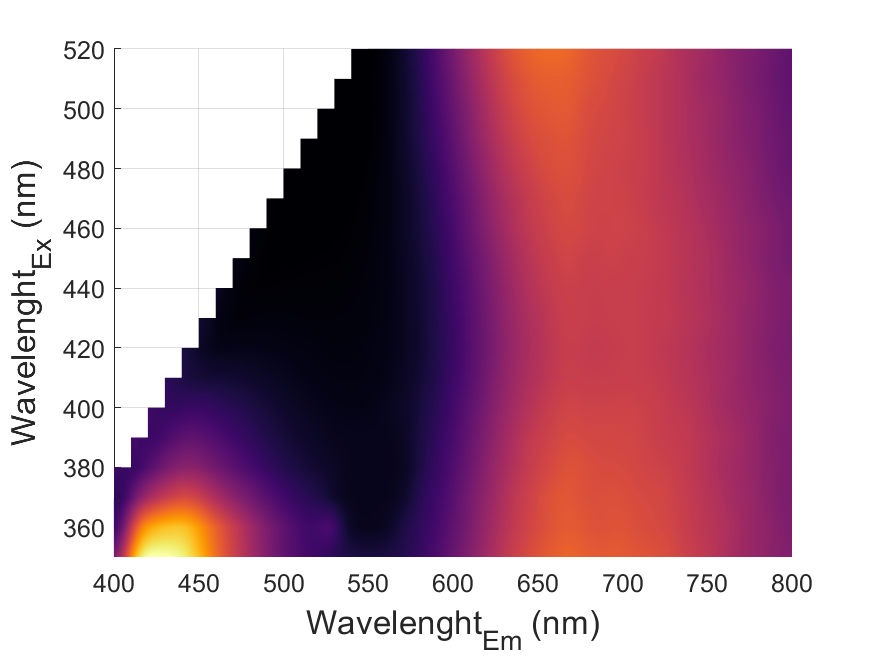}
			\put(-5,65){\color{black}\textbf{(D)}}
		\end{overpic}
	\end{subfigure}
	
	\caption{Fluorescence excitation--emission correlation diagrams obtained from steady-state fluorescence spectra. \textbf{(A)} UiO--66--NH$_2$, \textbf{(B)} UiO--66--Anisole, \textbf{(C)} UiO--66--AN ($\alpha$--naphthol), \textbf{(D)} UiO--66--BN ($\beta$--naphthol), \textbf{(E)} UiO--66--DFA (diphenylamine), and \textbf{(F)} UiO--66--NNDMA (N,N-dimethyaniline).}
	\label{fig:Correlation_SI}
\end{figure}

\section{Electronic characterization}	

\subsection{EPR spectra}

\begin{figure}[H]
	\centering
	% --- Subfigura A ---
	\begin{subfigure}[b]{0.48\linewidth}
		\centering
		\begin{overpic}[width=\linewidth]{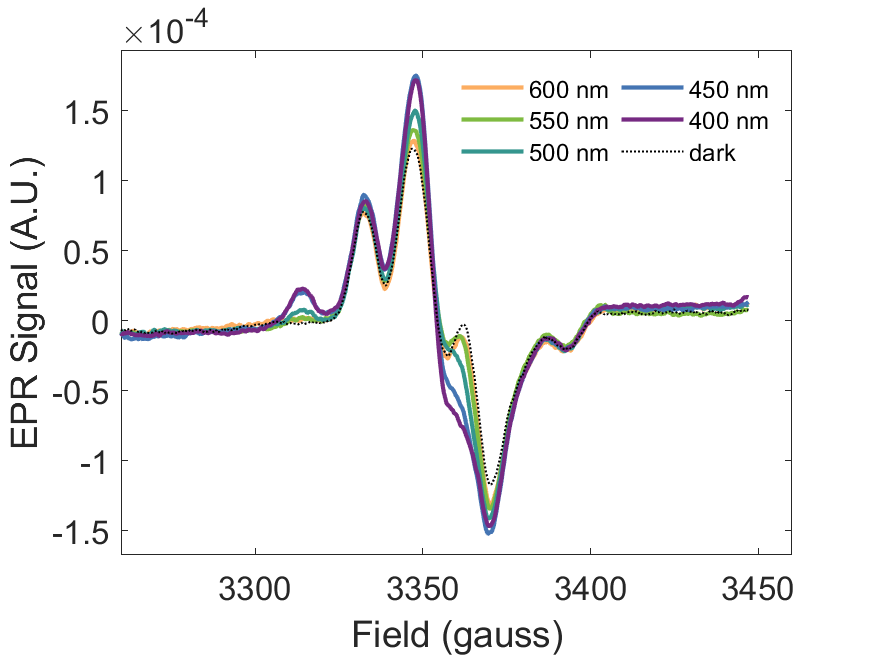}
			\put(0,65){\color{black}\textbf{(A)}}
		\end{overpic}
	\end{subfigure}
	\begin{subfigure}[b]{0.48\linewidth}
		\centering
		\begin{overpic}[width=\linewidth]{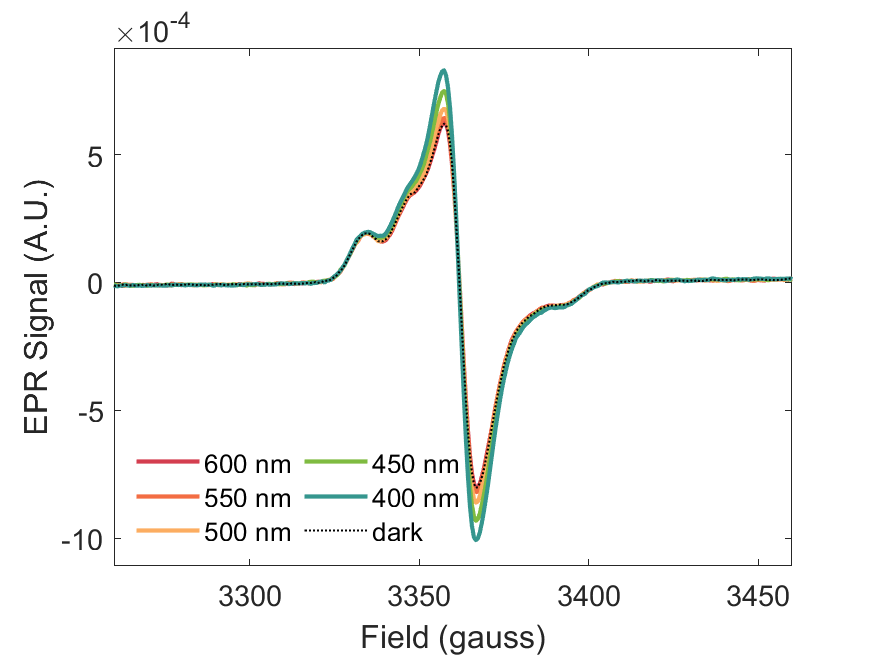}
			\put(0,65){\color{black}\textbf{(B)}}
		\end{overpic}
	\end{subfigure}

	% --- Subfigura B ---
	\begin{subfigure}[b]{0.48\linewidth}
		\centering
		\begin{overpic}[width=\linewidth]{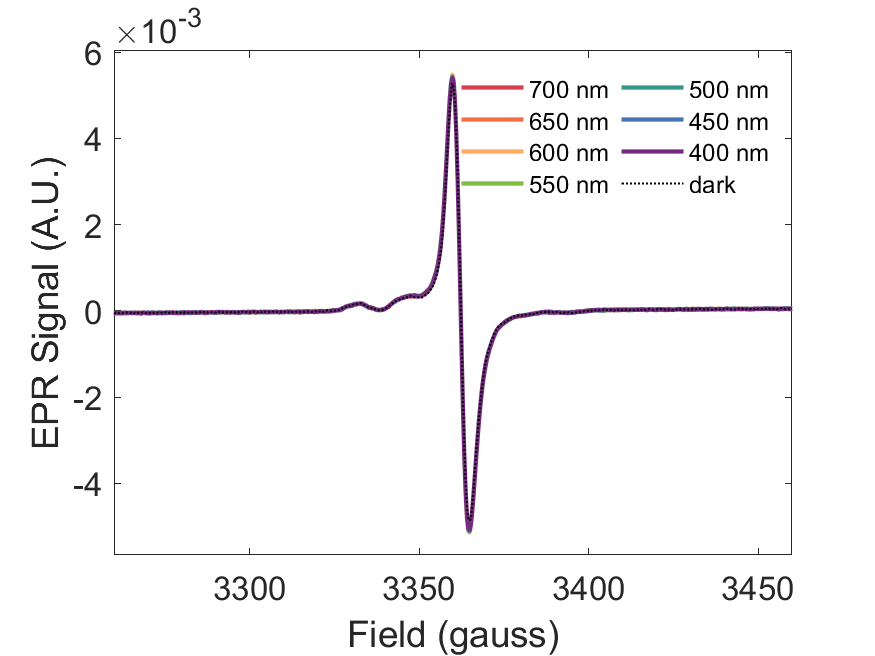}
			\put(0,65){\color{black}\textbf{(C)}}
		\end{overpic}
	\end{subfigure}
	\begin{subfigure}[b]{0.48\linewidth}
		\centering
		\begin{overpic}[width=\linewidth]{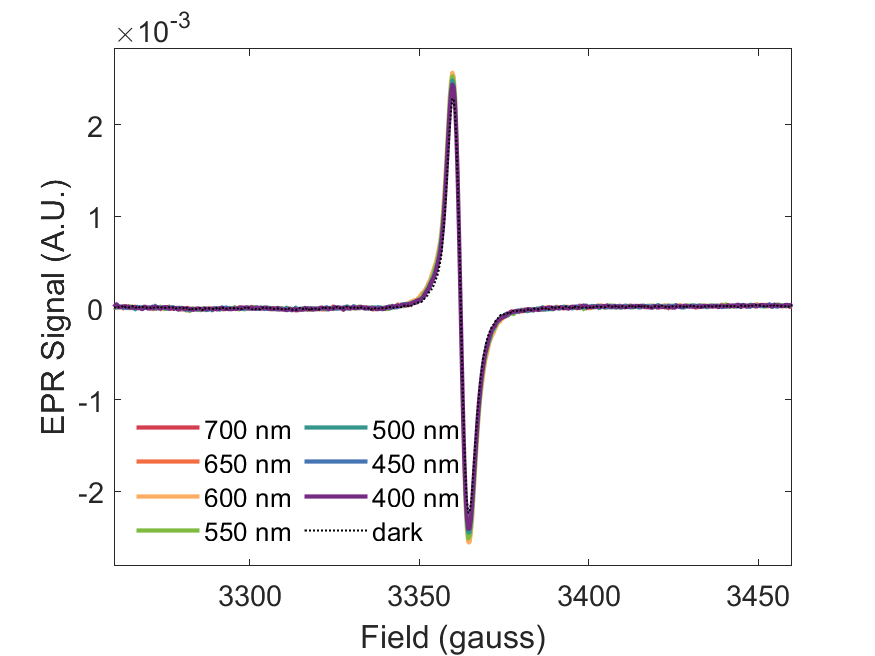}
			\put(0,65){\color{black}\textbf{(D)}}
		\end{overpic}
	\end{subfigure}
	% --- Subfigura C ---
	
	\vspace{0.5cm}
	
	\begin{subfigure}[b]{0.48\linewidth}
		\centering
		\begin{overpic}[width=\linewidth]{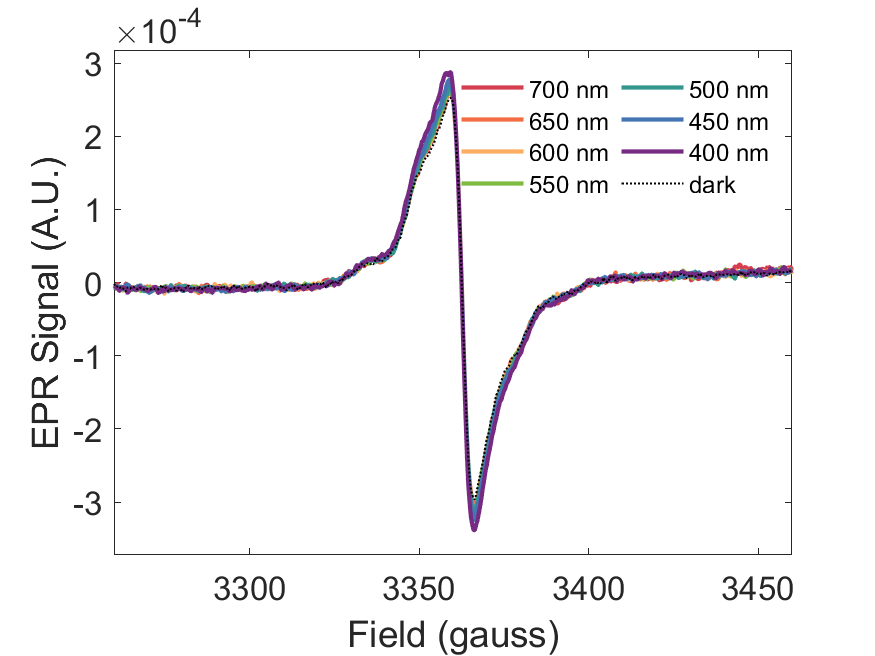}
			\put(0,65){\color{black}\textbf{(E)}}
		\end{overpic}
	\end{subfigure}
	\begin{subfigure}[b]{0.48\linewidth}
		\centering
		\begin{overpic}[width=\linewidth]{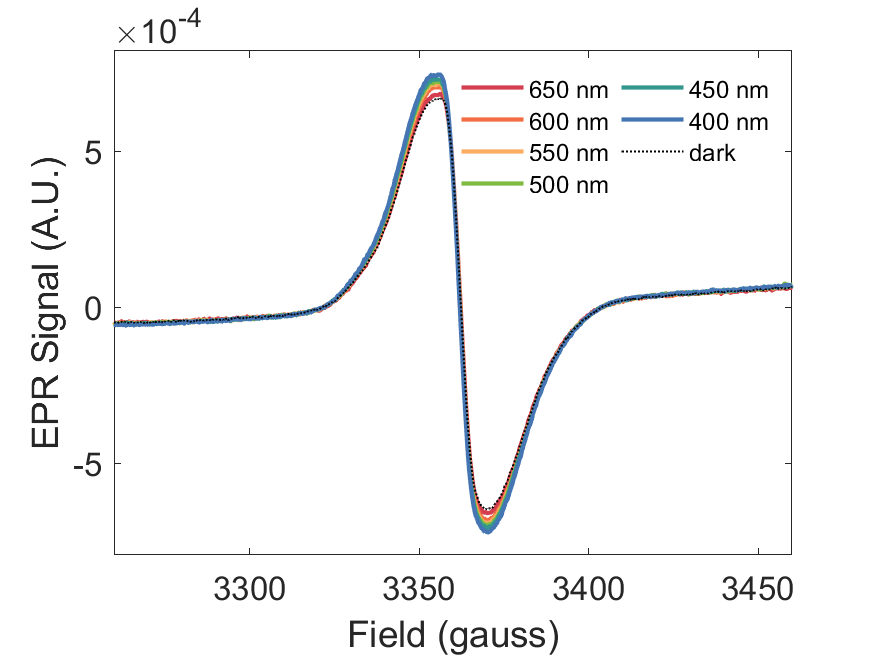}
			\put(0,65){\color{black}\textbf{(F)}}
		\end{overpic}
	\end{subfigure}
	
	\vspace{0.5cm}
	% --- Subfigura A ---

	\caption{Continuous-wave X-band EPR spectra of UiO-66 composites under monochromatic illumination. \textbf{(A)} UiO--66--NH$_2$, \textbf{(B)} UiO--66--Anisole, \textbf{(C)} UiO--66--AN ($\alpha$--naphthol), \textbf{(D)} UiO--66--BN ($\beta$--naphthol), \textbf{(E)} UiO--66--DFA (diphenylamine), and \textbf{(F)} UiO--66--NNDMA (N,N-dimethyaniline).}
	\label{fig:EPR_SI}
\end{figure}

\subsection{EPR differential spectra}

\begin{figure}[H]
	\centering
	% --- Subfigura A ---
	\begin{subfigure}[b]{0.48\linewidth}
		\centering
		\begin{overpic}[width=\linewidth]{Diff_Spectra_UiO-66-NH2.png}
			\put(0,65){\color{black}\textbf{(A)}}
		\end{overpic}
	\end{subfigure}
	% --- Subfigura A ---
	\begin{subfigure}[b]{0.48\linewidth}
		\centering
		\begin{overpic}[width=\linewidth]{Diff_Spectra_UiO-66-Anisole.png}
			\put(0,65){\color{black}\textbf{(B)}}
		\end{overpic}
	\end{subfigure}
	
	\vspace{0.5cm}
	
	% --- Subfigura B ---
	\begin{subfigure}[b]{0.48\linewidth}
		\centering
		\begin{overpic}[width=\linewidth]{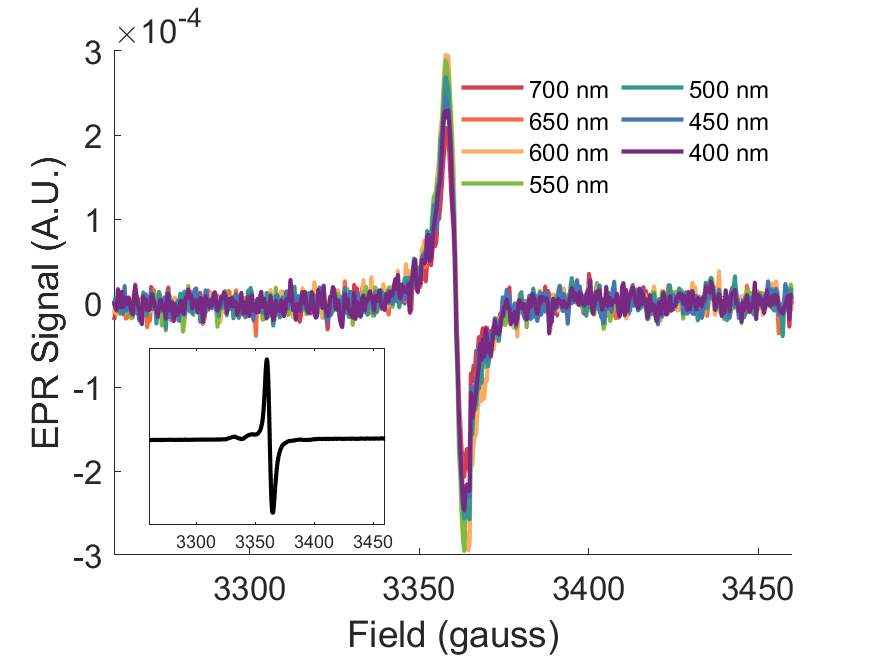}
			\put(0,65){\color{black}\textbf{(C)}}
		\end{overpic}
	\end{subfigure}
	% --- Subfigura C ---
	\begin{subfigure}[b]{0.48\linewidth}
		\centering
		\begin{overpic}[width=\linewidth]{Diff_Spectra_UiO-66-BN.png}
			\put(0,65){\color{black}\textbf{(D)}}
		\end{overpic}
	\end{subfigure}
	
	\vspace{0.5cm}
	
	% --- Subfigura C ---
	\begin{subfigure}[b]{0.48\linewidth}
		\centering
		\begin{overpic}[width=\linewidth]{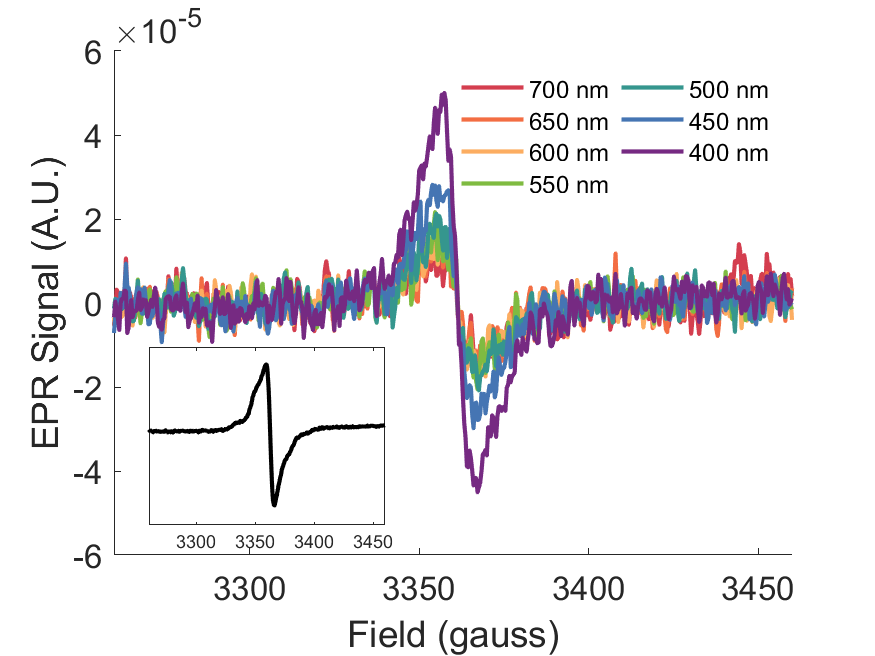}
			\put(0,65){\color{black}\textbf{(E)}}
		\end{overpic}
	\end{subfigure}
	\begin{subfigure}[b]{0.48\linewidth}
		\centering
		\begin{overpic}[width=\linewidth]{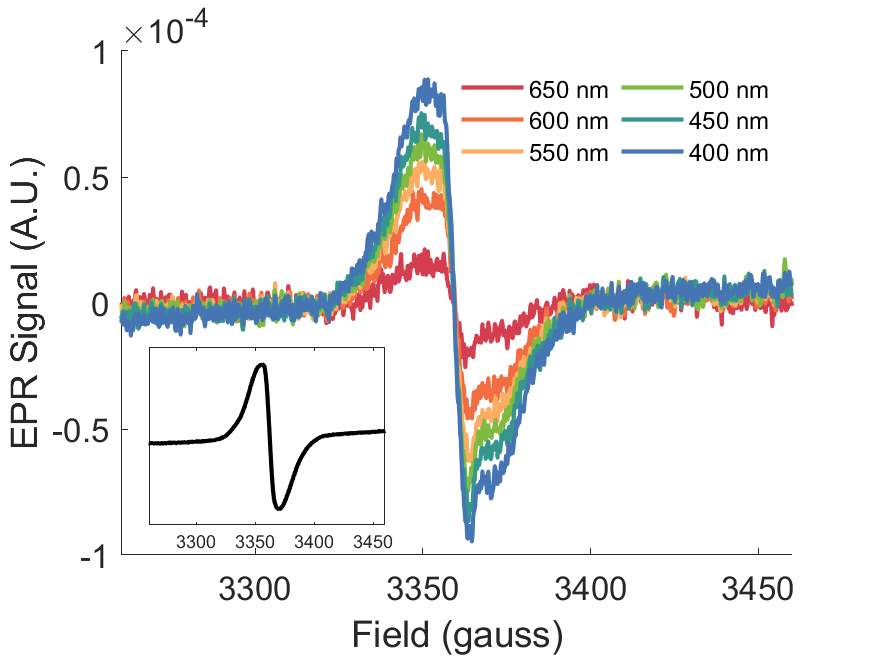}
			\put(0,65){\color{black}\textbf{(F)}}
		\end{overpic}
	\end{subfigure}
	
	\caption{Continuous-wave X-band EPR differential spectra of UiO-66 composites under monochromatic illumination. \textbf{(A)} UiO--66--NH$_2$, \textbf{(B)} UiO--66--Anisole, \textbf{(C)} UiO--66--AN ($\alpha$--naphthol), \textbf{(D)} UiO--66--BN ($\beta$--naphthol), \textbf{(E)} UiO--66--DFA (diphenylamine), and \textbf{(F)} UiO--66--NNDMA (N,N-dimethyaniline).}
	\label{fig:EPR_diff_SI}
\end{figure}

\subsection{EPR Integration and Absorbance}

\begin{figure}[H]
	\centering
	% --- Subfigura A ---
	\begin{subfigure}[b]{0.48\linewidth}
		\centering
		\begin{overpic}[width=\linewidth]{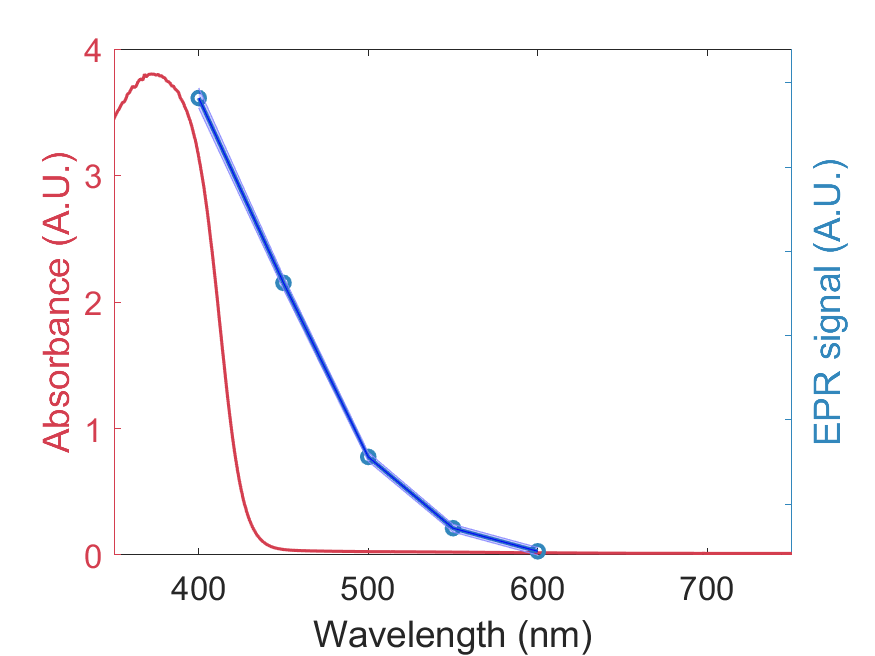}
			\put(0,65){\color{black}\textbf{(A)}}
		\end{overpic}
	\end{subfigure}
	% --- Subfigura A ---
	\begin{subfigure}[b]{0.48\linewidth}
		\centering
		\begin{overpic}[width=\linewidth]{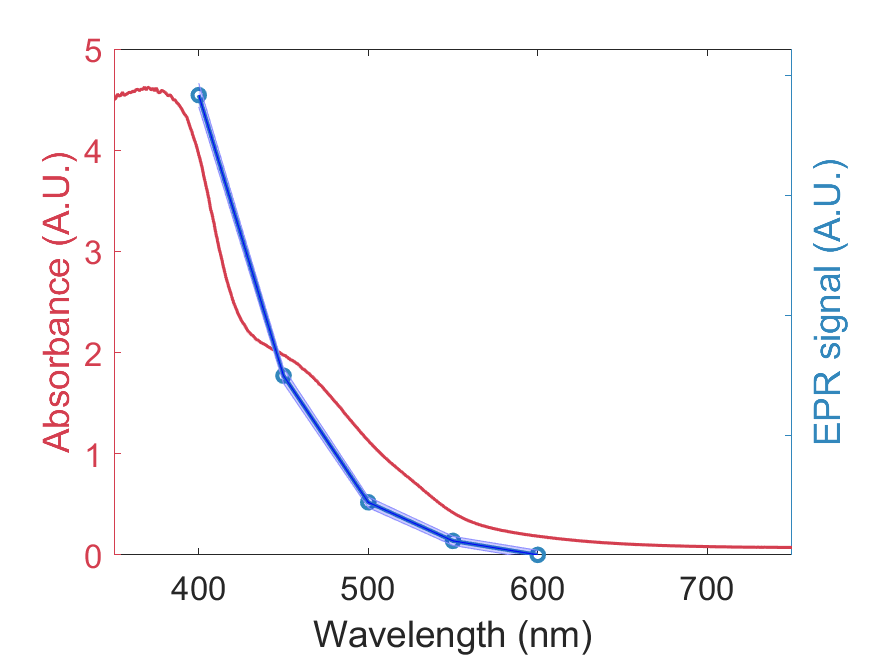}
			\put(0,65){\color{black}\textbf{(B)}}
		\end{overpic}
	\end{subfigure}
	
	\vspace{0.5cm}
	
	% --- Subfigura B ---
	\begin{subfigure}[b]{0.48\linewidth}
		\centering
		\begin{overpic}[width=\linewidth]{Spectra_UiO-66-AN_linear}
			\put(0,65){\color{black}\textbf{(C)}}
		\end{overpic}
	\end{subfigure}
	% --- Subfigura C ---
	\begin{subfigure}[b]{0.48\linewidth}
		\centering
		\begin{overpic}[width=\linewidth]{Spectra_UiO-66-BN_linear}
			\put(0,65){\color{black}\textbf{(D)}}
		\end{overpic}
	\end{subfigure}
	
	\vspace{0.5cm}
	
	% --- Subfigura C ---
	\begin{subfigure}[b]{0.48\linewidth}
		\centering
		\begin{overpic}[width=\linewidth]{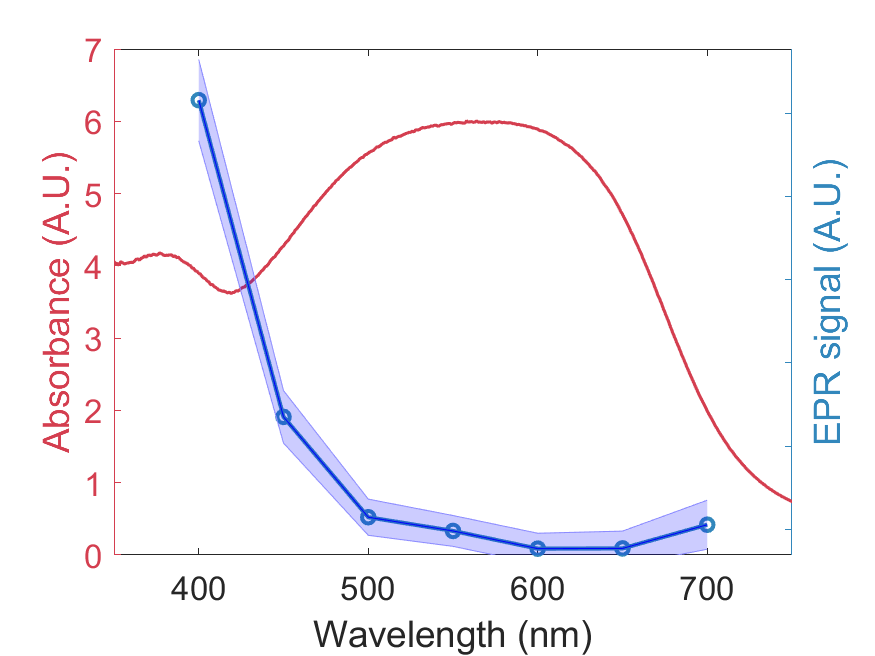}
			\put(0,65){\color{black}\textbf{(E)}}
		\end{overpic}
	\end{subfigure}
	\begin{subfigure}[b]{0.48\linewidth}
		\centering
		\begin{overpic}[width=\linewidth]{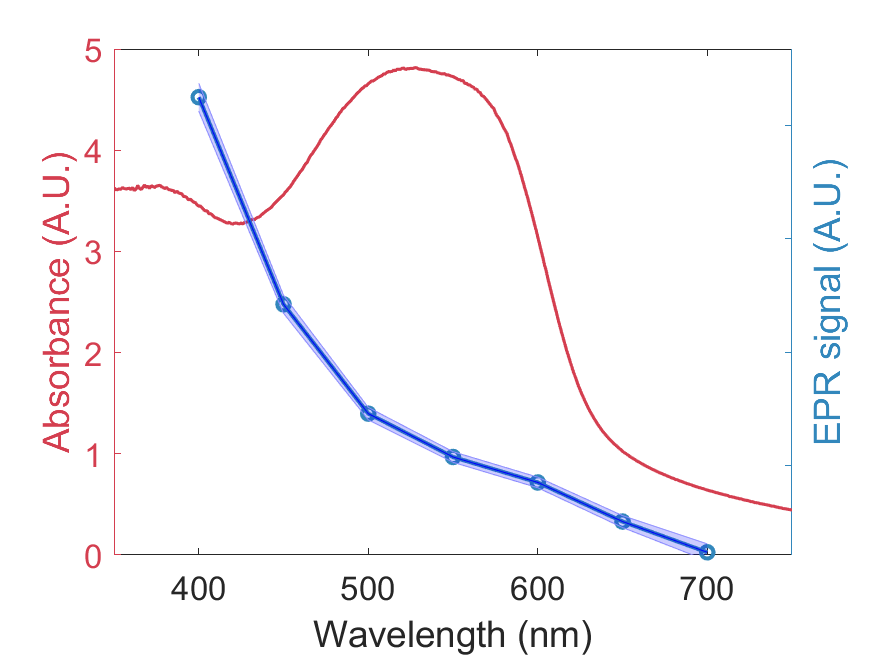}
			\put(0,65){\color{black}\textbf{(F)}}
		\end{overpic}
	\end{subfigure}
	
	\caption{Continuous-wave X-band EPR differential spectra integration of UiO-66 composites under monochromatic illumination and Absorbance spectra.  \textbf{(A)} UiO--66--NH$_2$, \textbf{(B)} UiO--66--Anisole, \textbf{(C)} UiO--66--AN ($\alpha$--naphthol), \textbf{(D)} UiO--66--BN ($\beta$--naphthol), \textbf{(E)} UiO--66--DFA (diphenylamine), and \textbf{(F)} UiO--66--NNDMA (N,N-dimethyaniline).}
	\label{fig:EPR_int_Abs_SI}
\end{figure}

\newpage

\newpage

\subsection{Spectroscopic Characterization of UiO-66-AN ($\alpha$-Naphthol)}

\begin{figure}[H]
	\centering
	\begin{subfigure}[b]{0.48\linewidth}
		\centering
		\begin{overpic}[width=\linewidth]{Correlation_AN_flat.png}
			\put(0,65){\color{black}\textbf{(A)}}
		\end{overpic}
	\end{subfigure}
	\begin{subfigure}[b]{0.48\linewidth}
		\centering
		\begin{overpic}[width=\linewidth]{Correlation_AN.png}
			\put(0,65){\color{black}\textbf{(B)}}
		\end{overpic}
	\end{subfigure}
	
	\vspace{0.5cm}
	\begin{subfigure}[b]{0.48\linewidth}
		\centering
		\begin{overpic}[width=\linewidth]{Diff_Spectra_UiO-66-AN.png}
			\put(0,65){\color{black}\textbf{(C)}}
		\end{overpic}
	\end{subfigure}
	\begin{subfigure}[b]{0.48\linewidth}
		\centering
		\begin{overpic}[width=\linewidth]{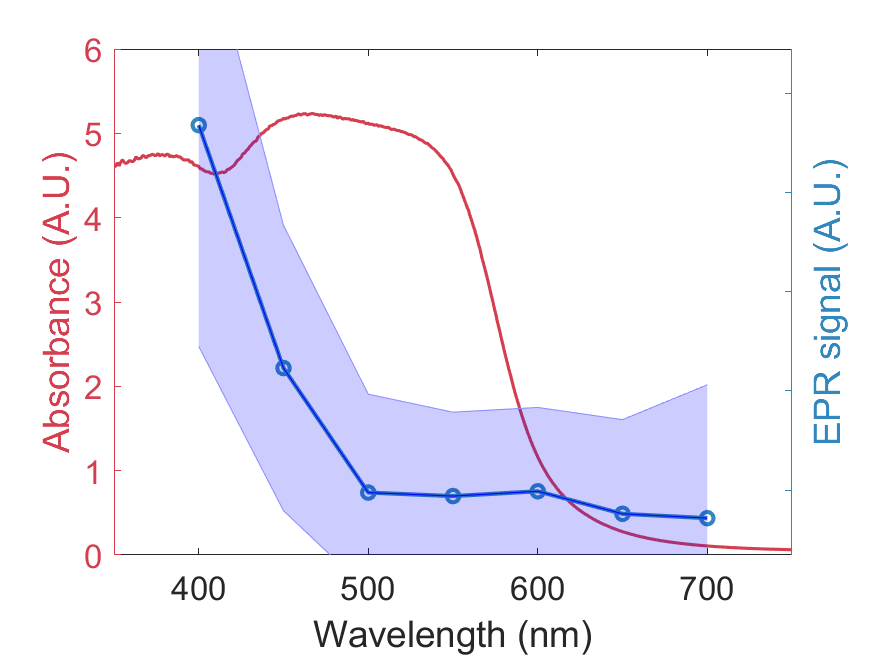}
			\put(0,65){\color{black}\textbf{(D)}}
		\end{overpic}
	\end{subfigure}
	\caption{Spectroscopic characterization for UiO-66-AN ($\alpha$-naphthol): \textbf{(A)} Blue: absorbance spectrum; Red: fluorescence excitation spectrum; Orange: fluorescence emission spectrum recorded at the indicated emission wavelengths. \textbf{(B)} 2D fluorescence excitation--emission correlation map. \textbf{(C)} Continuous-wave X-band EPR differential spectra under monochromatic illumination (inset: dark reference spectrum). \textbf{(D)} Integrated photo-EPR response under monochromatic illumination overlaid with the absorbance spectrum. Shaded area represents experimental error.}
	\label{fig:Patchwork-AN}
\end{figure}

The multi-spectroscopic characterization for UiO-66-AN ($\alpha$-naphthol) is compiled in \Cref{fig:Patchwork-AN}. In contrast to $\beta$-naphthol, the absence of a geometrically rigid intramolecular hydrogen bond in UiO-66-AN allows a significant fraction of the excited population to bypass the non-conductive hydrazone trap, resulting in moderate photocatalytic degradation of methylene blue (58\%).

\newpage

\subsubsection{UiO-66-Dyphenilamine and UiO-66-N,N-Dimethylaniline}

\begin{figure}[H]
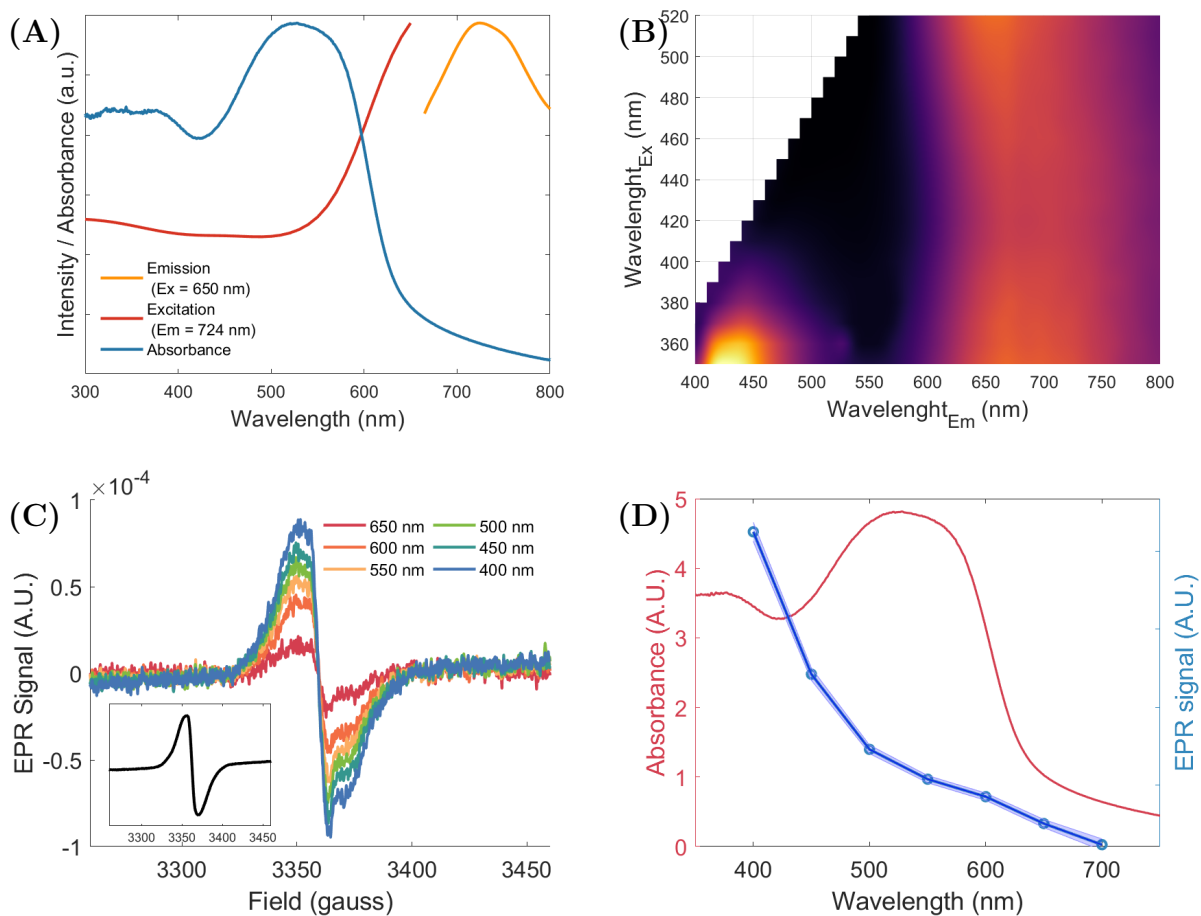

	\centering
	\begin{subfigure}[b]{0.48\linewidth}
		\centering
		\begin{overpic}[width=\linewidth]{Correlation_NNDMA_flat.png}
			\put(0,65){\color{black}\textbf{(A)}}
		\end{overpic}
	\end{subfigure}
	\begin{subfigure}[b]{0.48\linewidth}
		\centering
		\begin{overpic}[width=\linewidth]{Correlation_NNDMA53.png}
			\put(0,65){\color{black}\textbf{(B)}}
		\end{overpic}
	\end{subfigure}
	
	\vspace{0.5cm}
	\begin{subfigure}[b]{0.48\linewidth}
		\centering
		\begin{overpic}[width=\linewidth]{Diff_Spectra_UiO-66-NNDMA.png}
			\put(0,65){\color{black}\textbf{(C)}}
		\end{overpic}
	\end{subfigure}
	\begin{subfigure}[b]{0.48\linewidth}
		\centering
		\begin{overpic}[width=\linewidth]{Spectra_UiO-66-NNDMA_linear}
			\put(0,65){\color{black}\textbf{(D)}}
		\end{overpic}
	\end{subfigure}
	
	\caption{Spectroscopic characterization for UiO-66-NNDMA : \textbf{(A)} Blue: absorbance spectrum; Red: fluorescence excitation spectrum; Orange: fluorescence emission spectrum recorded at the indicated emission wavelengths. \textbf{(B)} 2D fluorescence excitation--emission correlation map. \textbf{(C)} Continuous-wave X-band EPR differential spectra under monochromatic illumination (inset: dark reference spectrum). \textbf{(D)} Integrated photo-EPR response under monochromatic illumination overlaid with the absorbance spectrum. Shaded area represents experimental error.}
	\label{fig:Patchwork-NNDMA}
\end{figure}

\begin{figure}[H]
	\centering
	\begin{subfigure}[b]{0.48\linewidth}
		\centering
		\begin{overpic}[width=\linewidth]{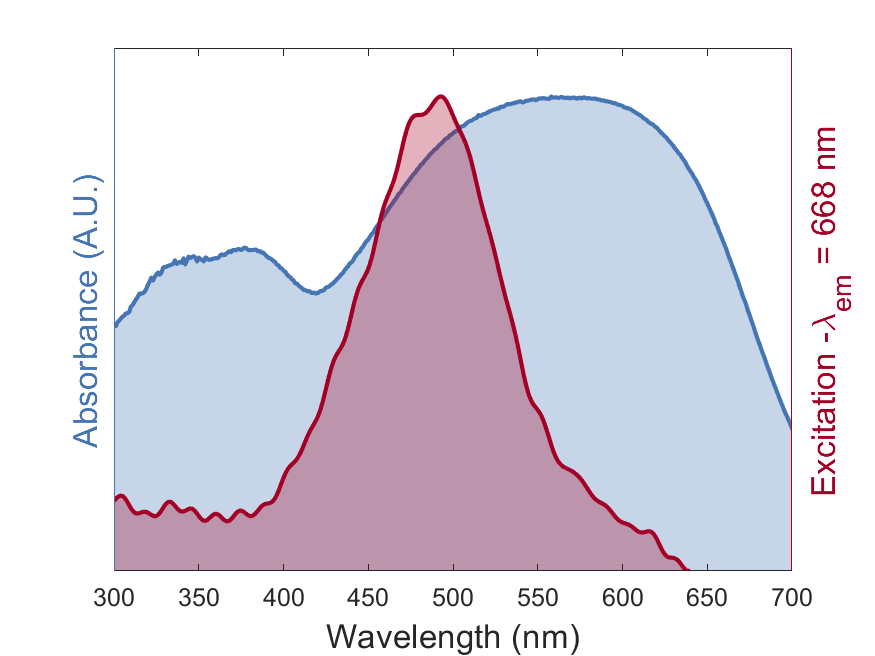}
			\put(0,65){\color{black}\textbf{(A)}}
		\end{overpic}
	\end{subfigure}
	\begin{subfigure}[b]{0.48\linewidth}
		\centering
		\begin{overpic}[width=\linewidth]{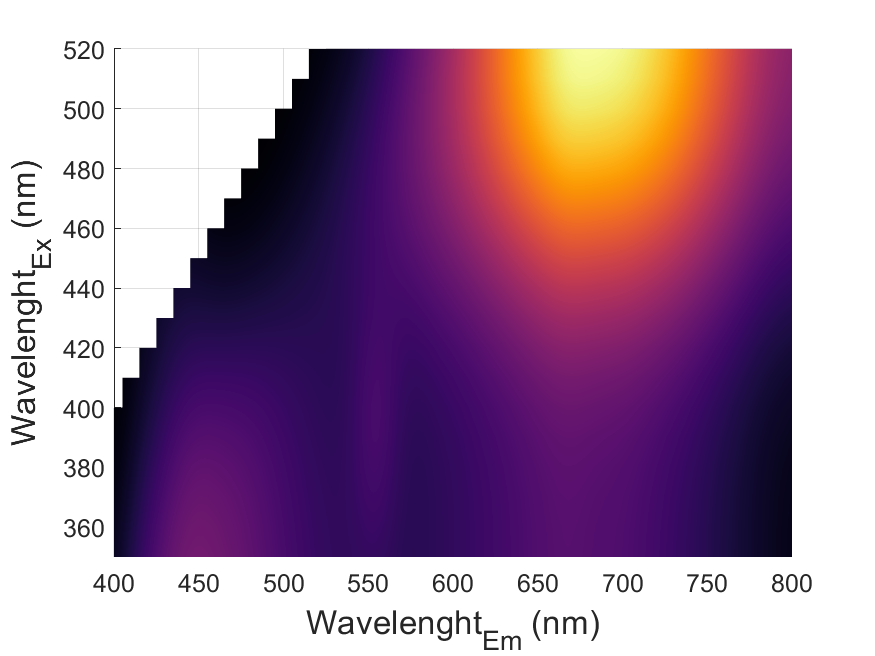}
			\put(0,65){\color{black}\textbf{(B)}}
		\end{overpic}
	\end{subfigure}
	
	\vspace{0.5cm}
	\begin{subfigure}[b]{0.48\linewidth}
		\centering
		\begin{overpic}[width=\linewidth]{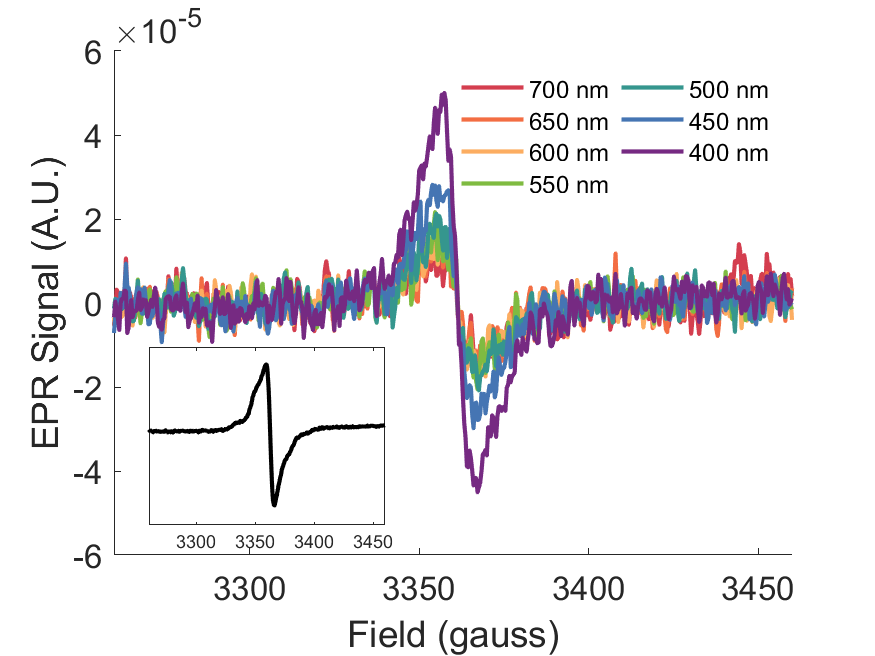}
			\put(0,65){\color{black}\textbf{(C)}}
		\end{overpic}
	\end{subfigure}
	\begin{subfigure}[b]{0.48\linewidth}
		\centering
		\begin{overpic}[width=\linewidth]{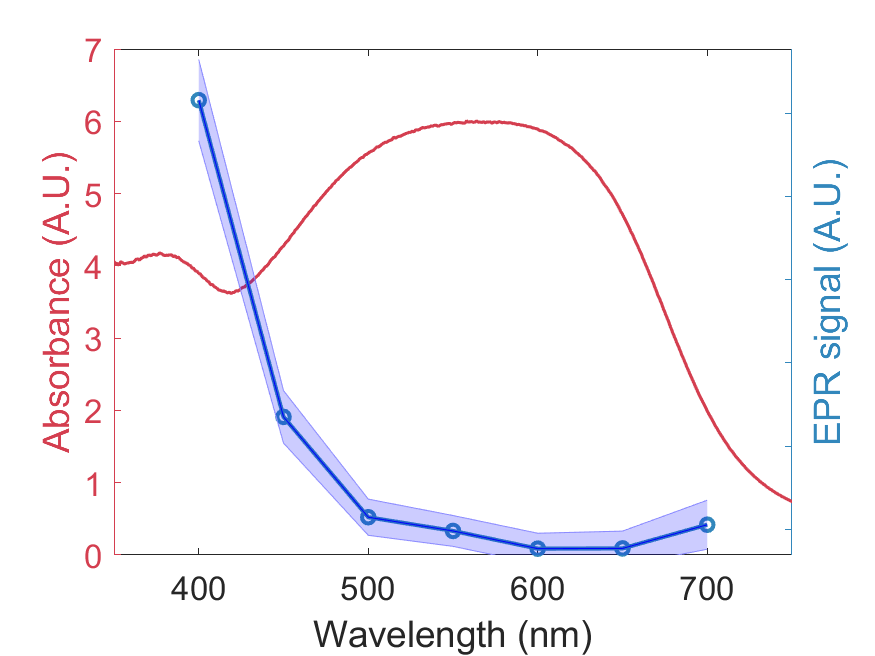}
			\put(0,65){\color{black}\textbf{(D)}}
		\end{overpic}
	\end{subfigure}
	\caption{Spectroscopic characterization for UiO-66-DPA : \textbf{(A)} Blue: absorbance spectrum; Red: fluorescence excitation spectrum; Orange: fluorescence emission spectrum recorded at the indicated emission wavelengths. \textbf{(B)} 2D fluorescence excitation--emission correlation map. \textbf{(C)} Continuous-wave X-band EPR differential spectra under monochromatic illumination (inset: dark reference spectrum). \textbf{(D)} Integrated photo-EPR response under monochromatic illumination overlaid with the absorbance spectrum. Shaded area represents experimental error.}
	\label{fig:Patchwork-DFA}
\end{figure}

UiO-66-Diphenylamine (UiO-66-DPA) and UiO-66-N,N-dimethylaniline (UiO-66-NNDMA) exhibit absorption spectra consistent with protonation of the grafted azo dye after the AcOH washing step (see \Cref{fig:Patchwork-DFA}-D and \Cref{fig:Patchwork-NNDMA}-D). For UiO-66-DPA, this interpretation is supported by the halochromic response reported for related diphenylamine-based azo dyes upon acidification \cite{aksungur_photophysical_2015}; the analogous spectral changes observed for UiO-66-NNDMA suggest a similar effect.

The X-band EPR spectra of UiO-66-DPA (\Cref{fig:Patchwork-DFA}-C) and UiO-66-NNDMA (\Cref{fig:Patchwork-NNDMA}-C) show the appearance of a broad signal that can be reasonably reproduced with a slightly anisotropic $g$-tensor ($g_x = 2.003$, $g_y = 2.003$, $g_z = 2.006$). These $g$ values are not characteristic of typical oxygen-centered radicals and are instead more consistent with delocalized organic radicals derived from the grafted amine-containing moieties. The absence of resolved hyperfine structure further suggests that the unpaired spin is significantly delocalized, in agreement with previous reports on diphenylamine-derived radical species \cite{garcia_generation_2001}. In the case of UiO-66-DPA, the spectrum is more complex, containing the anisotropic signal and  additional contribution related to the radical species already observed for UiO-66-AN; a more detailed discussion of this system has been presented by Kultaeva \textit{et al.} \cite{kultaeva_mechanistic_2026}.

A plausible origin of these radicals is nitrosative side chemistry occurring during diazotization, promoted by nitrite accumulated in the MOF pores and near the outer crystal region, as previously inferred for the UiO-66-AN and UiO-66-BN samples. Under these conditions, secondary amines such as diphenylamine can be nitrosated directly, whereas tertiary amines such as N,N-dimethylaniline require an initial nitrosative dealkylation step before nitrosamine formation becomes possible \cite{smith_nitrosative_1967,gowenlock_nitrosative_1979,ashworth_consideration_2023}. This makes the formation of nitrosated intermediates chemically plausible in both UiO-66-DPA and UiO-66-NNDMA, although through different reaction pathways (See \ref{sec:nitrosation}). These intermediates are relevant because they provide a chemically reasonable precursor to stable, delocalized radical species with slightly anisotropic EPR signatures, consistent with the observed spectra. Simple nitrosamines have long been known to undergo photochemical N--N bond cleavage under UV excitation, producing NO$\bullet$ together with amine-derived radical species \cite{bamford_3_1939,geiger_photodissociation_1981}. Although our samples were not intentionally exposed to UV light after synthesis, visible-light-triggered NO$\bullet$ release has also been demonstrated for suitably activated \textit{N}-nitrosated chromophores and for \textit{N}-pyramidal nitrosamines \cite{he_ring-restricted_2018,karaki_visible-light-triggered_2012}. Therefore, if nitrosated intermediates are formed during functionalization, their subsequent light-induced cleavage would provide a chemically plausible route to the broad, delocalized radical signals observed by EPR. While direct identification of such intermediates was not obtained in the present work, this pathway remains consistent with the available chemical precedent and with
the spectroscopic observations.

For UiO-66-DPA, the excitation spectrum is embedded within the absorption envelope but does not fully overlap with it (see \Cref{fig:Patchwork-DFA}-A). Three spectral regions can be distinguished. In the high-energy region, below approximately 450~nm, light is absorbed without a corresponding strong fluorescence response, while the integrated photo-EPR signal is prominent(\Cref{fig:Patchwork-DFA}-D), indicating that this part of the spectrum is the most effective for generating long-lived paramagnetic species. In the intermediate region, between about 450 and 550~nm, absorption is still significant but is accompanied by fluorescence emission, whereas the integrated EPR response becomes weak, consistent with stronger competition from radiative relaxation. Finally, in the low-energy region, from about 570~nm into the NIR, neither significant fluorescence nor an appreciable integrated EPR signal is observed, despite residual absorption. This behavior agrees well with the TDDFT results, which place the charge-transfer (CT) states responsible for charge-separated species mainly below 500~nm \cite{kultaeva_mechanistic_2026}. Although this partitioning is especially clear for UiO-66-DPA, the same general trend is observed across the full sample series: the integrated EPR signal rises preferentially at higher excitation energies, particularly below 500~nm. This suggests that future MOF--dye sensitization strategies should prioritize stronger absorption in the region below 550~nm while suppressing competing fluorescence pathways.

UiO-66-NNDMA (\Cref{fig:Patchwork-NNDMA}-A) shows two distinct excitation features: an intense high-energy band centered near 340~nm and a broader, much weaker excitation contribution that extends across the measured spectral range with nearly constant intensity (see \Cref{fig:Patchwork-NNDMA}-A,B). In this sample the integrated EPR signal does not closely track the absorption profile and the slope in the high energy region is lower than in the case of the UiO-66-DPA, indicating the that the feature in the Excitation ($\lambda_{Excitation}$ = 425~nm) modulates the charge generation  (\Cref{fig:Patchwork-NNDMA}-D). 

Taken together, the DPA and NNDMA results reinforce the idea that only a restricted high-energy portion of the absorption manifold, mainly below 500~nm, is effective for generating persistent paramagnetic states. In light of the TDDFT results, this behavior is consistent with selective population of the productive charge-transfer (CT) manifold, whereas lower-energy absorption is more likely associated with excited states of predominantly localized-exciton (LE) character or with other relaxation channels that do not efficiently lead to long-lived charge separation \cite{kultaeva_mechanistic_2026}.

\section{Photocatalysis}

\begin{table}[htbp]
	\centering
	\caption{Absorption edge and methylene blue degradation for UiO-66-based samples. Degradation \% expressed relative to TiO$_2$ (reference = 100\%).}
	\label{tab:abs_edge_mb_deg}
	\small
	\setlength{\tabcolsep}{5pt}
	\renewcommand{\arraystretch}{1.15}
	\begin{tabular}{lccc}
		\toprule
		Sample name & Absorption edge (nm) & Absorption edge (eV) & Methylene blue degradation (\%) \\
		\midrule
		UiO-66         		& 335 & 3.70 & 42 \\
		UiO-66-NH$_2$  		& 432 & 2.87 & 63 \\
		Anisole        		& 496 & 2.50 & 97 \\
		$\alpha$-naphthol   & 639 & 1.94 & 58 \\
		$\beta$-naphthol    & 620 & 2.00 & 4  \\
		DFA            		& 725 & 1.71 & 91 \\
		NNDMA          		& 656 & 1.89 & 75 \\
		\bottomrule
	\end{tabular}
\end{table}

\begin{figure}[H]
	\centering
	% --- Subfigura B ---
	\begin{subfigure}[b]{0.75\linewidth}
		\centering
		\begin{overpic}[width=\linewidth]{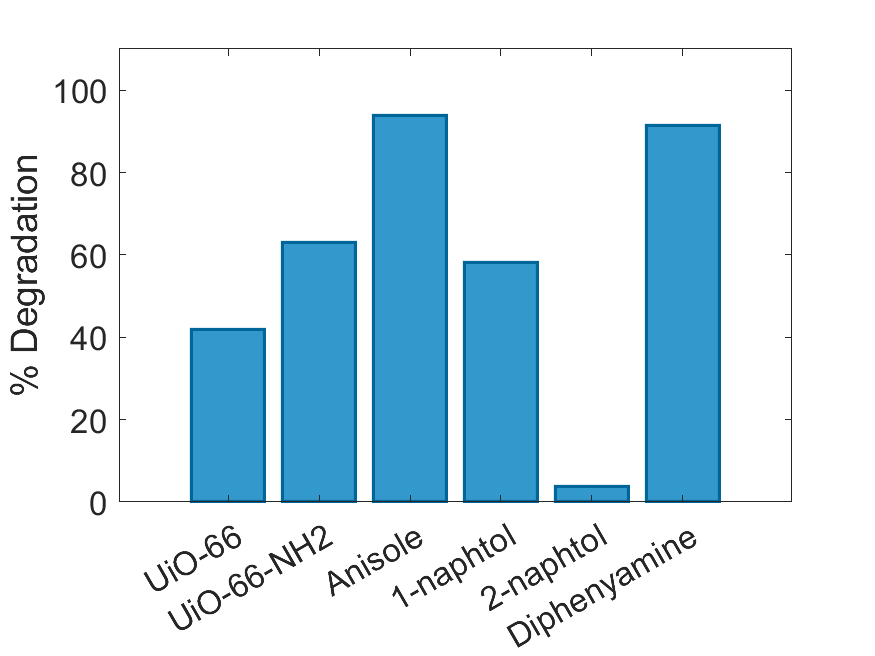}
			\put(0,65){\color{black}\textbf{ }}
		\end{overpic}
	\end{subfigure}
	\caption{Photocatalytic degradation of methylene blue (MB) under visible irradiation for the investigated MOF series, expressed relative to commercial anatase $\mathrm{TiO}_2$ reference (100\%).}
	\label{fig:PC-SI}
\end{figure}

% C:\Users\eugen\Dropbox\000My_Drafts\002-EPR-EMPA\photocat

\section{Tautomeric Equilibrium}

\begin{figure}[H]
	\centering
	% --- Subfigura B ---
	\begin{subfigure}[b]{0.75\linewidth}
		\centering
		\begin{overpic}[width=\linewidth]{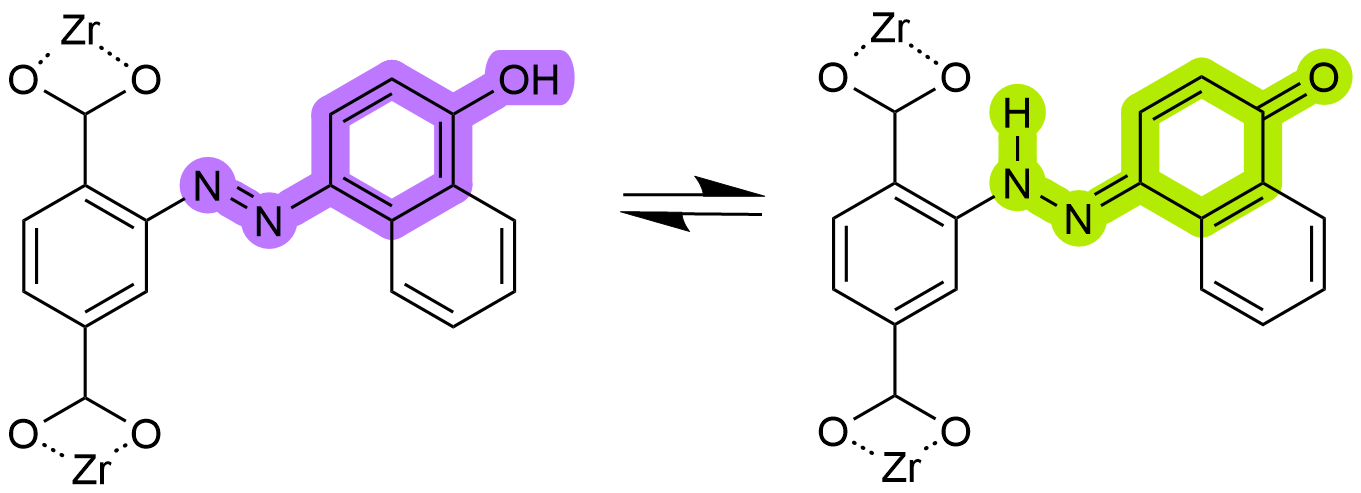}
			\put(-5,35){\color{black}\textbf{A1}}
			\put(55,35){\color{black}\textbf{A2}}
		\end{overpic}
	\end{subfigure}
	
	\vspace{1cm}
	
	\begin{subfigure}[b]{0.75\linewidth}
		\centering
		\begin{overpic}[width=\linewidth]{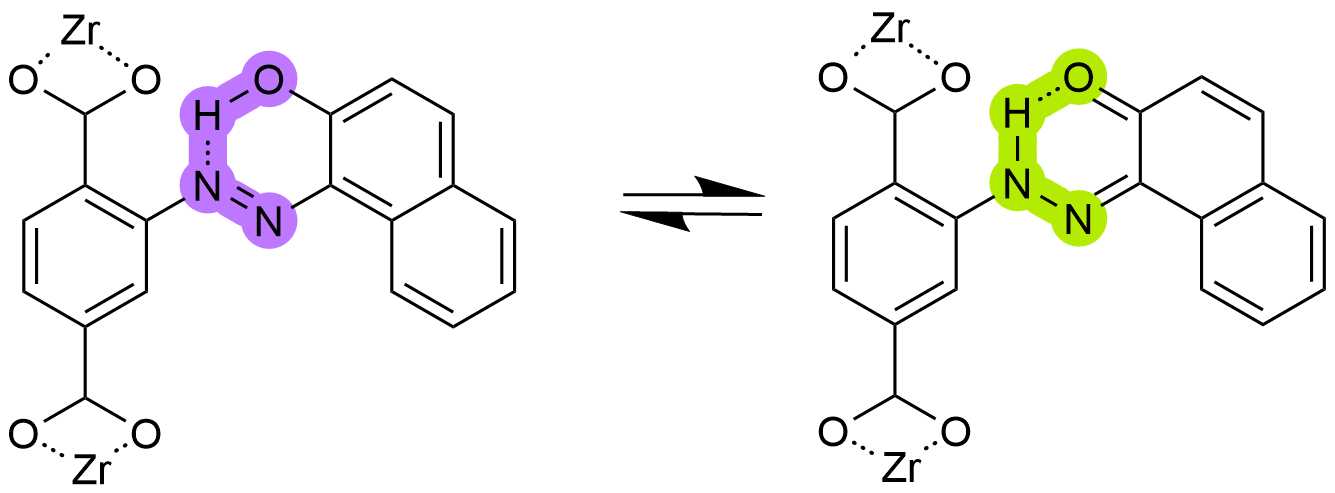}
			\put(-5,35){\color{black}\textbf{B1}}
			\put(55,35){\color{black}\textbf{B2}}
		\end{overpic}
	\end{subfigure}
	\caption{Keto and enol equilibrium in \textbf{(A)} UiO-66--AN (alpha-naphthol) and \textbf{(B)} UiO-66--BN (beta-naphthol). The purple trace corresponds to the enol conformer, and the green trace corresponds to the keto conformer.}
	\label{fig:keto_enol-SI}
\end{figure}

\section{Intrinsic radical formation}

\subsection{NO$^+$ formation} 
\label{sec:NOplus_formation}

In aqueous acidic media, nitrite is protonated to nitrous acid, which can undergo further protonation to the nitrous acidium ion (often written as $\mathrm{H_2ONO^+}$). This species readily dehydrates to yield nitrosonium ($\mathrm{NO^+}$), a highly reactive nitrosating and oxidizing electrophile. The key equilibria and the dehydration step can be summarized as:

\begin{equation}
	\mathrm{NO_2^- + H^+ \rightleftharpoons HNO_2} \qquad (pK_a \approx 3.4)
	\label{eq:HNO2_form}
\end{equation}

\begin{equation}
	\mathrm{HNO_2 + H^+ \rightleftharpoons H_2ONO^+} \qquad (pK_a \approx 1.7)
	\label{eq:H2ONO_form}
\end{equation}

\begin{equation}
	\mathrm{H_2ONO^+ \rightarrow NO^+ + H_2O} \qquad \text{(dehydration)}
	\label{eq:NOplus_form}
\end{equation}

These acid--base steps and the subsequent dehydration are commonly invoked to rationalize the presence of $\mathrm{NO^+}$ under strongly acidic conditions and its role in nitrosation chemistry.\cite{wu_transformation_2025}

\subsection{Anisole nitrosation and iminoxyl radical formation}
\label{sec:anisole_NO}

\begin{figure}[H]
	\centering
	\begin{subfigure}[b]{0.75\linewidth}
		\centering
		\begin{overpic}[width=\linewidth]{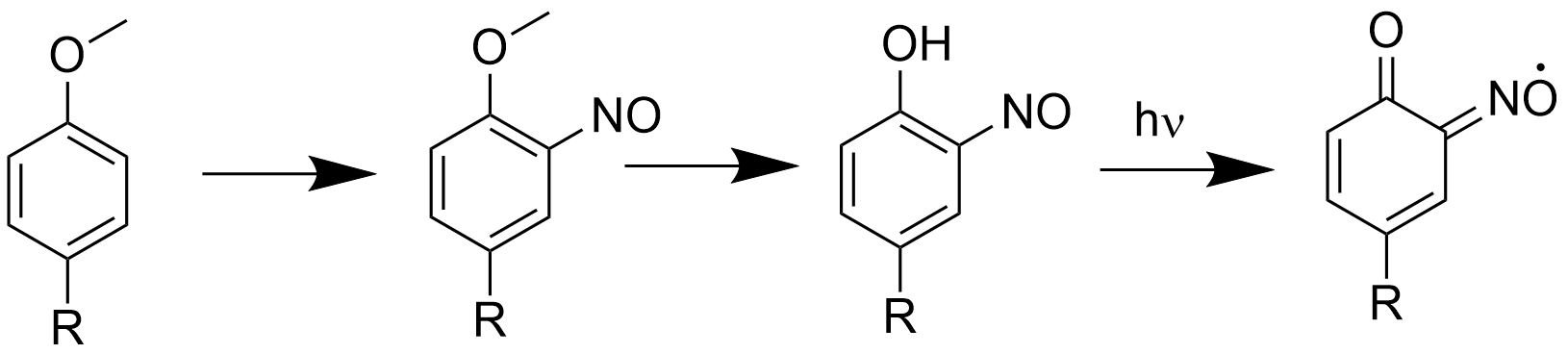}
			\put(-5,20){\color{black}\textbf{ }}
		\end{overpic}
	\end{subfigure}
	\caption{Proposed nitrosation and photoinduced radical-forming pathway of the anisole functionality in UiO-66--anisole. Electrophilic $C$-nitrosation by $\mathrm{NO^+}$ is followed by formation of a nitrosophenol/quinone-monoxime-type functionality and subsequent generation of an iminoxyl-type radical under irradiation. R represents the connection of the aromatic ring to the MOF framework.}
	\label{fig:anisole_nitrosation}
\end{figure}

The methoxy substituent of the anisole-derived linker strongly activates the aromatic ring toward electrophilic substitution. Therefore, the $\mathrm{NO^+}$ generated under acidic nitrite conditions (\Cref{sec:NOplus_formation}) can promote electrophilic aromatic nitrosation. In free anisole, nitrosation occurs preferentially at the \textit{para} position with respect to the methoxy group; however, in the present linker this position is occupied by the connection to the MOF, represented as R in \Cref{fig:anisole_nitrosation}. Consequently, nitrosation at an available \textit{ortho} position is a plausible pathway.

The proposed sequence is shown in \Cref{fig:anisole_nitrosation}. Initial electrophilic $C$-nitrosation gives an \textit{ortho}-nitrosoanisole-type functionality. Under the acidic reaction conditions, cleavage of the methyl ether can generate the corresponding nitrosophenol. Nitroso-substituted phenols are particularly relevant because their electronic structure can also be represented by the corresponding quinone-monoxime form. This conjugated motif provides a pathway toward stabilization of an unpaired electron over the $\mathrm{C=N\text{--}O}$--quinonoid framework.\cite{challis_chemistry_1971,atherton_nitrosation_2000}

Upon irradiation, photoinduced electron-transfer processes can lead to the formation of an iminoxyl-type radical, schematically represented in \Cref{fig:anisole_nitrosation}. The extensive delocalization of the unpaired electron over the nitroso/quinone-monoxime system increases the lifetime of the radical and favors its detection by EPR. The pathway presented here should be regarded as a mechanistically plausible explanation for the observed dark and photo-EPR response rather than an unambiguous identification of individual reaction intermediates.

\subsection{Aryloxyl radical formation via single-electron transfer}
\label{sec:NOplus_SET}

In addition to electrophilic nitrosation, $\mathrm{NO^+}$ can act as a one-electron oxidant toward sufficiently electron-rich aromatic species via single-electron transfer (SET). For a phenolic or naphtholic functionality, the overall process can be represented schematically as:

\begin{equation}
	\mathrm{ArOH + NO^+ \rightarrow ArO^\bullet + NO^\bullet + H^+}
	\label{eq:SET_ArOH}
\end{equation}

where reduction of $\mathrm{NO^+}$ produces nitric oxide while one-electron oxidation of the phenolic/naphtholic substrate generates an oxygen-centered aryloxyl radical ($\mathrm{ArO^\bullet}$). Equation~\eqref{eq:SET_ArOH} represents the overall redox balance and does not imply that electron transfer and proton transfer necessarily occur as fully separated elementary steps.

\subsection{Anilines nitrosation and radical formation}
\label{sec:nitrosation}

\begin{figure}[H]
	\centering
	\begin{subfigure}[b]{0.75\linewidth}
		\centering
		\begin{overpic}[width=\linewidth]{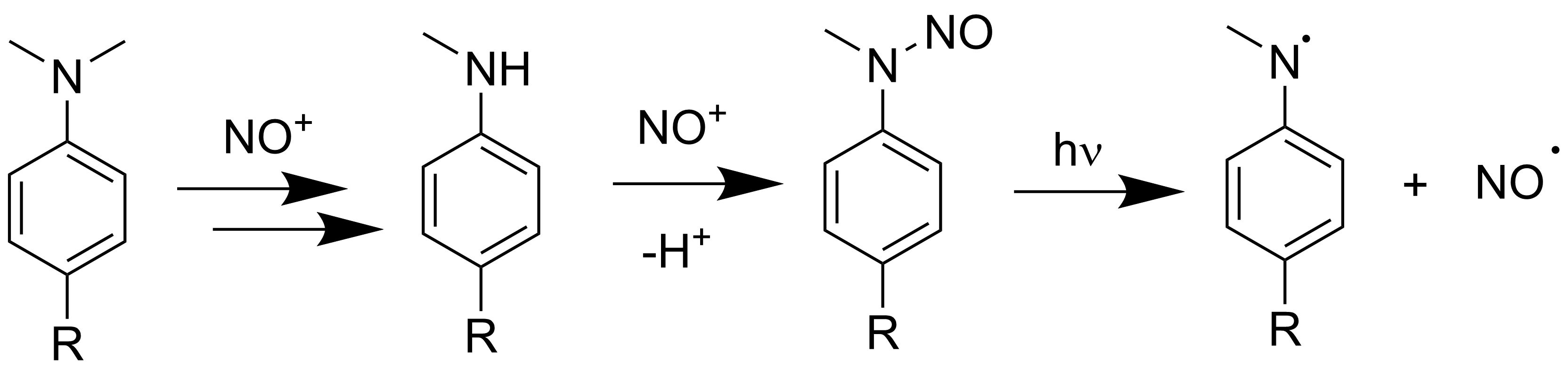}
			\put(-5,20){\color{black}\textbf{A}}
		\end{overpic}
	\end{subfigure}
	
	\vspace{1cm}
	
	\begin{subfigure}[b]{0.6\linewidth}
		\centering
		\begin{overpic}[width=\linewidth]{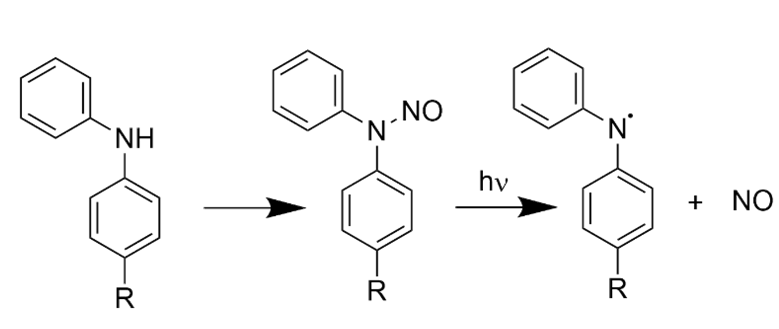}
			\put(-5,35){\color{black}\textbf{B}}
		\end{overpic}
	\end{subfigure}
	
	\caption{Proposed nitrosation and radical-forming pathways for \textbf{(A)} UiO-66--NNDMA and \textbf{(B)} UiO-66--DPA. For DPA, electrophilic $N$-nitrosation by $\mathrm{NO^+}$ produces an $N$-nitrosodiphenylamine intermediate, whose $\mathrm{N\text{--}N}$ bond can undergo photoinduced homolytic cleavage to yield a diphenylaminyl radical and $\mathrm{NO^\bullet}$. For a detailed discussion of the UiO-66--NNDMA pathway, see Ashworth et al.\cite{ashworth_consideration_2023}}
	\label{fig:aniline_nitrosation}
\end{figure}

A different radical-forming pathway is available for the amine-containing linkers. Secondary aromatic amines are readily nitrosated at nitrogen by nitrosonium, producing $N$-nitrosamines. This mechanism is particularly straightforward for the diphenylamine (DPA) functionality shown in \Cref{fig:aniline_nitrosation}B:

\begin{equation}
	\mathrm{Ar_2NH + NO^+ \rightarrow Ar_2N\text{--}NO + H^+}
	\label{eq:DPA_nitrosation}
\end{equation}

The resulting $N$-nitrosodiphenylamine contains a relatively weak $\mathrm{N\text{--}N}$ bond and constitutes a photochemically accessible radical precursor. Upon irradiation, homolytic cleavage of this bond can generate a diarylaminyl radical together with nitric oxide:

\begin{equation}
	\mathrm{Ar_2N\text{--}NO} \xrightarrow{h\nu} \mathrm{Ar_2N^\bullet + NO^\bullet}
	\label{eq:DPA_photolysis}
\end{equation}

providing a direct chemical connection between nitrite treatment, photoexcitation, and formation of the paramagnetic species detected by EPR.

For UiO-66--NNDMA (\Cref{fig:aniline_nitrosation}A), the chemistry is more complex because the tertiary amine functionality does not undergo direct $N$-nitrosation. Nitrosation and subsequent radical formation instead require an initial oxidative dealkylation step prior to formation of an $N$-nitroso intermediate, as described in detail by Ashworth et al.\cite{ashworth_consideration_2023}

\subsection{Definition and meaning of the overlap descriptors}
\label{sec:SExc_definition}

\subsection{Excitation-overlap descriptor}
\label{sec:SExc_SI}

The descriptor is defined from Eqs.~\eqref{eq:SExc_original_set_a}--\eqref{eq:SExc_original_set_c} and can be written in compact form as Eq.~\eqref{eq:SExc_combined}.

\begin{subequations}
	
	\label{eq:SExc_original_set}
	
	\begin{align}
		S_{\mathrm{Exc}} &= \frac{Abs \cap Exc}{Abs_{\mathrm{DRS}}}
		\label{eq:SExc_original_set_a} \\
		Abs \cap Exc &=
		\int_{\lambda_{\min}}^{\lambda_{\max}}
		\min\!\left[
		Exc_{\mathrm{norm}}(\lambda),
		Abs_{\mathrm{norm}}(\lambda)
		\right]\, d\lambda
		\label{eq:SExc_original_set_b} \\
		Abs_{\mathrm{DRS}} &=
		\int_{\lambda_{\min}}^{\lambda_{\max}}
		Abs_{\mathrm{norm}}(\lambda)\, d\lambda
		\label{eq:SExc_original_set_c}
	\end{align}
	
\end{subequations}

Substituting Eqs.~\eqref{eq:SExc_original_set_b} and \eqref{eq:SExc_original_set_c} into Eq.~\eqref{eq:SExc_original_set_a} gives

\begin{equation}
	S_{\mathrm{Exc}} =
	\frac{
		\displaystyle
		\int_{\lambda_{\min}}^{\lambda_{\max}}
		\min\!\left[
		Exc_{\mathrm{norm}}(\lambda),
		Abs_{\mathrm{norm}}(\lambda)
		\right]\, d\lambda
	}{
		\displaystyle
		\int_{\lambda_{\min}}^{\lambda_{\max}}
		Abs_{\mathrm{norm}}(\lambda)\, d\lambda
	}.
	\label{eq:SExc_combined}
\end{equation}

Since the numerator in Eq.~\eqref{eq:SExc_combined} is the common area between the two normalized spectra, $S_{\mathrm{Exc}}$ represents the fraction of the normalized absorption band that is covered by the normalized excitation profile within the interval $[\lambda_{\min},\lambda_{\max}]$.

\bibliography{references}

@misc{leone_il_1967,
	title = {Il buono, il brutto, il cattivo},
	publisher = {Produzioni Europee Associate (PEA), Arturo González Producciones Cinematográficas, Constantin Film},
	author = {Leone, Sergio},
	collaborator = {Vincenzoni, Luciano and Leone, Sergio and Incrocci, Agenore and Eastwood, Clint and Wallach, Eli and Cleef, Lee Van},
	month = dec,
	year = {1967},
	note = {Translated title: The Good, the Bad and the Ugly
	IMDb ID: tt0060196
	event-place: Italy, Spain, West Germany, United States},
}

@article{lakkaraju_epr_1994,
	title = {{EPR} {Spectra} of {Iminoxyl} {Radicals} in {Restricted} {Media}: {Direct} {Evidence} for the {Assignment} of {Z} and {E} {Isomers}},
	volume = {98},
	issn = {0022-3654},
	shorttitle = {{EPR} {Spectra} of {Iminoxyl} {Radicals} in {Restricted} {Media}},
	url = {https://doi.org/10.1021/j100062a002},
	doi = {10.1021/j100062a002},
	number = {11},
	urldate = {2026-04-03},
	journal = {The Journal of Physical Chemistry},
	publisher = {American Chemical Society},
	author = {Lakkaraju, Prasad S. and Zhang, Junxiong and Roth, Heinz D.},
	month = mar,
	year = {1994},
	pages = {2722--2725},
}

@article{long_amine-functionalized_2012,
	title = {Amine-functionalized zirconium metal–organic framework as efficient visible-light photocatalyst for aerobic organic transformations},
	volume = {48},
	issn = {1364-548X},
	url = {https://pubs.rsc.org/en/content/articlelanding/2012/cc/c2cc34620f},
	doi = {10.1039/C2CC34620F},
	language = {en},
	number = {95},
	urldate = {2023-08-29},
	journal = {Chemical Communications},
	publisher = {The Royal Society of Chemistry},
	author = {Long, Jinlin and Wang, Sibo and Ding, Zhengxin and Wang, Shuchao and Zhou, Yangen and Huang, Ling and Wang, Xuxu},
	month = nov,
	year = {2012},
	pages = {11656--11658},
}

@article{nasalevich_electronic_2016,
	title = {Electronic origins of photocatalytic activity in d0 metal organic frameworks},
	volume = {6},
	copyright = {2016 The Author(s)},
	issn = {2045-2322},
	url = {https://www.nature.com/articles/srep23676},
	doi = {10.1038/srep23676},
	language = {en},
	number = {1},
	urldate = {2023-08-29},
	journal = {Scientific Reports},
	publisher = {Nature Publishing Group},
	author = {Nasalevich, Maxim A. and Hendon, Christopher H. and Santaclara, Jara G. and Svane, Katrine and van der Linden, Bart and Veber, Sergey L. and Fedin, Matvey V. and Houtepen, Arjan J. and van der Veen, Monique A. and Kapteijn, Freek and Walsh, Aron and Gascon, Jorge},
	month = mar,
	year = {2016},
	note = {Number: 1},
	pages = {23676},
}

@article{linsebigler_photocatalysis_1995,
	title = {Photocatalysis on {TiO2} {Surfaces}: {Principles}, {Mechanisms}, and {Selected} {Results}},
	volume = {95},
	issn = {0009-2665},
	shorttitle = {Photocatalysis on {TiO2} {Surfaces}},
	url = {https://doi.org/10.1021/cr00035a013},
	doi = {10.1021/cr00035a013},
	number = {3},
	urldate = {2026-03-31},
	journal = {Chemical Reviews},
	publisher = {American Chemical Society},
	author = {Linsebigler, Amy L. and Lu, Guangquan and Yates, John T.  Jr.},
	month = may,
	year = {1995},
	pages = {735--758},
}

@article{cavka_new_2008,
	title = {A {New} {Zirconium} {Inorganic} {Building} {Brick} {Forming} {Metal} {Organic} {Frameworks} with {Exceptional} {Stability}},
	volume = {130},
	issn = {0002-7863},
	url = {https://doi.org/10.1021/ja8057953},
	doi = {10.1021/ja8057953},
	number = {42},
	urldate = {2025-10-31},
	journal = {Journal of the American Chemical Society},
	publisher = {American Chemical Society},
	author = {Cavka, Jasmina Hafizovic and Jakobsen, Søren and Olsbye, Unni and Guillou, Nathalie and Lamberti, Carlo and Bordiga, Silvia and Lillerud, Karl Petter},
	month = oct,
	year = {2008},
	pages = {13850--13851},
}

@article{ashworth_consideration_2023,
	title = {A {Consideration} of the {Extent} {That} {Tertiary} {Amines} {Can} {Form} {N}-{Nitroso} {Dialkylamines} in {Pharmaceutical} {Products}},
	volume = {27},
	issn = {1083-6160},
	url = {https://doi.org/10.1021/acs.oprd.3c00073},
	doi = {10.1021/acs.oprd.3c00073},
	number = {10},
	urldate = {2026-03-30},
	journal = {Organic Process Research \& Development},
	publisher = {American Chemical Society},
	author = {Ashworth, Ian W. and Curran, Timothy and Dirat, Olivier and Zheng, Jinjian and Whiting, Matthew and Lee, Daniel},
	month = oct,
	year = {2023},
	pages = {1714--1718},
}

@article{wu_transformation_2025,
	title = {The transformation of 1-naphthol in the presence of nitrite in ice},
	volume = {285},
	issn = {0013-9351},
	url = {https://www.sciencedirect.com/science/article/pii/S0013935125018250},
	doi = {10.1016/j.envres.2025.122573},
	urldate = {2026-03-30},
	journal = {Environmental Research},
	author = {Wu, Kai and Wang, Wenlong and Yang, Shubing and Zhang, Xu and Yang, Peizeng and Ji, Yuefei and Chen, Jing and Lu, Junhe},
	month = nov,
	year = {2025},
	pages = {122573},
}

@article{kultaeva_mechanistic_2026,
	title = {Mechanistic origin of charge separation and enhanced photocatalytic activity in {D}–pi–{A}-functionalized {UiO}-66-{NH2} {MOFs}},
	issn = {2051-6355},
	url = {https://pubs.rsc.org/en/content/articlelanding/2026/mh/d5mh02418h},
	doi = {10.1039/D5MH02418H},
	language = {en},
	urldate = {2026-05-28},
	journal = {Materials Horizons},
	publisher = {The Royal Society of Chemistry},
	author = {Kultaeva, Anastasiia and Vasylkovskyi, Volodymyr and Sperlich, Andreas and Otal, Eugenio and Teshima, Katsuya and Schmidt, Wolf Gero and Biktagirov, Timur},
	month = apr,
	year = {2026},
}

@article{alif_photochemistry_1991,
	title = {Photochemistry and environment {XIII}: {Phototransformation} of 2-nitrophenol in aqueous solution},
	volume = {59},
	issn = {1010-6030},
	shorttitle = {Photochemistry and environment {XIII}},
	url = {https://www.sciencedirect.com/science/article/pii/101060309187009K},
	doi = {10.1016/1010-6030(91)87009-K},
	number = {2},
	urldate = {2026-02-25},
	journal = {Journal of Photochemistry and Photobiology A: Chemistry},
	author = {Alif, Admin and Pilichowski, Jean-François and Boule, Pierre},
	month = jul,
	year = {1991},
	pages = {209--219},
}

@article{sturgeon_tyrosine_2001,
	title = {Tyrosine {Iminoxyl} {Radical} {Formation} from {Tyrosyl} {Radical}/{Nitric} {Oxide} and {Nitrosotyrosine}*},
	volume = {276},
	issn = {0021-9258},
	url = {https://www.sciencedirect.com/science/article/pii/S0021925819373326},
	doi = {10.1074/jbc.M106835200},
	number = {49},
	urldate = {2026-02-25},
	journal = {Journal of Biological Chemistry},
	author = {Sturgeon, Bradley E. and Glover, Richard E. and Chen, Yeong-Renn and Burka, Leo T. and Mason, Ronald P.},
	month = dec,
	year = {2001},
	pages = {45516--45521},
}

@article{krylov_oxime_2020,
	title = {Oxime radicals: generation, properties and application in organic synthesis},
	volume = {16},
	copyright = {© 2020 Krylov et al.; licensee Beilstein-Institut.},
	issn = {1860-5397},
	shorttitle = {Oxime radicals},
	url = {https://www.beilstein-journals.org/bjoc/articles/16/107},
	doi = {10.3762/bjoc.16.107},
	language = {en},
	number = {1},
	urldate = {2026-02-25},
	journal = {Beilstein Journal of Organic Chemistry},
	publisher = {Beilstein-Institut},
	author = {Krylov, Igor B. and Paveliev, Stanislav A. and Budnikov, Alexander S. and Terent’ev, Alexander O.},
	month = jun,
	year = {2020},
	pages = {1234--1276},
}

@article{cudic_transformations_2000,
	title = {Transformations of 2,6-{Diisopropylphenol} by {NO}-{Derived} {Nitrogen} {Oxides}, {Particularly} {Peroxynitrite}},
	volume = {4},
	issn = {1089-8603},
	url = {https://www.sciencedirect.com/science/article/pii/S1089860300902775},
	doi = {10.1006/niox.2000.0277},
	number = {2},
	urldate = {2026-02-25},
	journal = {Nitric Oxide},
	author = {Cudic, Mare and Ducrocq, Claire},
	month = apr,
	year = {2000},
	pages = {147--156},
}

@article{challis_chemistry_1971,
	title = {The chemistry of nitroso-compounds. {Part} {II}. {The} nitrosation of phenol and anisole},
	issn = {0045-6470},
	url = {https://pubs.rsc.org/en/content/articlelanding/1971/j2/j29710000770},
	doi = {10.1039/J29710000770},
	language = {en},
	number = {0},
	urldate = {2026-02-25},
	journal = {Journal of the Chemical Society B: Physical Organic},
	publisher = {The Royal Society of Chemistry},
	author = {Challis, B. C. and Lawson, A. J.},
	month = jan,
	year = {1971},
	pages = {770--775},
}

@article{aksungur_photophysical_2015,
	title   = {Photophysical and theoretical studies on newly synthesized \textit{N},\textit{N}-diphenylamine based azo dye},
	journal = {Journal of Molecular Structure},
	volume  = {1099},
	pages   = {543--550},
	year    = {2015},
	month   = nov,
	issn    = {0022-2860},
	doi     = {10.1016/j.molstruc.2015.07.010},
	url     = {https://www.sciencedirect.com/science/article/pii/S0022286015301216},
	urldate = {2026-02-24},
	author  = {Aksungur, Tu{\u g}\c{c}e and Arslan, {\"O}mer and Sefero{\u g}lu, Nurg{\"u}l and Sefero{\u g}lu, Zeynel}
}

@article{karaki_visible-light-triggered_2012,
	title = {Visible-{Light}-{Triggered} {Release} of {Nitric} {Oxide} from {N}-{Pyramidal} {Nitrosamines}},
	volume = {18},
	copyright = {Copyright © 2012 WILEY-VCH Verlag GmbH \& Co. KGaA, Weinheim},
	issn = {1521-3765},
	url = {https://onlinelibrary.wiley.com/doi/abs/10.1002/chem.201101427},
	doi = {10.1002/chem.201101427},
	language = {en},
	number = {4},
	urldate = {2026-02-24},
	journal = {Chemistry – A European Journal},
	author = {Karaki, Fumika and Kabasawa, Yoji and Yanagimoto, Takahiro and Umeda, Nobuhiro and {Firman} and Urano, Yasuteru and Nagano, Tetsuo and Otani, Yuko and Ohwada, Tomohiko},
	year = {2012},
	note = {\_eprint: https://chemistry-europe.onlinelibrary.wiley.com/doi/pdf/10.1002/chem.201101427},
	pages = {1127--1141},
}

@article{he_ring-restricted_2018,
	series = {{SI}: {Fluorescence} imaging},
	title = {Ring-restricted \textit{{N}}-nitrosated rhodamine as a green-light triggered, orange-emission calibrated and fast-releasing nitric oxide donor},
	volume = {29},
	issn = {1001-8417},
	url = {https://www.sciencedirect.com/science/article/pii/S1001841718303450},
	doi = {10.1016/j.cclet.2018.08.019},
	number = {10},
	urldate = {2026-02-24},
	journal = {Chinese Chemical Letters},
	author = {He, Haihong and He, Tingting and Zhang, Ziqian and Xu, Xiu and Yang, Huibin and Qian, Xuhong and Yang, Youjun},
	month = oct,
	year = {2018},
	pages = {1497--1499},
}

@article{geiger_photodissociation_1981,
	title = {Photodissociation of dimethylnitrosamine},
	volume = {79},
	issn = {0009-2614},
	url = {https://www.sciencedirect.com/science/article/pii/0009261481850294},
	doi = {10.1016/0009-2614(81)85029-4},
	number = {3},
	urldate = {2026-02-24},
	journal = {Chemical Physics Letters},
	author = {Geiger, G. and Stafast, H. and Brühlmann, U. and Huber, J. Robert},
	month = may,
	year = {1981},
	pages = {525--528},
}

@article{bamford_3_1939,
	title = {3. {A} study of the photolysis of organic nitrogen compounds. {Part} {I}. {Dimethyl}- and diethyl-nitrosoamines},
	issn = {0368-1769},
	url = {https://pubs.rsc.org/en/content/articlelanding/1939/jr/jr9390000012},
	doi = {10.1039/JR9390000012},
	language = {en},
	number = {0},
	urldate = {2026-02-24},
	journal = {Journal of the Chemical Society (Resumed)},
	publisher = {The Royal Society of Chemistry},
	author = {Bamford, C. H.},
	month = jan,
	year = {1939},
	pages = {12--17},
}

@article{gowenlock_nitrosative_1979,
	title = {Nitrosative dealkylation of some symmetrical tertiary amines},
	issn = {1364-5471},
	url = {https://pubs.rsc.org/en/content/articlelanding/1979/p2/p29790001110},
	doi = {10.1039/P29790001110},
	language = {en},
	number = {8},
	urldate = {2026-02-24},
	journal = {Journal of the Chemical Society, Perkin Transactions 2},
	publisher = {The Royal Society of Chemistry},
	author = {Gowenlock, Brian G. and Hutchison, Roderick J. and Little, Janet and Pfab, Josef},
	month = jan,
	year = {1979},
	pages = {1110--1114},
}

@article{smith_nitrosative_1967,
	title = {Nitrosative {Cleavage} of {Tertiary} {Amines}},
	volume = {89},
	issn = {0002-7863},
	url = {https://doi.org/10.1021/ja00981a021},
	doi = {10.1021/ja00981a021},
	number = {5},
	urldate = {2026-02-24},
	journal = {Journal of the American Chemical Society},
	publisher = {American Chemical Society},
	author = {Smith, Peter A. S. and Loeppky, Richard N.},
	month = mar,
	year = {1967},
	pages = {1147--1157},
}

@article{garcia_generation_2001,
	title = {Generation and conversions of aromatic amine radical cations in acid zeolites},
	volume = {3},
	issn = {1463-9084},
	url = {https://pubs.rsc.org/en/content/articlelanding/2001/cp/b101383l},
	doi = {10.1039/B101383L},
	language = {en},
	number = {14},
	urldate = {2026-02-24},
	journal = {Physical Chemistry Chemical Physics},
	publisher = {The Royal Society of Chemistry},
	author = {García, Hermenegildo and Martí, Vicente and Casades, Isabel and Fornés, Vicente and Roth, Heinz D.},
	month = jan,
	year = {2001},
	pages = {2955--2960},
}

@article{katz_facile_2013,
	title = {A facile synthesis of {UiO}-66, {UiO}-67 and their derivatives},
	volume = {49},
	issn = {1364-548X},
	url = {https://pubs.rsc.org/en/content/articlelanding/2013/cc/c3cc46105j},
	doi = {10.1039/C3CC46105J},
	language = {en},
	number = {82},
	urldate = {2023-08-24},
	journal = {Chemical Communications},
	author = {Katz, Michael J. and Brown, Zachary J. and Colón, Yamil J. and Siu, Paul W. and Scheidt, Karl A. and Snurr, Randall Q. and Hupp, Joseph T. and Farha, Omar K.},
	month = sep,
	year = {2013},
	note = {Publisher: The Royal Society of Chemistry},
	pages = {9449--9451},
}

@article{joshi_temperature_2001,
	title = {Temperature dependent absorption spectroscopy of some tautomeric azo dyes and {Schiff} bases},
	issn = {1364-5471},
	url = {https://pubs.rsc.org/en/content/articlelanding/2001/p2/b106241g},
	doi = {10.1039/B106241G},
	language = {en},
	number = {12},
	urldate = {2026-02-20},
	journal = {Journal of the Chemical Society, Perkin Transactions 2},
	author = {Joshi, Hem and Kamounah, Fadhil S. and Zwan, Gert van der and Gooijer, Cees and Antonov, Liudmil},
	month = nov,
	year = {2001},
	note = {Publisher: The Royal Society of Chemistry},
	pages = {2303--2308},
}

@article{roys_efficiency_2025,
	title = {Efficiency {Trends} in {Metal}-{Free} {Heterocyclic} {Sensitizers} in {Dye}-{Sensitized} {Solar} {Cell} {Application}},
	volume = {8},
	url = {https://doi.org/10.1021/acsaem.5c02470},
	doi = {10.1021/acsaem.5c02470},
	number = {22},
	urldate = {2026-02-20},
	journal = {ACS Applied Energy Materials},
	author = {Roys, Krupa Elsa and Manju, S L},
	month = nov,
	year = {2025},
	note = {Publisher: American Chemical Society},
	pages = {16376--16415},
}

@article{haessner_1h-nmr-spektroskopische_1985,
	title = {{1H}-{NMR}-spektroskopische {Untersuchungen} zur {Azo}-{Hydrazon}-{Tautomerie} in substituierten 1-{Phenylazo}-2-naphtholen},
	volume = {327},
	issn = {1521-3897},
	url = {https://onlinelibrary.wiley.com/doi/abs/10.1002/prac.19853270405},
	doi = {10.1002/prac.19853270405},
	language = {de},
	number = {4},
	urldate = {2026-02-20},
	journal = {Journal für Praktische Chemie},
	author = {Haessner, R. and Mustroph, H. and Borsdorf, R.},
	year = {1985},
	note = {\_eprint: https://onlinelibrary.wiley.com/doi/pdf/10.1002/prac.19853270405},
	pages = {555--566},
}

@article{gegiou_emission_1971,
	title = {The emission properties of ortho-phenylazo-phenols and naphthols},
	volume = {10},
	issn = {0009-2614},
	url = {https://www.sciencedirect.com/science/article/pii/0009261471804347},
	doi = {10.1016/0009-2614(71)80434-7},
	number = {2},
	urldate = {2026-02-20},
	journal = {Chemical Physics Letters},
	author = {Gegiou, D. and Fischer, E.},
	month = jul,
	year = {1971},
	pages = {99--101},
}

@article{tang_evaluation_2024,
	title = {Evaluation of {Key} {Intermediates} in {Azo} {Dye} {Degradation} by {Advanced} {Oxidation} {Processes}: {Comparing} {Anilines} and {Phenols}},
	volume = {4},
	shorttitle = {Evaluation of {Key} {Intermediates} in {Azo} {Dye} {Degradation} by {Advanced} {Oxidation} {Processes}},
	url = {https://doi.org/10.1021/acsestwater.4c00532},
	doi = {10.1021/acsestwater.4c00532},
	number = {11},
	urldate = {2026-02-20},
	journal = {ACS ES\&T Water},
	author = {Tang, Zhengkun and Xu, Chenye and Shen, Chensi and Sun, Songmei and Meng, Xiang-Zhou and Wang, Shaoxian and Li, Fang},
	month = nov,
	year = {2024},
	note = {Publisher: American Chemical Society},
	pages = {4872--4880},
}

@article{taboada-puig_activation_2013,
	title = {Activation of {Kraft} {Lignin} by an {Enzymatic} {Treatment} with a {Versatile} {Peroxidase} from {Bjerkandera} sp. {R1}},
	volume = {169},
	issn = {1559-0291},
	url = {https://doi.org/10.1007/s12010-012-0023-z},
	doi = {10.1007/s12010-012-0023-z},
	language = {en},
	number = {4},
	urldate = {2026-02-20},
	journal = {Applied Biochemistry and Biotechnology},
	author = {Taboada-Puig, R. and Lú-Chau, T. A. and Moreira, M. T. and Feijoo, G. and Lema, J. M.},
	month = feb,
	year = {2013},
	pages = {1262--1278},
}

@article{biktagirov_unveiling_2025,
	title = {Unveiling {Linker}-{Born} {Electron} {Spin} {Centers} in {UiO}-66-{NH2} {MOF}},
	issn = {1932-7447},
	url = {https://doi.org/10.1021/acs.jpcc.5c06905},
	doi = {10.1021/acs.jpcc.5c06905},
	urldate = {2025-12-16},
	journal = {The Journal of Physical Chemistry C},
	author = {Biktagirov, Timur and Otal, Eugenio and Trost, Leopold and Dörflinger, Patrick and Teshima, Katsuya and Trukhina, Olga and Klose, Daniel and Schmidt, Wolf Gero and Kultaeva, Anastasiia},
	month = dec,
	year = {2025},
	note = {Publisher: American Chemical Society},
}

@article{otal_panchromatic_2016,
	title = {A panchromatic modification of the light absorption spectra of metal–organic frameworks},
	volume = {52},
	issn = {1364-548X},
	url = {https://pubs.rsc.org/en/content/articlelanding/2016/cc/c6cc02319c},
	doi = {10.1039/C6CC02319C},
	language = {en},
	number = {40},
	urldate = {2022-07-14},
	journal = {Chemical Communications},
	author = {Otal, E. H. and Kim, M. L. and Calvo, M. E. and Karvonen, L. and Fabregas, I. O. and Sierra, C. A. and Hinestroza, J. P.},
	month = may,
	year = {2016},
	note = {Publisher: The Royal Society of Chemistry},
	pages = {6665--6668},
}

@article{ahn_fluorescence_nodate,
	title = {Fluorescence {Modulation} in {UiO}-66-{NH2} via {Photooxidation} and {Selective} {Reduction}},
	volume = {10},
	issn = {2470-1343},
	url = {https://pmc.ncbi.nlm.nih.gov/articles/PMC12461390/},
	doi = {10.1021/acsomega.5c05521},
	number = {37},
	urldate = {2025-11-14},
	journal = {ACS Omega},
	author = {Ahn, Yongdeok and Park, Minsoo and Son, Younghu and Cho, Juhyeong and Park, Min Jeong and Seo, Daeha and Yoon, Minyoung},
	pmid = {41018611},
	pmcid = {PMC12461390},
	pages = {42882--42891},
}

@article{Hagfeldt2010,
	title = {Dye-{Sensitized} {Solar} {Cells}},
	volume = {110},
	issn = {0009-2665},
	url = {https://doi.org/10.1021/cr900356p},
	doi = {10.1021/cr900356p},
	number = {11},
	urldate = {2025-11-18},
	journal = {Chemical Reviews},
	author = {Hagfeldt, Anders and Boschloo, Gerrit and Sun, Licheng and Kloo, Lars and Pettersson, Henrik},
	month = nov,
	year = {2010},
	note = {Publisher: American Chemical Society},
	pages = {6595--6663},
}

@article{ORegan1991,
	author    = {O'Regan, Brian},
	title     = {Trap states in nanoporous TiO$_2$ electrodes},
	journal   = {J. Photochem. Photobiol. A},
	year      = {1991},
	volume    = {58},
	pages     = {223--229},
	doi       = {10.1016/1010-6030(91)80022-8}
}

@article{Boschloo2009,
	author    = {Boschloo, Göran and Hagfeldt, Anders},
	title     = {Stability and Recombination in Dye-Sensitized Solar Cells},
	journal   = {Acc. Chem. Res.},
	year      = {2009},
	volume    = {42},
	number    = {11},
	pages     = {1819--1828},
	doi       = {10.1021/ar900055m}
}

@article{Eddaoudi2002,
	title = {Systematic {Design} of {Pore} {Size} and {Functionality} in {Isoreticular} {MOFs} and {Their} {Application} in {Methane} {Storage}},
	volume = {295},
	url = {https://www.science.org/doi/10.1126/science.1067208},
	doi = {10.1126/science.1067208},
	number = {5554},
	urldate = {2025-11-18},
	journal = {Science},
	author = {Eddaoudi, Mohamed and Kim, Jaheon and Rosi, Nathaniel and Vodak, David and Wachter, Joseph and O'Keeffe, Michael and Yaghi, Omar M.},
	month = jan,
	year = {2002},
	note = {Publisher: American Association for the Advancement of Science},
	pages = {469--472},
}

@article{Wang2018,
	author    = {Wang, Chang and Lin, Wenbin},
	title     = {MOF-Catalyzed Photochemical Reactions},
	journal   = {Nat. Rev. Chem.},
	year      = {2018},
	volume    = {2},
	pages     = {65--81},
	doi       = {10.1038/s41570-018-0062-6}
}

@article{Cohen2012,
title = {Postsynthetic {Methods} for the {Functionalization} of {Metal}–{Organic} {Frameworks}},
volume = {112},
issn = {0009-2665},
url = {https://doi.org/10.1021/cr200179u},
doi = {10.1021/cr200179u},
number = {2},
urldate = {2025-10-31},
journal = {Chemical Reviews},
author = {Cohen, Seth M.},
month = feb,
year = {2012},
note = {Publisher: American Chemical Society},
pages = {970--1000},
}

@article{KalajCohen2020,
title = {Postsynthetic {Modification}: {An} {Enabling} {Technology} for the {Advancement} of {Metal}–{Organic} {Frameworks}},
volume = {6},
issn = {2374-7943},
shorttitle = {Postsynthetic {Modification}},
url = {https://doi.org/10.1021/acscentsci.0c00690},
doi = {10.1021/acscentsci.0c00690},
number = {7},
urldate = {2025-10-15},
journal = {ACS Central Science},
author = {Kalaj, Mark and Cohen, Seth M.},
month = jul,
year = {2020},
note = {Publisher: American Chemical Society},
pages = {1046--1057},
}

@article{Mondal2023,
	author    = {Mondal, Sahil and Mondal, Abhishek and Banerjee, Rahul},
	title     = {Reversible Postsynthetic Modification in a Metal–Organic Framework},
	journal   = {Angew. Chem. Int. Ed.},
	year      = {2023},
	volume    = {62},
	number    = {10},
	pages     = {e202213456},
	doi       = {10.1002/anie.202213456}
}

@article{Zhang2015,
	author    = {Zhang, Honglin and Kaye, Steven S. and Snurr, Randall Q.},
	title     = {Investigation of Ti-Based MOFs for Photocatalytic Water Splitting},
	journal   = {J. Am. Chem. Soc.},
	year      = {2015},
	volume    = {137},
	number    = {5},
	pages     = {1428--1434},
	doi       = {10.1021/ja511334u}
}

@article{Karimi2017,
	author    = {Karimi, Bita and Civalleri, Bruno and García, Armand and et al.},
	title     = {Charge Separation in Ti-MOFs Studied by EPR Spectroscopy},
	journal   = {J. Mater. Chem. A},
	year      = {2017},
	volume    = {5},
	number    = {12},
	pages     = {6046--6054},
	doi       = {10.1039/C6TA09854J}
}

@article{atherton_nitrosation_2000,
	title = {Nitrosation of anisole, stability constants of complexes of the nitrosonium ion with aromatic nitroso compounds, and {NMR} studies of restricted rotation in the complexes},
	volume = {2000},
	issn = {1472-779X},
	doi = {10.1039/A908441J},
	number = {4},
	journal = {J. Chem. Soc., Perkin Trans. 2},
	author = {Atherton, John H. and Moodie, Roy B. and Noble, Darren R.},
	year = {2000},
	pages = {699--705},
}

\end{document}